\documentclass[a4paper,11pt]{article}
\pdfoutput=1
\usepackage{jheppub}
\usepackage{amsthm}
\usepackage{amsmath}
\usepackage{graphicx}
\usepackage{multirow}
\usepackage{tabularx}
\usepackage{makecell}
\usepackage{booktabs}
\usepackage{wrapfig}
\usepackage{subcaption}
\usepackage{slashed}
\usepackage{amssymb}
\usepackage{enumitem}
\usepackage{braket}
\usepackage{bbm}
\usepackage{bm}
\usepackage{physics}
\usepackage{comment}

\usepackage{framed}
\definecolor{shadecolor}{rgb}{0.95,0.95,0.95}

\newcommand*\diff{\mathop{}\!\mathrm{d}}

\newcommand{\nn}{\nonumber}

\newcommand{\be}{\begin{eqnarray}}
\newcommand{\ee}{\end{eqnarray}}
\newcommand{\ma}{\mathrm}
\newcommand{\ml}{\mathcal}
\newcommand{\bs}{\boldsymbol}

\def\O{{\mathcal O}}

\def\p{{\bs p}}
\def\q{{\bs q}}
\def\k{{\bs k}}

\def\D{{\mathbb{D}}}

\begin{document}
\title{Non-Abelian Electric Field Correlator at NLO for Dark Matter Relic Abundance and Quarkonium Transport}

\author[1]{Tobias Binder}
\affiliation[1]{Kavli IPMU (WPI), UTIAS, The University of Tokyo, Kashiwa, Chiba 277-8583, Japan}

\author[2,3]{Kyohei Mukaida}
\affiliation[2]{Theory Center, IPNS, KEK, 1-1 Oho, Tsukuba, Ibaraki 305-0801, Japan}
\affiliation[3]{Graduate University for Advanced Studies (Sokendai), 1-1 Oho, Tsukuba, Ibaraki 305-0801, Japan}

\author[4]{Bruno Scheihing-Hitschfeld}
\affiliation[4]{Center for Theoretical Physics, Massachusetts Institute of Technology, Cambridge, MA 02139, USA}

\author[4]{Xiaojun Yao}

\emailAdd{tobias.binder@ipmu.jp}
\emailAdd{kyohei.mukaida@kek.jp}
\emailAdd{bscheihi@mit.edu}
\emailAdd{xjyao@mit.edu}

\preprint{KEK-TH-2334, MIT-CTP/5313}

\abstract{We perform a complete next-to-leading order calculation of the non-Abelian electric field correlator in a SU($N_c$) plasma, which encodes properties of the plasma relevant for heavy particle bound state formation and dissociation and is different from the correlator for the heavy quark diffusion coefficient.
The calculation is carried out in the real-time formalism of thermal field theory and includes both vacuum and finite temperature contributions. By working in the $R_\xi$ gauge, we explicitly show the results are gauge independent, infrared and collinear safe. The renormalization group equation of this electric field correlator is determined by that of the strong coupling constant.
Our next-to-leading order calculation can be directly applied to any dipole singlet-adjoint transition of heavy particle pairs. For example, it can be used to describe dissociation and (re)generation of heavy quarkonia inside the quark-gluon plasma well below the melting temperature, as well as heavy dark matter pairs (or charged co-annihilating partners) in the early universe.}

\maketitle
\flushbottom

\section{Introduction}
\label{sect:intro}

Pairs of heavy particles inside a relativistic plasma are intriguing systems from both theoretical and experimental points of view. One interesting example is heavy quarkonium (bound state of a heavy quark-antiquark pair) production in relativistic heavy ion collisions, where a quark-gluon plasma (QGP) near thermal equilibrium is produced shortly after the collision. Currently, two relativistic heavy ion collision experiments are operating to study properties of the QGP. One is the Relativistic Heavy ion Collider (RHIC) at Brookhaven National Laboratory and the other is the Large Hadron Collider (LHC) at the European Organization for Nuclear Research (CERN). Both experiments have observed suppression of quarkonium production in heavy ion collisions, compared with that in proton-proton collisions~\cite{Adare:2011yf,Abelev:2012rv,Khachatryan:2016xxp,Aaboud:2018quy,Adam:2019rbk}. Most of the suppression is originated from the thermal medium effects, which include plasma screening effects~\cite{Matsui:1986dk,Karsch:1987pv}, medium-induced dissociation~\cite{Laine:2006ns,Beraudo:2007ky,Brambilla:2008cx} and (re)generation~\cite{Thews:2000rj,Andronic:2003zv,Andronic:2007bi}. Recent reviews can be found in Refs.~\cite{Mocsy:2013syh, Rothkopf:2019ipj,Akamatsu:2020ypb,Chapon:2020heu,Sharma:2021vvu,Yao:2021lus,Muller:2021ygo}.

A further example of such a system is the production of hypothetical dark matter (DM) particles in the early universe plasma. In analogy to quarkonia, long-range interactions between DM pairs (or co-annihilating partners) can lead to the existence of metastable bound states. Prominent examples where this is indeed the case are electroweak and colored coannihilation scenarios~\cite{Hisano:2006nn,Ellis:2015vna,Liew:2016hqo,Asadi:2016ybp,Johnson:2016sjs,Kim:2016zyy,Mitridate:2017izz,Harz:2017dlj,Harz:2018csl,Biondini:2018pwp,Biondini:2018ovz,Biondini:2018xor,Fukuda:2018ufg,Harz:2019rro,Biondini:2019int,Beneke:2020vff}, and self-interacting DM~\cite{Buckley:2009in,Aarssen:2012fx,Tulin:2013teo,Pearce:2013ola,Petraki:2014uza,Pearce:2015zca,Bringmann:2016din,An:2016gad,Cirelli:2016rnw,Baldes:2017gzw,Baldes:2017gzu,Kamada:2018zxi,Kamada:2018kmi,Matsumoto:2018acr,Kamada:2019jch,Ko:2019wxq,Geller:2018biy}. Bound state effects make it challenging to predict the DM relic abundance on a precise level that matches with the observational percent accuracy~\cite{Ade:2015xua}. Medium-induced corrections to formation and dissociation have been studied for an Abelian gauge theory in Ref.~\cite{Binder:2020efn}. 

A precise theoretical understanding of the bound state formation and dissociation in a plasma is important for studying quarkonium production in heavy ion collisions and predicting the dark matter thermal relic abundance. In this work, we consider medium-induced bound state formation and dissociation in non-Abelian gauge theories. In order to compute these processes precisely in a proper thermal field theoretical approach, we apply the open quantum system framework, which has been intensively used to investigate quarkonium transport~\cite{Akamatsu:2011se,Akamatsu:2012vt,Akamatsu:2014qsa,Blaizot:2015hya,Katz:2015qja,Brambilla:2016wgg,Brambilla:2017zei,Kajimoto:2017rel,Yao:2018nmy,Blaizot:2018oev,Akamatsu:2018xim,Miura:2019ssi,Sharma:2019xum,Yao:2020eqy,Brambilla:2020qwo} and jet quenching~\cite{Vaidya:2020cyi,Vaidya:2020lih,Vaidya:2021mjl} in the QGP. In particular, we consider a subsystem consisting of a pair of heavy particles, interacting weakly with a thermal environment, i.e., the hot plasma. We assume the size of the heavy particle pair is much smaller than the inverse of the environment temperature.
With the environment traced out, the time evolution of the subsystem is non-unitary and can be written as a Lindblad equation in certain limits. The Lindblad equation can be transformed into a Boltzmann equation by making further simplifications, which were reviewed in Ref.~\cite{Yao:2021lus} and will be briefly explained here later.
As we shall see, the resulting bound state formation and dissociation terms in the Boltzmann equation depend on the plasma property encoded by a (non-Abelian) \emph{electric field correlator}, {given by
\begin{align} 
\label{eq:correlator-intro}
\left\langle {E}^a_i(t) \ml{W}^{ab}(t,0)  {E}^b_i(0) \right\rangle_T \,,
\end{align}
where $E_i$ is the (non-Abelian) electric field with the spatial index $i=1,2,3$, the subscript $T$ is the temperature and indicates the expectation value is taken at thermal equilibrium, ${\ml{W}}$ represents a path-ordered Wilson line in the adjoint representation, and $a,b$ are color indices that are summed over.}


{
At this stage, we would like to emphasize that the correlator in Eq.~(\ref{eq:correlator-intro}) for bound state formation and dissociation is different from the well-studied electric field correlator for the heavy quark diffusion coefficient~\cite{Casalderrey-Solana:2006fio,Eller:2019spw,CaronHuot:2007gq,Burnier:2010rp},
\begin{align} \label{eq:correlator-intro-Eller}
\left\langle {\rm Tr}_{\rm color} \left[ U(-\infty,t) E_i(t) U(t,0) E_i(0) U(0,-\infty) \right] \right\rangle_T \,,
\end{align} 
where $U$ denotes path-ordered Wilson lines in the fundamental representation. In detail, these two correlators differ in the configuration of the Wilson lines.} 
{ In the calculation of the heavy quark diffusion coefficient, hard-thermal-loop (HTL) resummation~\cite{Braaten:1989mz,Braaten:1989kk,Braaten:1991gm} is required to cope with the infrared divergence that occurs when zero energy and momentum are transferred in the \emph{elastic} scattering. Bound state formation and dissociation, however, are \emph{inelastic} processes with a minimum energy (which is the binding energy) transferred, which serves as an infrared regulator. When the Debye screening mass $m_D$ is much smaller than the binding energy, resummation has little effect.}

As the main part of this work, we will perform a complete next-leading order (NLO) calculation of this correlator for the SU($N_c$) gauge theory, with $n_f$ effectively massless fermion fields in the plasma. The computation is carried out in the real-time formalism, with a general $R_{\xi}$ gauge. The gauge invariance of our result in perturbation theory is verified carefully by taking into account all contributions, in particular those from the Wilson lines. Moreover, we will prove that our result is both infrared and collinear finite, even if there is no screening mass regulator when finite energy is transferred.

This paper is organized as follows. In Section~\ref{sect:scope}, we will briefly review the open quantum system framework, the potential nonrelativistic effective theory that we use to describe the pair of heavy particles, and the derivation of the Boltzmann equation and the corresponding rate equation that are used in studies of quarkonia transport in the QGP and DM relic abundance. We will explain how the electric field correlator defined in Eq.~(\ref{eq:correlator-intro}) appears in the collision terms of the Boltzmann equation and the reaction rates. Then we turn to the full computation of the SU($N_c$) electric field correlator at NLO in Section~\ref{sect:nlo}, and present our numerical results in Section~\ref{sec:results}. Finally, a summary and conclusions will be given in Section~\ref{sec:conclusion}.

\section{Transport equations and electric field correlators}
\label{sect:scope}
In this section, we briefly explain the formalism that we adopt to describe the dynamics of heavy particle pairs inside a thermal environment. Section~\ref{subsect:open} introduces the concept of the open quantum system framework, that we will use to describe the in-medium dynamics of the subsystem, i.e., a pair of heavy particles, which will be specified in Section~\ref{subsect:subsystem}. In Section~\ref{subsect:dipole}, we introduce the interactions between the subsystem and the environment in an effective field theoretical framework. We will explicitly show the collision terms in the Boltzmann equation and the rate equation in Section~\ref{subsect:elcol}, in which the electric field correlator appears. We will also briefly discuss the electric field correlator.

\subsection{Open quantum system formalism}
\label{subsect:open}
Our starting point is an interacting quantum system consisting of a subsystem and a thermal environment, where the full Hamiltonian $H$ can be written as 
\begin{align}
H = H_S +H_E + H_I\,,
\end{align}
where $H_S$ is the \emph{subsystem} Hamiltonian (e.g., quarkonium, dark matter), $H_E$ denotes the \emph{environment} Hamiltonian (e.g., quark-gluon plasma, early universe plasma), and $H_I$ contains the \emph{interactions} between the subsystem and the environment. The time evolution of the density matrix of the full system is given by
\begin{align}
\frac{\diff \rho(t)}{\diff t} = -i[H,\rho(t)] \,.
\end{align}
In the interaction picture $\rho^{(\ma{int})}(t) = e^{i(H_S+H_E)t} \rho(t) e^{-i(H_S+H_E)t}$, the formal solution can be written as
\begin{align}
\rho^{(\ma{int})}(t) = U(t) \rho^{(\ma{int})}(0)  U^\dagger(t) \,,
\end{align}
where the time evolution is given by
\begin{align}
U(t) &= \ml{T}\exp\Big( -i\int_0^t \diff t' H_I^{(\ma{int})}(t')\Big)\\
H_I^{(\ma{int})}(t) &= e^{i(H_S+H_E)t} H_I(t) e^{-i(H_S+H_E)t} \,.
\end{align}
The time evolution of the subsystem can be written as
\begin{align}
\label{eqn:rho_S(t)}
\rho_S^{(\ma{int})}(t) = \Tr_E \big[ \rho^{(\ma{int})}(t)  \big] = \Tr_E \big[  U(t) \rho^{(\ma{int})}(0)  U^\dagger(t)  \big] \,.
\end{align}
When the subsystem and the environment are weakly interacting, we can assume the density matrix of the full system factorizes
\begin{align}
\rho(t) = \rho_S(t) \otimes \rho_E \,,
\end{align}
where the environment density matrix is set to be thermal $\rho_E = \frac{1}{Z} e^{-\beta H_E}$ and thus independent of time. Under the assumption of factorization, it is known that Eq.~(\ref{eqn:rho_S(t)}) can be written as a Lindblad equation in two limits: the limit of the quantum Brownian motion and the quantum optical limit. These two limits are specified by the hierarchies of time scales. Relevant time scales include the environment correlation time $\tau_E$, the subsystem intrinsic time scale $\tau_S$ and the subsystem relaxation time $\tau_R$. The limit of the quantum Brownian motion is valid if $\tau_R \gg \tau_E$ and $\tau_S \gg \tau_E$ while the quantum optical limit is valid when $\tau_R \gg \tau_E$ and $\tau_R \gg \tau_S$. In this work, we will focus on the quantum optical limit. The hierarchies of time scales in the quantum optical limit are closely related to the separation of energy scales that we assume $M\gg \frac{1}{r} \gg E, T \gtrsim m_D$, where $M$ is the heavy particle mass, $r$ the typical relative distance between the two heavy particles in the pair, $E$ the binding energy, $T$ the temperature of the thermal environment and $m_D$ the Debye screening mass. The subsystem intrinsic time scale can be estimated as $\tau_S \sim \frac{1}{E}$, since the energy transferred in the transition between bound and unbound states is at least $E$. Furthermore, the environment correlation time can be estimated by the inverse of the temperature $\tau_E\sim\frac{1}{T}$. Finally, when the subsystem and the environment are weakly-coupled, the subsystem relaxation time can be estimated as $\tau_R \sim \frac{T}{(H_I)^2}$. For sufficiently heavy particles with the separation $\frac{1}{r}\gg T$, the interaction between the subsystem and the environment is given by a dipole interaction as shown in Section~\ref{subsect:dipole}, which implies $\tau_R\sim\frac{1}{(rT)^2T}$.
Since we focus on the regime of $\frac{1}{r} \gg T$, the hierarchy $\tau_R \gg \tau _E$ is automatically fulfilled.
Strictly speaking, the second inequality $\tau_R \gg \tau_S$ indicates a slightly stronger condition than $\frac{1}{r} \gg T$, namely $\frac{v^{1/3}}{r} \gg T$.
However, since we expect $\frac{1}{r}\sim Mv$ and $E\sim Mv^2$ where $v\sim \alpha_s(Mv)$ is the typical relative velocity between the heavy particle pair, this difference would be minor for the SM gauge couplings.
In addition, in order for our fixed order calculations to be valid, we have assumed $E\gtrsim m_D$.

As shown later, the dipole interaction between the heavy particle pair and the plasma can be matched into the form of
\be
\label{eqn:HI}
H_I = \sum_{\alpha} O_{\alpha}^{S} \otimes  O_{\alpha}^{E}\,,
\ee
where $\alpha$ denotes all continuous (such as position) and discrete (such as spin and SU($N_c$) indexes) variables that the subsystem $O_{\alpha}^{S}$ and environment $O_{\alpha}^{E}$ operators depend on, the Lindblad equation in the quantum optical limit can be written as
\be
\label{eqn:lindblad}
\frac{\diff \rho_S^{(\text{int})}(t)}{\diff t} &=& -i \sum_{n,k} \sigma_{nk} \Big[ |n\rangle \langle k|, \rho_S^{(\text{int})}(t) \Big] \nn\\
&+& \sum_{n,m,k,l} \gamma_{nm,kl} \Big( |n\rangle \langle m| \rho_S^{(\text{int})}(t) |k\rangle \langle l| - \frac{1}{2}\big\{ |k\rangle \langle l|n\rangle \langle m| , \rho_S^{(\text{int})}(t) \big\} \Big) \\
\sigma_{nk} &=& \frac{1}{2} \sum_{\alpha,\beta} \sum_m  \Sigma_{\alpha\beta}(E_n-E_m) \delta_{E_n,E_k}  \langle n| O^{(S)}_\alpha |m \rangle \langle m| O^{(S)}_\beta |k \rangle \\
\gamma_{nm,kl} &=& \sum_{\alpha,\beta} D_{\alpha\beta}(E_m-E_n) \delta_{E_k-E_l,E_m-E_n}
\langle k| O^{(S)}_\alpha |l \rangle \langle n| O^{(S)}_\beta |m \rangle
\,,
\ee
where $|n\rangle$ denotes the eigenstate of the subsystem Hamiltonian: $H_S|n\rangle = E_n|n\rangle$. The environment correlators are defined as
\be
D_{\alpha\beta}(q_0) &=& \int \diff t\,e^{iq_0t} D_{\alpha\beta}(t,0) = \int \diff t\,e^{iq_0t} \Tr_E\big(\rho_E O^E(t) O^E(0) \big) \\
\Sigma_{\alpha\beta}(q_0) &=& -i\int \diff t\,e^{iq_0t} {\rm sgn}(t) D_{\alpha\beta}(t,0)\,,
\ee
where we assume the environment is invariant under time translation.

To derive the Boltzmann equation describing the transport of heavy particle pairs, we need to specify the subsystem eigenstates and eigenenergies, and the interaction between the subsystem and the environment, which will be discussed in the next two subsections. We also need to make a Wigner transform and apply a semiclassical expansion, which will be explained in Section~\ref{subsect:elcol}. Details of the construction can be found in Ref.~\cite{Yao:2021lus}.

\subsection{Subsystem}
\label{subsect:subsystem}

We consider a subsystem that contains heavy pairs, consisting of a particle denoted by $\chi$ in a representation ${\bs R}$ of SU($N_c$) and another particle $\bar{\chi}$ in the conjugate representation $\bar{\bs {R}}$.
Their tensor product  describes two-particle states and can always be decomposed into (e.g., Ref.~\cite{Ellis:2015vna, Harz:2018csl}):
\begin{align}
{\bs R} \otimes \bs{\bar{R}} = {\bs 1} \oplus \bs{adj} \oplus \cdots \,,
\label{eq:group}
\end{align}
where ${\bs 1}$ denotes the trivial representation and ${\bs{adj}}$ represents the adjoint representation. The interaction between the $\chi\bar{\chi}$ is mediated via the gauge vector boson of the SU($N_c$) theory. Among all these irreducible representations on the right hand side, the singlet configuration, ${\bs 1}$, has the most attractive two-body potential and can form the most deeply bound state. Therefore, the singlet configuration can decay the fastest (here decay means annihilation of the $\chi\bar{\chi}$ state, which then becomes particles other than $\chi\bar{\chi}$, and is thus different from dissociation). All the other representations may accommodate shallower bound states with larger typical sizes.

Through the interaction with the non-Abelian gauge field quanta (on- or off-shell as induced by the environment interaction), the following transitions in the subsystem (bound state formation, dissociation, level transitions) are of our interest:
\be
\label{eq:transitions}
\mathcal{S}(\chi \bar{\chi})_{\rm{adj}} \rightleftharpoons  \mathcal{B}(\chi \bar{\chi})_{\rm{1}} \,,\quad
\mathcal{S}(\chi \bar{\chi})_{\rm{1}} \rightleftharpoons  \mathcal{B}(\chi \bar{\chi})_{\rm{adj}} \,,\quad
\mathcal{S}(\chi \bar{\chi})_{\rm{adj}} \rightleftharpoons  \mathcal{B}(\chi \bar{\chi})_{\rm{adj}} \,,
\ee
where $\mathcal{S}$ denotes a two-particle scattering state and $\mathcal{B}$ a bound state, both of which can be in either the singlet or the adjoint representations. We restrict ourselves to transitions mediated by just one gauge field dipole interaction, which will be explained in the next subsection. This is why we omit possible higher dimensional representations in Eq.~\eqref{eq:transitions}, which might appear in the decomposition in Eq.~(\ref{eq:group}).
These higher dimensional representations can be \emph{directly} converted to the singlet state through multiple gauge field dipole interactions that we neglect since they are suppressed by powers of $rT$ in the multipole expansion. We note that, however, they can also reach the ground state \emph{indirectly}, by consecutively transforming into different states in the network, each of which is mediated by one gauge field dipole interaction. Since the bound states in the higher dimensional representations are shallowly bound, it is interesting to note already at this stage that the interaction with the thermal environment (at NLO) can strongly enhance the rapidness of these otherwise slow (in vacuum) transformations, which might have physical consequences.
From now on, we will, however, for simplicity restrict our discussions to transitions listed in  Eq.~(\ref{eq:transitions}).

Particularly, Eq.~(\ref{eq:transitions}) fully captures the essence of quarkonium transitions in the QGP, where the two heavy particles are fermionic particle and antiparticle in the fundamental and antifundamental representations of the SU($3$) group:
\begin{align}
{\bs 3} \otimes \bs{\bar{3}} = {\bs 1} \oplus {\bs 8} \,.\label{eq:dec}
\end{align}
The static octet potential, induced by gluon exchanges, is exclusively repulsive, implying that no octet bound states exist in the solution of the corresponding Schr\"odinger equation. Therefore, the last two transitions in Eq.~(\ref{eq:transitions}) do not exist.

Regarding DM, one famous example of the $\chi$ particle is a complex scalar triplet in the Standard Model (SM) gauge group SU($3$), obeying the same decomposition as in Eq.~(\ref{eq:dec}). These color charged particles are the \emph{co-annihilating partners} of the actual (color neutral) DM particles. If the DM is a Majorana fermion, then the complex scalar in the co-annihilation scenario corresponds to the famous squark in the context of the Minimal Supersymmetric Standard Model (MSSM), e.g., see Ref.~\cite{Ellis:2015vna}. Here, the octet potential can have bound state solutions if the attractive interactions mediated by the SM Higgs particles are included~\cite{Harz:2019rro}.
Another interesting example of the $\chi$ particle is a Majorana fermion octet state in the SM gauge group SU($3$), which corresponds to the gluino co-annihilation scenario.
In this case, we have two adjoint representations in the decomposition:
\begin{align}
  {\bs 8} \otimes {\bs 8} = {\bs 1} \oplus {\bs 8}_A \oplus {\bs 8}_S \oplus \cdots,\label{eq:adjadj}
\end{align}
where the subscripts $A$ and $S$ indicate the two ways of connecting gluinos via the totally antisymmetric tensor and symmetric tensor, respectively. Both octet potentials are attractive and support bound state solutions.
Still, as we will see later in a more general discussion, only one of them can be directly converted to the singlet state through the electric dipole transition.
In these two DM example, all processes in Eq.~(\ref{eq:transitions}) are allowed.

\subsection{Interactions in pNREFT}
\label{subsect:dipole}
For sufficiently heavy particles in the hierarchy $M\gg \frac{1}{r}\gg E, T$, transitions between different states of the $\chi\bar{\chi}$ pairs are governed by the Lindblad equation in the quantum optical limit, as explained in Section~\ref{subsect:open}.
To write out the Lindblad equation for the pair explicitly, we need to specify the interaction between the subsystem and the environment in the form of Eq.~(\ref{eqn:HI}). In the aforementioned hierarchy, the interaction can be described by \emph{potential nonrelativistic effective field theory} (pNREFT) \cite{Pineda:1997bj,Brambilla:1999xf,Brambilla:2004jw}, which is an effective field theory for nonrelativistic two-body states. If the scale $\frac{1}{r}\sim Mv$ is perturbative, the pNREFT can be constructed perturbatively, under systematic nonrelativistic and multipole expansions, which are expansions in terms of $v$ and $r$ respectively. At the leading order in the nonrelativistic expansion and linear order in the multipole expansion, the Lagrangian density for the \emph{subsystem} consisting of a $\chi$-$\bar{\chi}$ pair and its \emph{interaction} with the environment via the gauge field is given by:
\begin{align}
\ml{L}_\ma{pNREFT} \supset& \int \diff^3r\, \Tr\Big[  \ma{S}^{\dagger}(i\partial_0-H_s)\ma{S} + \mathrm{Adj}^{\dagger}( iD_0-H_\mathrm{adj} )\mathrm{Adj}  \nn\\
\label{eq:lagr}
&- V_A( \ma{Adj}^{\dagger}\bs r \cdot g{\bs E} \ma{S} + \ma{h.c.})  - \frac{V_B}{2}\ma{Adj}^{\dagger}\{ \bs r\cdot g\bs E, \ma{Adj}  \} +\cdots \Big]\,.
\end{align}
The subsystem degrees of freedom include the singlet configuration of the $\chi$-$\bar{\chi}$ pair, $\ma{S}({\bs x}_{\rm cm}, \bs r, t)$, and the adjoint configuration $\ma{Adj}({\bs x}_{\rm cm}, \bs r, t)$,\footnote{
  If the decomposition involves multiple adjoint representations, e.g., Eq.~\eqref{eq:adjadj}, an appropriate summation over these adjoint states should be taken. 
  See also discussions around Eq.~\eqref{eq:adjadj_Edipole}.
} consistent with the group decomposition in Eq.~(\ref{eq:group}), where the center-of-mass (c.m.) position of the heavy particle pair is denoted by ${\bs x}_{\rm cm}$  while $\bs r$ is the relative position. These operators contain the two-body bound and scattering states. 
At linear order in the multipole expansion, these states interact through the (non-Abelian) \emph{electric dipole operator}, $\bm{r} \cdot g \bm{E}$, shown in the second line of the Lagrangian density. Concretely, this operator describes the bound state formation, dissociation, and level transitions, represented here as any dipole transition compatible with Eq.~(\ref{eq:transitions}). 

The singlet and adjoint two-body fields are expressed as $d_{\bs R} \times d_{\bs R}$ matrices with $d_{\bs R}$ being the dimension of a representation ${\bs R}$:
\begin{align}
  \mathrm{S} = \frac{{\bs 1}_{d_{\bs R}}}{\sqrt{d_{\bs R}}} S\,, 
  \quad \mathrm{Adj} = \frac{T_{\bs R}^a}{\sqrt{C({\bs R})}} \mathrm{Adj}^a \,.
\end{align}
Here a unit matrix of size $d_{\bs R}$ is ${\bs 1}_{d_{\bs R}}$,
an SU($N_c$) generator acting on a representation ${\bs R}$ is $T_{\bs R}$,
and its normalization is $\Tr (T^a_{\bs R} T^b_{\bs R}) = C({\bs R}) \delta^{ab}$.
If two-body fields are made from fermions, a summation over spin indices are implicit in the trace.
Under the ultrasoft gauge transformation of $e^{ig \theta^a T^a_{\bs R}}$, they transform as
\begin{align}
  S (\bm{R},\bm{r},t) \mapsto S (\bm{R}, \bm{r}, t), \quad
  \mathrm{Adj} (\bm{R},\bm{r},t) \mapsto e^{ig \theta^a (\bm{R},t) T^a_{\bs R}} \mathrm{Adj} (\bm{R},\bm{r},t) e^{- ig \theta^a (\bm{R},t) T^a_{\bs R}}\,.
\end{align}
Hence, the effective Lagrangian (\ref{eq:lagr}) is invariant under the ultrasoft gauge transformation.

In cases where the bound states are formed by a representation whose conjugate is itself, e.g., ${\bs R} = \overline{{\bs R}} = \bs{adj}$ as in Eq.~(\ref{eq:adjadj}),
the two-body states may involve a symmetry with respect to an interchange of the two particles, i.e., $\bm{r} \mapsto - \bm{r}$ (and spin indices: $(s_1,s_2) \mapsto (s_2,s_1)$).
This symmetry property suppresses some terms in our effective Lagrangian.
We take two primary examples here to illustrate this: (i) ${\bs R} = {\bs N}$ and (ii) ${\bs R} = \bs{adj}$.
In the former case (i), the singlet and adjoint two-body states are composed of a particle $\eta$ ($\eta$ annihilates the particle) and an antiparticle $\xi$ ($\xi$ creates the antiparticle) which reside in different representations.
This implies
\begin{align}
  S (\bm{r}, t) &\sim 
  \frac{1}{\sqrt{d_{{\bs N}}}} \xi^\dagger \qty(\frac{\bm{r}}{2}, t) W_{\qty[\qty(\frac{\bm{r}}{2},t),(-\frac{\bm{r}}{2},t)]} \eta \qty(- \frac{\bm{r}}{2}, t)\,, \\
  \mathrm{Adj}^a (\bm{r}, t) &\sim 
  \frac{1}{\sqrt{C({\bs N})}}
  \xi^\dagger \qty(\frac{\bm{r}}{2}, t) W_{\qty[\qty(\frac{\bm{r}}{2},t),(\bm{0},t)]} T^a_{{\bs N}} W_{\qty[\qty(\bm{0},t),(-\frac{\bm{r}}{2},t)]} \eta \qty(- \frac{\bm{r}}{2}, t)\,, 
\end{align}
where we take the c.m.~position to be $\bm{x}_{\text{cm}} = 0$ without loss of generality and suppress spin indices for notational brevity.
The Wilson line in a representation ${\bs R}$ connecting two spatially different points at the same time is given by
\begin{align}
  W_{[(\bm{x}_1,t),(\bm{x}_2,t)]}
  =
  \mathcal{P} \exp \qty[ - i g \int^{\bm{x}_1}_{\bm{x}_2} \dd \bm{x} \cdot \bm{A}^a(\bm{x},t) T^a_{{\bs R}} ]\,,
\end{align}
where $\mathcal{P}$ denotes path ordering.
On the other hand, in the latter case (ii), the two-body fields can be made of the same particles $\lambda$.
This is, in particular, the case of a real scalar and a Majorana fermion in the adjoint representation.
In this case, two-body fields can be written as
\begin{align}
  S (\bm{r}, t) &\sim 
  \frac{1}{\sqrt{2 d_{\bs{adj}}}} \lambda \qty(\frac{\bm{r}}{2}, t) W_{\qty[\qty(\frac{\bm{r}}{2},t),(-\frac{\bm{r}}{2},t)]} \lambda \qty(- \frac{\bm{r}}{2}, t)\,, \\
  \mathrm{Adj}^a_A (\bm{r}, t) &\sim 
  \frac{1}{\sqrt{2 C(\bs{adj})}}
  \lambda \qty(\frac{\bm{r}}{2}, t) W_{\qty[\qty(\frac{\bm{r}}{2},t),(\bm{0},t)]} T^a_{\bs{adj}} W_{\qty[\qty(\bm{0},t),(-\frac{\bm{r}}{2},t)]} \lambda \qty(- \frac{\bm{r}}{2}, t)\,,  \\
  \mathrm{Adj}^a_S (\bm{r}, t) &\sim 
  \frac{1}{\sqrt{2 C(\bs{adj})}}
  \lambda \qty(\frac{\bm{r}}{2}, t) W_{\qty[\qty(\frac{\bm{r}}{2},t),(\bm{0},t)]} D^a W_{\qty[\qty(\bm{0},t),(-\frac{\bm{r}}{2},t)]} \lambda \qty(- \frac{\bm{r}}{2}, t)\,.
\end{align}
In general ($N_c \geq 3$), we have two adjoint states in the decomposition, reflecting the fact that there are two rank-three invariant tensors, i.e., the totally anti-symmetric one $(T_{\bs{adj}}^a)_{bc} = i f^{bac} = 2 \Tr (T^b_{{\bs N}}[T^a_{{\bs N}},T^c_{{\bs N}}])$ and the totally symmetric one $(D^a)_{bc} = 4d^{bac} = 2 \Tr (T^b_{{\bs N}}\{T^a_{{\bs N}}, T^c_{{\bs N}}\})$.
By interchanging two $\lambda$ fields in the above expression, one may readily see that the following properties should be fulfilled (for the real scalar case, only the result for Spin=$0$ applies):
\begin{align}
  S(\bm{x}_{\text{cm}},\bm{r},t) & = 
  \begin{cases}
    + S(\bm{x}_{\text{cm}},-\bm{r},t) &\text{for} \quad \mathrm{Spin} = 0\,, \\
    - S(\bm{x}_{\text{cm}},-\bm{r},t) &\text{for} \quad \mathrm{Spin} = 1\,,
  \end{cases} \\
  \mathrm{Adj}^a_{A/S}(\bm{x}_{\text{cm}},\bm{r},t) & = 
  \begin{cases}
    \mp \mathrm{Adj}^a_{A/S}(\bm{x}_{\text{cm}},-\bm{r},t) &\text{for} \quad \mathrm{Spin} = 0\,, \\
    \pm \mathrm{Adj}^a_{A/S}(\bm{x}_{\text{cm}},-\bm{r},t) &\text{for} \quad \mathrm{Spin} = 1\,.
  \end{cases}
\end{align}
Therefore, only particular combinations of two-body fields with the electric dipole operators are allowed by parity symmetry, which include
\begin{align}
  \mathrm{Adj}^\dag_A \bm{r} \cdot g\bm{E} S\, , \quad
  \mathrm{Adj}^\dag_A \bm{r} \cdot g \bm{E} \mathrm{Adj}_S \,,
  \label{eq:adjadj_Edipole}
\end{align}
and their Hermitian conjugates.\footnote{
  Note that the electric dipole operator does not involve spin-changing transitions.
}
Therefore, the first term in the second line in Eq.~\eqref{eq:lagr} only involves $\mathrm{Adj}_A$ while the second term in the same line describes the cross term between $\mathrm{Adj}_A$ and $\mathrm{Adj}_S$. This completes the discussion of Dirac fermion, Majorana fermion and scalar pairs.

The equations of motion of the free singlet and adjoint fields are Schr\"odinger equations with the Hamiltonians organized by powers of $\frac{1}{M}$ or equivalently, $v$:
\begin{align}
H_{s} &= \frac{(i\bs \nabla_\ma{cm})^2}{4M} + \frac{(i\bs \nabla_\ma{rel})^2}{M} + V_{s}^{(0)}(r) + \frac{V_{s}^{(1)}(r)}{M} + \frac{V_{s}^{(2)}(r)}{M^2} + \cdots\\
\label{eqn:Ho}
H_\mathrm{adj} &= \frac{(i\bs D_\ma{cm})^2}{4M} + \frac{(i\bs \nabla_\ma{rel})^2}{M} + V_\mathrm{adj}^{(0)}(r) + \frac{V_\mathrm{adj}^{(1)}(r)}{M} + \frac{V_\mathrm{adj}^{(2)}(r)}{M^2} + \cdots\,.
\end{align}
At leading order in the nonrelativistic expansion, the Hamiltonians can be simplified as
\begin{align}
\label{eqn:hamiltonian}
H_{s,\mathrm{adj}}  =  \frac{(i\bs \nabla_\ma{rel})^2}{M} + V_{s,\mathrm{adj}}^{(0)}(r)\,.
\end{align}
Here $V_{s,\mathrm{adj}}^{(0)}$, $V_A$ and $V_B$ are Wilson coefficients. Perturbatively, at leading order in $\alpha_s(Mv)$ ($\alpha_s\equiv g^2/4\pi$), we have
\begin{align}
\label{eqn:match}
V_{s}^{(0)}(r) = -C_2 ({\bs R})\frac{\alpha_s}{r}\,,\quad V_\mathrm{adj}^{(0)}(r) = - \qty[ C_2 ({\bs R}) - \frac{C_2 (\bs{adj})}{2} ]\frac{\alpha_s}{r}\,,\quad V_A=V_B=1\,,
\end{align}
where the quadratic Casimir $C_2({\bs R})$ is defined by $T^a_{{\bs R}}T^a_{{\bs R}}=C_2({\bs R}) {\bs 1}_{d_{\bs R}}$.
One may readily see that, while the potential for the adjoint two-body field is repulsive for ${\bs R} = {\bs N}$, it is attractive for ${\bs R} = \bs{adj}$.

The main difference to the Abelian case, i.e., the U($1$) gauge theory that was recently studied in Ref.~\cite{Binder:2020efn}, is that the covariant derivatives $D_0$ and ${\bs D}_\ma{cm}$\footnote{
At leading order in the $v$ expansion, the c.m.\ covariant derivative term $\frac{{\bs D}_\ma{cm}^2}{4M}$ is suppressed in powers of $v$ for ultrasoft modes, compared with the $D_0$ term. However, for Coulomb modes that mediate the Coulomb interaction between the c.m.\ motion of the adjoint field and the gauge field, the $\frac{{\bs D}_\ma{cm}^2}{4M}$ term is at the same leading order in $v$ as the $D_0$ term.
The $\frac{{\bs D}_\ma{cm}^2}{4M}$ term does not affect dynamics at finite time in a non-singular gauge such as the $R_\xi$ gauge. However, it affects dynamics at infinite time and is crucially important for the construction of a gauge invariant electric field correlator in general~\cite{Yao:2020eqy}.
} introduce in the Lagrangian density (\ref{eq:lagr}) extra couplings between the subsystem of the two-body fields and the non-Abelian plasma. Complete accounting of these extra couplings will be crucial in showing the full gauge invariance of our NLO result later.
To take the $D_0$ term into account in an elegant way, we apply a field redefinition
\begin{align}
\label{eq:O_redef}
\mathrm{Adj}(\bm{x}_{\text{cm}}, \bm{r}, t) \to W_{[(\bm{x}_{\text{cm}}, t_0),(\bm{x}_{\text{cm}},t)]}
\widetilde{\mathrm{Adj}}(\bm{x}_{\text{cm}}, \bm{r}, t)
W_{[(\bm{x}_{\text{cm}},t),(\bm{x}_{\text{cm}}, t_0)]}  \,,
\end{align}
where $t_0$ is an arbitrary constant that cancels out in the end. The Wilson line $W$ for a representation ${\bs R}$ which connects $(\bm{x}_{\text{cm}}, t_f)$ and $(\bm{x}_{\text{cm}}, t_i)$ is defined by
\begin{align}
 W_{[(\bm{x}_{\text{cm}}, t_f),(\bm{x}_{\text{cm}}, t_i)]} = \ml{P}\exp \left[ ig\int^{t_f}_{t_i} \diff s A^a_0(\bm{x}_{\text{cm}}, s) T^a_{{\bs R}} \right] \,.
\end{align}
After this field redefinition, the $D_0$ covariant derivative term of the Lagrangian becomes canonical, and the EFT under consideration can be expressed as
\begin{align}
\ml{L}_\ma{pNREFT} \supset \int \diff^3r \Tr \Big[ & \ma{S}^{\dagger}(i\partial_0-H_s)\ma{S} +\widetilde{\ma{Adj}}^{\dagger}( i\partial_0 - H_\mathrm{adj} )\widetilde{\ma{Adj}} \nn\\
&- g ( \widetilde{\ma{Adj}}^{\dagger} r_i     \widetilde{E}_i  \ma{S} + \ma{S}^\dagger r_i  \widetilde{E}_i \widetilde{\ma{Adj}} ) 
-  \frac{g}{2}\widetilde{\ma{Adj}}^{\dagger}\{ r_i  \widetilde{E}_i, \widetilde{\ma{Adj}}\}
\Big] \,, \label{eq:action}
\end{align}
where
\begin{align}
\widetilde{E}_i(\bm{x}_{\text{cm}},t) = W_{[(\bm{x}_{\text{cm}}, t_0),(\bm{x}_{\text{cm}},t)]}
{E}_i(\bm{x}_{\text{cm}},t)
W_{[(\bm{x}_{\text{cm}}, t),(\bm{x}_{\text{cm}},t_0)]}\,.
\end{align}
Inclusion of the Coulomb interaction between the c.m.~motion of the adjoint field and the gauge field is more involved, but can be calculated by a diagram-by-diagram approach to all orders~\cite{Yao:2020eqy}. These Coulomb interactions introduce the Wilson lines at infinite time that will be shown in the next subsection.

By identifying the subsystem Hamiltonian, $H_S$, as the free Hamiltonian of the singlet and adjoint states and the interaction Hamiltonian, $H_I$, as the electric dipole operator,
we can write down the explicit form of the Lindblad equation (\ref{eqn:lindblad}) for the bound state formation and dissociation, while carefully taking into account the original $D_0$ (now stored in the Wilson lines $W$) and ${\bs D}_{\text{cm}}$ contributions. 
Such a derivation was explicitly done in Ref.~\cite{Yao:2020eqy}. From the Lindblad equation we can also derive a Boltzmann equation describing the evolution of the phase space distribution of each bound state, which has been shown in Ref.~\cite{Yao:2020eqy} and reviewed in Ref.~\cite{Yao:2021lus} in detail for quarkonium. We will briefly explain the derivation in the next subsection. By integrating over the momentum, we can further obtain an ordinary differential equation for the number density, as somewhat similarly done for dark matter in the Abelian case in Ref~\cite{Binder:2020efn}. We will generalize this equation to the non-Abelian case, taking into account the Wilson lines that are absent in the U($1$) case.

\subsection{Transport equations and electric field correlator}
\label{subsect:elcol}
We have discussed the subsystem $H_S$ and the interaction $H_I$ Hamiltonians in the previous subsection. The subsystem operators $O_S$ in (\ref{eqn:HI}) include $(\widetilde{\ma{Adj}}^{a\dagger} r_i  \ma{S} + \ma{S}^\dagger r_i   \widetilde{\ma{Adj}}^a )$ and $
d^{abc}\widetilde{\ma{Adj}}^{a\dagger} r_i \widetilde{\ma{Adj}}^b $ while the environment operators $O_E$ contain $\widetilde{E}_i^a$. To write down the Lindblad equation (\ref{eqn:lindblad}) explicitly, we also need the eigenstates of the subsystem Hamiltonian.
For a bound state of a heavy particle pair, the eigenstate and the associated eigenenergy are $|{\bs k}, \ml{B}\rangle$ and $-|E_{\ml{B}}|$ respectively where ${\bs k}$ is the three c.m.~momentum of the bound state specified by the quantum number $\ml{B}$. An example of the quantum number can be $\ml{B}=n l m_l s$, where $n$, $l$, $m_l$ and $s$ denote the principle quantum number for the radial excitation, the orbital angular momentum, its third component and the spin. For an unbound pair, the eigenstate is labeled by $|{\bs p}_{\text{cm}},{\bs p}_{\text{rel}} \rangle$ with the eigenenergy given by $\frac{{\bs p}_{\text{rel}}^2}{M}$. At leading order in the nonrelativistic expansion, the c.m.~momentum does not contribute to the eigenenergy. The eigenenergy of the unbound state is continuous, which causes complications in justifying the hierarchy of time scales that is required to apply the Lindblad equation in the quantum optical limit. However, the associated hierarchy of time scales can be justified in the semiclassical limit~\cite{Yao:2021lus}. To take the semiclassical limit, we take a Wigner transformation of the Lindblad equation~(\ref{eqn:lindblad}), defined by
\be
f_{\ml{B}}({\bs x},{\bs k},t) = \int \diff^3 k' \, e^{i{\bs k}'\cdot{\bs x}} \Big\langle {\bs k}+\frac{{\bs k}'}{2}, \ml{B} \Big| \rho(t) \Big| {\bs k}-\frac{{\bs k}'}{2}, \ml{B} \Big\rangle\,,
\ee
which transforms the quantum density matrix into the phase-space distribution, and apply a semiclassical expansion, which corresponds to a gradient expansion. At leading order of the semiclassical expansion, we can show that the Lindblad equation in the quantum optical limit turns to a Boltzmann equation. Details of the derivation of this semiclassical Boltzmann equation from the quantum evolution and justifications of the hierarchy of time scales for heavy particle bound states inside a thermal plasma can be found in Ref.~\cite{Yao:2021lus}.

After these procedures, the Boltzmann equation for the bound state phase space distribution can be written for singlet-adjoint transitions as
\begin{align}
\label{eqn:Boltzmann}
\frac{\partial}{\partial t} f_{\mathcal{B}}({\bs x}, {\bs k}, t) + \frac{{\bs k}}{2M} \cdot \nabla_{\bs x} f_{\mathcal{B}}({\bs x}, {\bs k}, t)
= \ml{C}_{\mathcal{B}}^+({\bs x}, {\bs k}, t) - \ml{C}_{\mathcal{B}}^-({\bs x}, {\bs k}, t) \,,
\end{align}
where $\ml{C}_{\ml{B}}^+$ and $\ml{C}_{\ml{B}}^-$ are the collision terms for bound state formation and dissociation. More explicitly, the collision terms can be written as
\be
\ml{C}_{\ml{B}}^-({\bs x}, {\bs k}, t) &=& g^2 \frac{C({\bs R})}{d_{{\bs R}}} \sum_{i_1,i_2} \int\frac{\diff^3p_{\ma{cm}}}{(2\pi)^3} \frac{\diff^3p_{\ma{rel}}}{(2\pi)^3} \frac{\diff^4q}{(2\pi)^4} (2\pi)^4\delta^3({\bs k} - {\bs p}_\ma{cm} + {\bs q}) \delta\Big(E_{\ml{B}}- \frac{p_{\ma{rel}}^2}{M} - q_0\Big) \nn\\
\label{eqn:C-B}
&\times& \langle \psi_{\ml{B}} | r_{i_1} | \Psi_{{\bs p}_\ma{rel}} \rangle 
\langle \Psi_{{\bs p}_\ma{rel}} | r_{i_2} | \psi_{\ml{B}} \rangle\, [g_E^{++}]^{>}_{i_1i_2}(q_0,{\bs q})\,
f_{\ml{B}}({\bs x}, {\bs k}, t) \\
\ml{C}_{\ml{B}}^+({\bs x}, {\bs k}, t) &=& g^2\frac{C({\bs R})}{d_{{\bs R}}} \sum_{i_1,i_2}
\int\frac{\diff^3p_{\ma{cm}}}{(2\pi)^3} \frac{\diff^3p_{\ma{rel}}}{(2\pi)^3}
\frac{\diff^4q}{(2\pi)^4} 
(2\pi)^4\delta^3({\bs k} - {\bs p}_{\ma{cm}} - {\bs q}) 
\delta\Big(E_{\ml{B}}- \frac{p_{\ma{rel}}^2}{M} + q_0\Big)  \nn\\
\label{eqn:C+B}
&\times&  \langle \psi_{\ml{B}} | r_{i_1} | \Psi_{{\bs p}_\ma{rel}} \rangle 
\langle \Psi_{{\bs p}_\ma{rel}} | r_{i_2} | \psi_{\ml{B}} \rangle\,
 [g_E^{--}]^>_{i_2i_1}(q_0,{\bs q})\, f_{\ml{S}}({\bs x}, {\bs p}_{\ma{cm}}, {\bs r}=0, {\bs p}_{\ma{rel}}, t ) \,,
\ee
where $|\psi_{\ml{B}}\rangle$ and $|\Psi_{{\bs p}_\ma{rel}}\rangle$ are the bound state and scattering state Schr\"odinger wave functions, and $f_{\ml{S}}({\bs x}, {\bs p}_{\ma{cm}}, {\bs r}=0, {\bs p}_{\ma{rel}}, t )$ is the phase space distribution of a two-particle scattering state with a specific adjoint group index, with the c.m.~position ${\bs x}$ and momenta ${\bs p}_{\text{rel}}$ and relative position ${\bs r}=0$ and momentum ${\bs p}_{\ma{rel}}$. The distribution function already includes the multiplicity factors of spins and colors. Since the electric field correlator shown in the above expressions also sums over colors, we need to divide $d_{{\bs R}} d_{\overline{\bs R}}$ in the bound state formation (recombination) term to avoid double counting.

The collision terms in the Boltzmann equation depend on the properties of the plasma via the \emph{electric field correlator}:
\be
[g_E^{++}]^{>}_{ji}(q) &=& \int\diff^4(y-x)\, e^{iq\cdot(y-x)} [g_E^{++}]^{>}_{ji}(y,x) \\
\,[g_E^{--}]^{>}_{ji}(q) &=& \int\diff^4(y-x)\, e^{iq\cdot(y-x)} [g_E^{--}]^{>}_{ji}(y,x) \\
\,[g_E^{++}]^{>}_{ji}(y,x) &\equiv&  \Big\langle  \big[{E}_j(y) \ml{W}_{[( y^0, {\bs y}), (+\infty, {\bs y})]} \ml{W}_{[(+\infty, {\bs y}), (+\infty, {\bs \infty})]} \big]^a \nn\\
&\times& \big[\ml{W}_{[(+\infty, {\bs \infty}), (+\infty, {\bs x})]}
 \ml{W}_{[(+\infty, {\bs x}),(x^0, {\bs x})]} {E}_i(x) \big]^a \Big\rangle_T \\
\,[g_E^{--}]^{>}_{ji}(y,x) &\equiv&  \Big\langle   \big[\ml{W}_{[(-\infty, {\bs \infty}), (-\infty, {\bs y})]}
 \ml{W}_{[(-\infty, {\bs y}),(y^0, {\bs y})]} {E}_j(y) \big]^a\nn\\
&\times& \big[{E}_i(x) \ml{W}_{[( x^0, {\bs x}), (-\infty, {\bs x})]} \ml{W}_{[(-\infty, {\bs x}), (-\infty, {\bs \infty})]} \big]^a \Big\rangle_T \,,
\ee
where $x=(x^0, {\bs x})$, $y=(y^0,{\bs y})$, $\langle O\rangle_T \equiv \Tr_E(\rho_EO)$ denotes the average of the operators with respect to the density matrix of the environment (being in thermal equilibrium), and we assume the environment is invariant under spacetime translation. The color index $a$ is summed over. The component of the electric field is given by $E_i^a = F_{0i}^a = \partial_0 A_i^a - \partial_i A_0^a + gf^{abc} A_0^b A_i^c$. The Wilson lines $\mathcal{W}$ here are in the adjoint representation. The Wilson lines along the time direction are originated from the interaction in the $D_0$ term of the adjoint field. The Wilson lines along the spatial direction at infinite time come from the Coulomb interaction between the c.m.~motion of the adjoint field and the plasma, as explained previously.

{
The two electric field correlators are different due to the orientation of the Wilson lines. For the subsystem to approach the proper thermal equilibrium, the two correlators must satisfy the Kubo-Martin-Schwinger (KMS) relation. This can be seen by first defining the lesser correlation functions:
\be
\,[g_E^{++}]^{<}_{ji}(y,x) &\equiv&  \Big\langle 
\big[\ml{W}_{[(+\infty, {\bs \infty}), (+\infty, {\bs x})]}
 \ml{W}_{[(+\infty, {\bs x}),(x^0, {\bs x})]} {E}_i(x) \big]^a  \nn\\
&\times& 
 \big[{E}_j(y) \ml{W}_{[( y^0, {\bs y}), (+\infty, {\bs y})]} \ml{W}_{[(+\infty, {\bs y}), (+\infty, {\bs \infty})]} \big]^a
 \Big\rangle_T \\
\,[g_E^{--}]^{<}_{ji}(y,x) &\equiv&  \Big\langle  \big[{E}_i(x) \ml{W}_{[( x^0, {\bs x}), (-\infty, {\bs x})]} \ml{W}_{[(-\infty, {\bs x}), (-\infty, {\bs \infty})]} \big]^a  \nn\\
&\times& 
\big[\ml{W}_{[(-\infty, {\bs \infty}), (-\infty, {\bs y})]}
 \ml{W}_{[(-\infty, {\bs y}),(y^0, {\bs y})]} {E}_j(y) \big]^a
 \Big\rangle_T \,.
\ee
The usual proof of the KMS relation leads to
\be
\label{eqn:KMS_for_g++}
\,[g_E^{++}]^{>}_{ji}(q) = e^{q^0/T} [g_E^{++}]^{<}_{ji} (q) \,, \ \ \ \ \ \ \ \ \,[g_E^{--}]^{>}_{ji}(q) = e^{q^0/T} [g_E^{--}]^{<}_{ji} (q) \,.
\ee
Then one can define the spectral functions:
\be
\,[\rho_E^{++}]_{ji}(q) &=& [g_E^{++}]^{>}_{ji}(q) - [g_E^{++}]^{<}_{ji} (q) \\
\,[\rho_E^{--}]_{ji}(q) &=& [g_E^{--}]^{>}_{ji}(q) - [g_E^{--}]^{<}_{ji} (q) \,,
\ee
and obtain the relation between the greater (lesser) correlation functions and the spectral functions
\be
[g_E^{++}]^{>}_{ji}(q) &=& \big(1+n_B(q^0)\big) [\rho_E^{++}]_{ji}(q) \\
\, [g_E^{--}]^{>}_{ji}(q) &=& \big(1+n_B(q^0)\big) [\rho_E^{--}]_{ji}(q) \,,
\ee
where $n_B(q^0) = (e^{q^0/T}-1)^{-1}$ is the Bose-Einstein distribution. However, these relations do not guarantee that the subsystem approaches proper thermal equilibrium. For proper thermalization, the subsystem needs to be invariant under parity and time reversal transformations.
These symmetries imply
\be
[g_E^{++}]^{>}_{ji}(q) = [g_E^{--}]^{<}_{ji}(-q)\,,
\ee
which is explained in Appendix~\ref{app:kms}. Using the above relations we find
\be
[g_E^{++}]^{>}_{ji}(q) = e^{q^0/T} [g_E^{--}]^{>}_{ji}(-q) \,,
\ee
which is the correct KMS relation that guarantees proper thermalization of the subsystem. (Note the differences in the spatial indexes of $[g_E^{++}]^{>}_{ji}$ and $[g_E^{--}]^{>}_{ji}$ and the signs of $q^0$ in the delta functions in Eqs.~(\ref{eqn:C-B}, \ref{eqn:C+B}).)
}
At leading order (LO) in the coupling constant $g$, the Wilson lines do not contribute and we have
\be
&& [\rho_E^{++}]_{ii}(q) = [\rho_E^{--}]_{ii}(q)  \nn\\
&=& (N_c^2-1) g_{\mu\nu} (q^0 g^{i \mu} - q^i g^{0 \mu})(-q^0 g^{i \nu} + q^i g^{0 \nu}) (2\pi){\rm sgn}(q^0) \delta(q_0^2 - {\bs q}^2) \nn\\
\label{eqn:lo_rho}
&=& (N_c^2-1) |{\bs q}| (2\pi){\rm sgn}(q^0) \big(  \delta(q^0-|{\bs q}|) + \delta(q^0+|{\bs q}|) \big) \,.
\ee
We will discuss the computation of the electric field correlator at NLO in Section~\ref{sect:nlo}. For non-singular gauges such as the $R_\xi$ gauge, the Wilson lines at infinite time do not matter and can be neglected. We explicitly checked this and find that the contributions from the Wilson lines at infinite time do not contribute to the total bound state formation and dissociation rates. (A similar case in the transverse momentum dependent parton distribution function has been shown in Ref.~\cite{Belitsky:2002sm}.) But the Wilson lines along the time direction start to contribute at NLO.

At LO, the non-Abelian gauge quanta in the emission or absorption are on-shell, while at NLO they can be both off- and on-shell, due to the interaction with the thermal environment. The collision terms derived are independent of the precise particle content of the environment, which can be either weakly-coupled or strongly-coupled. We will assume the environment consists of massless fermions and gauge bosons for our NLO calculation. We also note that the `internal' emission of non-Abelian gauge field quanta, as often called in the DM literature, is fully taken into account in the Lagrangian (\ref{eq:lagr}).

Finally, before we move on to the details of the NLO calculation, we discuss how the Boltzmann equation (\ref{eqn:Boltzmann}) is related to the rate equation. Integrating the Boltzmann equation (\ref{eqn:Boltzmann}) over the momentum ${\bs k}$ gives the rate equation for the number density, which is defined by
\be
n_{\ml{B}}({\bs x},t) = g_{\ml{B}} \int\frac{\diff^3k}{(2\pi)^3} f_{\ml{B}}({\bs x}, {\bs k}, t) \,.
\ee
After the integration, the left hand side of the Boltzmann equation becomes simply $\dot{n}_{\ml{B}} = \frac{\diff}{\diff t} n_{\ml{B}}$. On the right hand side, the collision terms become
\be
\int\frac{\diff^3k}{(2\pi)^3} \ml{C}_{\ml{B}}^-({\bs x}, {\bs k}, t) &=&g^2 \frac{C({\bs R})}{d_{{\bs R}}} \sum_{i_1,i_2}  \int \frac{\diff^3p_{\ma{rel}}}{(2\pi)^3}  
\langle \psi_{\ml{B}} | r_{i_1} | \Psi_{{\bs p}_\ma{rel}} \rangle 
\langle \Psi_{{\bs p}_\ma{rel}} | r_{i_2} | \psi_{\ml{B}} \rangle \nn\\ &\times& G^{>}_{i_1i_2}\Big(E_{\ml{B}}- \frac{p_{\ma{rel}}^2}{M}\Big)
n_{\ml{B}}({\bs x}, t) \ \ \ \\
\int\frac{\diff^3k}{(2\pi)^3} \ml{C}_{\ml{B}}^+({\bs x}, {\bs k}, t)&=& g^2 \frac{C({\bs R})}{d_{{\bs R}}} \sum_{i_1,i_2} \int\frac{\diff^3p_{\ma{cm}}}{(2\pi)^3} \frac{\diff^3p_{\ma{rel}}}{(2\pi)^3}
\langle \psi_{\ml{B}} | r_{i_1} | \Psi_{{\bs p}_\ma{rel}} \rangle 
\langle \Psi_{{\bs p}_\ma{rel}} | r_{i_2} | \psi_{\ml{B}} \rangle \nn\\ &\times&
G^>_{i_2i_1}\Big(\frac{p^2_\ma{rel}}{M}-E_{\ml{B}}\Big)\, f_{\ml{S}}({\bs x}, {\bs p}_{\ma{cm}}, {\bs r}=0, {\bs p}_{\ma{rel}}, t ) \,,
\ee
where the momentum independent electric field correlation is given by
{
\be
G^>_{i_1i_2}(q^0) &=& \int\frac{\diff^3 q}{(2\pi)^3} [g_E^{++}]_{i_1i_2}^>(q^0,{\bs q}) = \int\frac{\diff^3 q}{(2\pi)^3} [g_E^{--}]_{i_1i_2}^<(-q^0, -{\bs q}) \\
&=& \int \diff { t}\, e^{iq^0t} \left\langle {E}^a_{i_1}(t) \ml{W}^{ab}(t,0)  {E}^b_{i_2}(0) \right\rangle_T \,.
\ee
From the above equation, we see that the inclusive reaction rates only depend on the momentum independent electric field correlator which, in position space, has no spatial separation. In general, the differential reaction rates depend on the momentum dependent electric field correlator.
}
We can further simplify the collision terms by averaging over the third component of the orbital angular momentum quantum number $m_l$ of the bound state $| \psi_{\ml{B}} \rangle = | \psi_{nlm_l} \rangle$, which leads to~\cite{Yao:2018sgn}
\be
&& \frac{1}{2l+1} \sum_{m_l=-l}^l\int \frac{\diff^3p_{\ma{rel}}}{(2\pi)^3}
\langle \psi_{nlm_l} | r_{i_1} | \Psi_{{\bs p}_\ma{rel}} \rangle 
\langle \Psi_{{\bs p}_\ma{rel}} | r_{i_2} | \psi_{nlm_l} \rangle f(|{\bs p}_\ma{rel}|) \nn\\
\label{eqn:average_ml}
&\equiv& \frac{1}{3}\delta_{i_1i_2} \int \frac{\diff^3p_{\ma{rel}}}{(2\pi)^3} 
| \langle \psi_{\ml{B}} | {\bs r} | \Psi_{{\bs p}_\ma{rel}} \rangle |^2 f(|{\bs p}_\ma{rel}|) \,,\
\ee
where $f$ is an arbitrary smooth function. From now on, the averaging over $m_l$ will be implicit.
Then the Boltzmann equation, after the momentum is integrated over, can be written as a rate equation
\be
\label{eqn:rate}
\dot{n}_{\ml{B}} = -\Gamma^{\rm diss}\, n_{\ml{B}} + F \,,
\ee
where the dissociation rate $\Gamma^{\rm diss}$ and the recombination (bound state formation) contribution $F$ can be written as
\be
\label{eqn:disso}
\Gamma^{\rm diss} &=& g_{\ml{B}} \frac{g^2}{3}\frac{C({\bs R})}{d_{{\bs R}}} \int \frac{\diff^3p_{\ma{rel}}}{(2\pi)^3} 
| \langle \psi_{\ml{B}} | {\bs r} | \Psi_{{\bs p}_\ma{rel}} \rangle |^2 G^{>}_{ii}\Big(E_{\ml{B}} - \frac{p^2_\ma{rel}}{M}\Big) \\
\label{eqn:reco}
F &=& g_{\ml{B}} \frac{g^2}{3}\frac{C({\bs R})}{d_{{\bs R}}} \int \frac{\diff^3p_{\ma{cm}}}{(2\pi)^3}  \frac{\diff^3p_{\ma{rel}}}{(2\pi)^3} 
| \langle \psi_{\ml{B}} | {\bs r} | \Psi_{{\bs p}_\ma{rel}} \rangle |^2
G^{>}_{ii}\Big(\frac{p^2_\ma{rel}}{M}-E_{\ml{B}}\Big) f_{\ml{S}}({\bs x}, {\bs p}_{\ma{cm}}, {\bs r}=0, {\bs p}_{\ma{rel}}, t ) \,. \nn\\
\ee
Equations~(\ref{eqn:Boltzmann},~\ref{eqn:rate}) have been widely used in phenomenological studies of quarkonium production in heavy ion collisions~\cite{Grandchamp:2003uw,Grandchamp:2005yw,Yan:2006ve,Liu:2009nb,Zhao:2010nk,Liu:2010ej,Zhao:2011cv,Song:2011xi,Song:2011nu,Sharma:2012dy,Zhou:2014kka,Nendzig:2014qka,Du:2015wha,Petreczky:2016etz,Zhou:2016wbo,Chen:2017duy,Zhao:2017yan,Du:2017qkv,Aronson:2017ymv,Yao:2017fuc,Yao:2018zze,Ferreiro:2018vmr,Du:2019tjf,Chen:2019qzx,Yao:2020xzw}. From now on, we define $\Delta E = p^2_\ma{rel}/M-E_{\ml{B}}$ as short-hand notation.

As for the dark matter bound state formation, usually further simplifications are made based on more assumptions. The first assumption is the factorization of the two-particle distribution function into two single-particle distributions
\be
f_{\ml{S}}({\bs x}, {\bs p}_{\ma{cm}}, {\bs r}=0, {\bs p}_{\ma{rel}}, t ) = f_{\chi}({\bs x},{\bs p}_1,t) f_{\bar{\chi}}({\bs x},{\bs p}_2,t) \,,
\ee
which means no correlation between $\chi$ and $\bar{\chi}$ particles. Nonrelativistically, ${\bs p}_{\ma{cm}} = {\bs p}_1 + {\bs p}_2$ and ${\bs p}_{\ma{rel}} = \frac{{\bs p}_1 - {\bs p}_2}{2}$. Another assumption is that the dark matter particles are in kinetic equilibrium, rather than chemical equilibrium, which gives
\be
f_{\chi}({\bs x},{\bs p}_1,t) &=&  \frac{n_{\chi}}{n_{\chi}^{\rm eq}} f^{\rm eq}_{\chi}({\bs x},{\bs p}_1,t) \\
f_{\bar{\chi}}({\bs x},{\bs p}_2,t) &=&  \frac{n_{\bar{\chi}}}{n_{\bar{\chi}}^{\rm eq}} f^{\rm eq}_{\bar{\chi}}({\bs x},{\bs p}_2,t) \,,
\ee
where $f_i^{\rm eq}({\bs p})$ is the thermal Boltzmann distribution multiplied by the multiplicity factor $g_i$: $f_i^{\rm eq}({\bs p})=  e^{-E({\bs p})/T}$ with $E({\bs p}) = M+\frac{{\bs p}^2}{2M}$, $n_{\chi} = g_{\chi}\int\frac{\diff p}{(2\pi)^3} f_{\chi}({\bs x}, {\bs p}, t)$ and similarly for $\bar{\chi}$. The multiplicity factor $g_\chi$ is given by the product of $d_{\textbf{R}}$ and the number of spin degrees of freedom $s_{\chi}$ and similarly for $g_{\bar{\chi}}$.
Then the recombination contribution (\ref{eqn:reco}) can be written as
\be
F &=& g_{\ml{B}} \frac{g^2}{3}\frac{C({\bs R})}{d_{{\bs R}}} \frac{ n_{\chi} n_{\bar{\chi}} }{ n_{\chi}^{\rm eq} n_{\bar{\chi}}^{\rm eq} } 
\nn\\
&\times&
\label{eqn:reco2}
 \int \frac{\diff^3p_1}{(2\pi)^3}  \frac{\diff^3p_2}{(2\pi)^3} 
| \langle \psi_{\ml{B}} | {\bs r} | \Psi_{{\bs p}_\ma{rel}} \rangle |^2
G^{>}_{ii}\Big( \frac{p^2_\ma{rel}}{M} - E_{\ml{B}} \Big) f^{\rm eq}_{\chi}({\bs x},{\bs p}_1,t) f^{\rm eq}_{\bar{\chi}}({\bs x},{\bs p}_2,t) \,.
\ee
To match with the standard notation in the dark matter community, we write the recombination contribution as
\be
F = \langle \sigma v \rangle_{\mathcal{B}}\, n_{\chi}   n_{\bar{\chi}} \,,
\ee
where the thermally averaged bound state formation cross section is defined by
\be
\label{eqn:form_cross_section}
\langle \sigma v \rangle_{\mathcal{B}} = \big\langle (\sigma v_{\rm rel})_{\mathcal{B}} \big\rangle &=& g_{\ml{B}} \frac{g_{\chi} g_{\bar{\chi}}}{n_{\chi}^{\rm eq}   n_{\bar{\chi}}^{\rm eq}} \int \frac{\diff^3p_1}{(2\pi)^3}  \frac{\diff^3p_2}{(2\pi)^3} f^{\rm eq}_{\chi}({\bs x},{\bs p}_1,t) f^{\rm eq}_{\bar{\chi}}({\bs x},{\bs p}_2,t)   \nn\\
&\times&
 \frac{g^2}{3}\frac{C({\bs R})}{d_{{\bs R}}} \frac{1}{g_{\chi} g_{\bar{\chi}}}
| \langle \psi_{\ml{B}} | {\bs r} | \Psi_{{\bs p}_\ma{rel}} \rangle |^2
G^{>}_{ii}\Big( \frac{p^2_\ma{rel}}{M} - E_{\ml{B}}\Big) \,,\quad\quad
\ee
where the bound state formation cross section is defined by
\be
(\sigma v_{\rm rel})_{\mathcal{B}} = \frac{g^2}{3}\frac{C({\bs R})}{d_{{\bs R}}}
\frac{ s_{\chi} s_{\bar{\chi}} }{ g_{\chi} g_{\bar{\chi}} }
| \langle \psi_{\ml{B}} | {\bs r} | \Psi_{{\bs p}_\ma{rel}} \rangle |^2
G^{>}_{ii}\Big( \frac{p^2_\ma{rel}}{M} - E_{\ml{B}}\Big) \,,
\label{eq:BSFcrossx}
\ee
where we have summed over the spin degrees of freedom of the scattering state.
The process is independent of spin, so no average over spin is needed.
Finally, since the dark matter bound state formation occurs in the early universe, when the universe is expanding, we need to replace the ordinary time derivative with the cosmic time derivative. Under the assumption of spatial homogeneity and isotropy, we find that the rate equation for the dark matter bound state formation can be written as
\begin{align}
\dot{n}_{\mathcal{B}} + 3 H n_{\mathcal{B}} &= \langle \sigma v \rangle_{\mathcal{B}}\, n_{\chi}  n_{\bar{\chi}} - \Gamma^{\rm diss} n_{\mathcal{B}} 
+ \cdots \,,
\end{align}
where $H=\dot{a}/a$ is the Hubble expansion rate with scale factor $a$ from the Friedmann-Robertson-Walker metric, and the dots indicate other processes such as the dark matter bound state annihilation into standard model particles. These other processes are model-dependent and can be added in the Lagrangian of Eq.~(\ref{eq:action}) (which is outside the scope of this work).

To make the (Saha) \emph{ionization equilibrium} condition more explicit, we can relate the thermally averaged bound state formation cross section with the dissociation rate by considering the rate equation at full equilibrium, at which we expect:
\be
\langle \sigma v \rangle_{\mathcal{B}}\, n_{\chi}^{\rm eq}  n_{\bar{\chi}}^{\rm eq} - \Gamma^{\rm diss} n_{\mathcal{B}}^{\rm eq}  = 0\,.
\ee
Then the rate equation for dark matter bound states can be written as
\begin{align}
\dot{n}_{\mathcal{B}} + 3 H n_{\mathcal{B}} &= \langle \sigma v \rangle_{\mathcal{B}} \left( n_{\chi}  n_{\bar{\chi}} - n_{\mathcal{B}} \frac{ n_{\chi}^{\text{eq}}  n_{\bar{\chi}}^{\text{eq}}}
{n_{\mathcal{B}}^{\text{eq}}}
\right)
+ \cdots \,\label{eq:nbound}.
\end{align}
Similarly, the effect of DM bound state formation and dissociation on the single particle number density can be described by
\begin{align}
\dot{n}_{\chi} + 3 H n_{\chi} = -\sum_{\mathcal{B}} \langle \sigma v \rangle_{\mathcal{B}} \left( n_{\chi}  n_{\bar{\chi}} - n_{\mathcal{B}} \frac{ n_{\chi}^{\text{eq}}  n_{\bar{\chi}}^{\text{eq}}}
{n_{\mathcal{B}}^{\text{eq}}} \right) + \cdots \; ,\label{eq:nscatter}
\end{align}
where we have summed over all bound states $\ml{B}=nl$. For the case of real scalar or Majorana fermion, i.e., $\chi=\bar{\chi}$, the right hand side of Eq.~(\ref{eq:nscatter}) needs to be multiplied by a factor of 2.\footnote{The factor of $2$ can be easily seen from the following consistency check: If $\chi$ is a Dirac fermion, we will expect $\sum_{\ml{B}} n_{\ml{B}} + n_{\chi}$ to be conserved while if $\chi$ is a Majorana fermion or a real scalar, we will have $2\sum_{\ml{B}} n_{\ml{B}} + n_{\chi}$ conserved, since the bound state contains two $\chi$ particles. Overall, we are consistent with Refs.~\cite{Gondolo:1990dk,Liew:2016hqo}. We note that our conversion in Eq.~(\ref{eq:BSFcrossx}) is written in the point of view of Dirac fermions.} As can be seen from these two rate equations, the attractor solution of the transitions is given by:
\begin{align}
\frac{n_{\chi}  n_{\bar{\chi}}}{n_{\chi}^{\text{eq}}  n_{\bar{\chi}}^{\text{eq}}} = \frac{n_{\mathcal{B}}}{n_{\mathcal{B}}^{\text{eq}}} \;,\quad \forall \mathcal{B}\;. \label{eq:saha}
\end{align}
This is nothing but ionization equilibrium condition, relating the scattering state and the bound state number densities. For thermally produced DM, this relation is generically satisfied in the high temperature regime.
In such an equilibrium state, the collision term in each rate equation introduced vanishes and is thus independent of the actual value of the bound state formation cross section, as can be seen by inserting the relation in Eq.~(\ref{eq:saha}) into Eq.~(\ref{eq:nscatter}) or Eq.~(\ref{eq:nbound}).

{ In contrast, the scattering states start to decouple from the bound states typically when the Debye mass becomes comparable to the absolute value of the bound state binding energy. This is why we perform a fixed NLO computation of the electric correlator entering the bound state formation cross section~(\ref{eqn:form_cross_section}) without any resummation or hard-thermal-loop approximation, in order to precisely resolve the decoupling from ionization equilibrium around the expected regime where $m_D \lesssim E$.}

With this discussion, we have introduced the non-Abelian electric field correlator that determines both the quarkonium and DM bound state formation and dissociation rates. We will now proceed to explain the details of the NLO calculation of the correlator in the next section.

\section{Electric field correlator at NLO}
\label{sect:nlo}

In this section, we discuss the NLO calculation of the non-Abelian electric field correlator
\begin{align}\label{eq:elcorrelatorcalc}
[g_E^{++}]^{>,da}_{ji}(y,x) \equiv \Big\langle  \big[{E}_j(y) \ml{W}_{[( y^0, {\bs y}), (+\infty, {\bs y})]} \big]^d
\big[ \ml{W}_{[(+\infty, {\bs x}),(x^0, {\bs x})]} {E}_i(x) \big]^a \Big\rangle_T \,,
\end{align}
which, as discussed above, fully determines the rates of bound state formation/dissociation inside the thermal plasma in the introduced open quantum system treatment. We have omitted the Wilson lines at infinite time for our calculations in $R_\xi$ gauge, since they do not contribute in the end, as discussed below Eq.~(\ref{eqn:lo_rho}).

We define $\Tr_E(\rho_EO) \equiv \langle O \rangle_T$ as a short-hand notation for the correlation of operators in the thermal environment. Gauge fields written as $A = A^aT_{\bs N}^a$ are in the fundamental representation while those written as $\ml{A} = A^a T_{\bs{adj}}^a$ are in the adjoint representation. The fundamental representation is normalized by $\Tr_c(T_{\bs N}^a T_{\bs N}^b) = C({\bs N}) \delta^{ab}$ where $C({\bs N})=1/2$ and $\Tr_c$ denotes the trace in the color space.
The adjoint representation is given by $(T_{\bs{adj}}^a)^{bc} = -if^{abc}$.
The chromoelectric field is defined as $E_i^a = F_{0i}^a = \partial_0 A_i^a - \partial_i A_0^a + g f^{abc} A_0^b A_i^c$.

Since it is a rather lengthy calculation, we will focus on describing the main aspects of the calculation so that the interested reader can straightforwardly reproduce it. First, we discuss the formulation of the calculation and define the conventions we will use throughout. We then proceed to describe the contributing diagrams, and we explicitly verify that they give a gauge-invariant result by using the $R_\xi$ gauge. After that, we explain the calculation of the electric field correlator, discussing both the vacuum theory results ($T=0$) as well as the intrinsically finite-temperature pieces, and examining some aspects of the infrared and collinear limits of the relevant diagrams. We close this section by adding up all contributions so that we can readily apply this calculation to the computation of the bound state formation and dissociation rates.

\subsection{Formulation and conventions}

In this subsection we outline the formalism we use to perform our calculation, as well as making the appropriate definitions of our conventions regarding: field branches on the Schwinger-Keldysh contour, momentum flows in Feynman diagrams with the corresponding sign conventions in the Feynman rules, and introducing the main objects that are involved in these rules. As the final part of this subsection, we enumerate all Feynman diagrams that contribute to the non-Abelian electric field correlator~\eqref{eq:elcorrelatorcalc}, and therefore set up all of the groundwork needed to carry out the explicit calculation in later sections.

\begin{figure}
    \centering
    \includegraphics[width=0.9\textwidth]{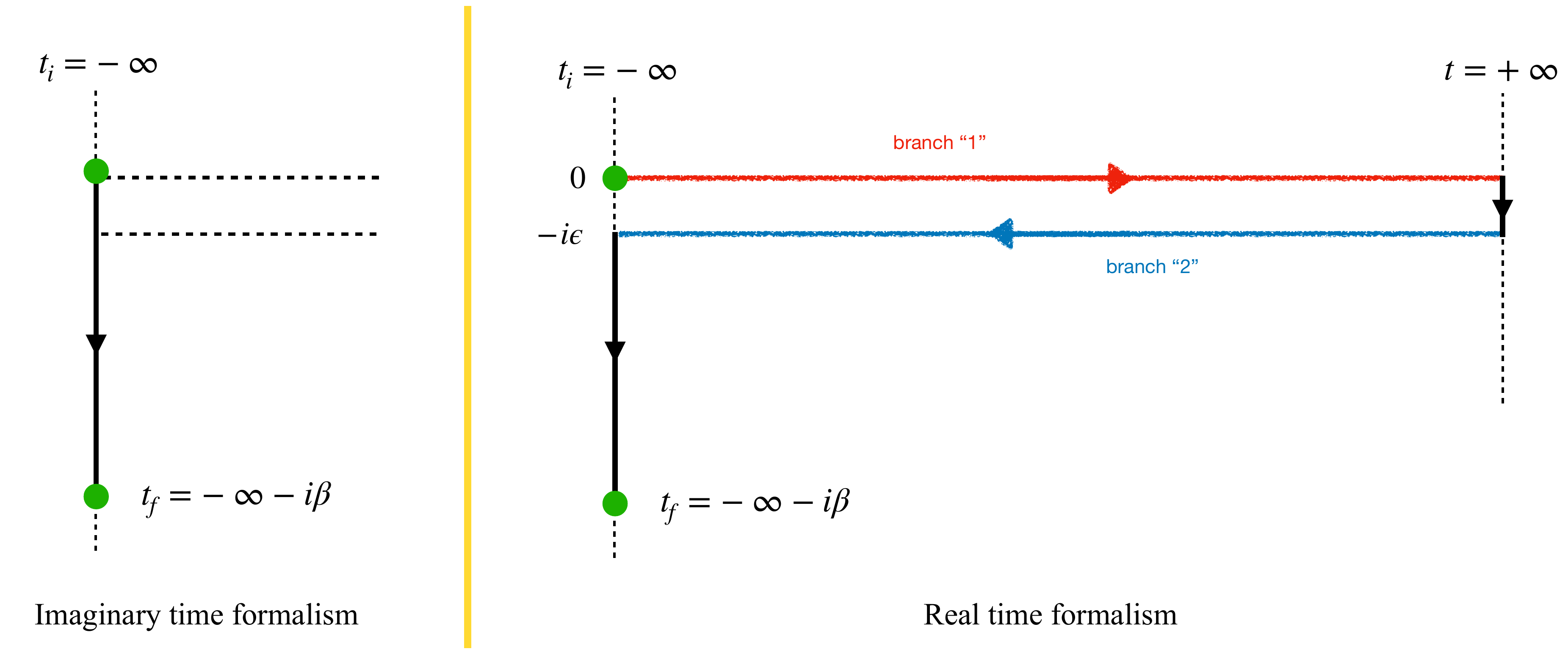}
    \caption{Left: (Imaginary) Time integration contour in the imaginary-time formalism, where the partition function $\int \ml{D} \phi\, e^{-\beta H} $ is calculated by writing $e^{-\beta H}$ using a path integral representation, as one does in quantum mechanics for $e^{-iHt}$, but with time going from $t_i= - \infty $ to $t_f = - \infty -i\beta $ (the reference real time, in this case ${\rm Re} \{t_i\} = {\rm Re} \{t_f\} = -\infty$ is arbitrary for time-independent observables). Right: The conventional Schwinger-Keldysh contour in the real-time formalism, where the imaginary-time path is deformed to allow for insertions of operators at arbitrary real times. To close the path and allow for insertions at an arbitrary real time, it is necessary to go from $t=-\infty$ to $t=+\infty$ and back. From the real time $t$ point of view, there are two copies of the fields: one in a time-ordered branch (branch 1, in red) and the other in an anti-time-ordered branch (branch 2, in blue). Both contours are ``closed'' in the sense that the field configurations at the initial $t_i$ and final time $t_f$ are identified when taking the trace~\eqref{eq:def-expectation}.}
    \label{fig:keldysh-contour}
\end{figure}

\subsubsection{Correlations on the Schwinger-Keldysh contour}

We first review the standard real-time formalism of thermal field theory. In this framework, the number of fields is doubled in order to be able to compute correlation functions at arbitrary (real) time separation. Formally speaking, one would write
\begin{equation} \label{eq:def-expectation}
    \langle \O \rangle = {\rm Tr}\left[ \rho \O \right]\,,
\end{equation}
with $\O$ some operator of which we want to know the expectation value, and in thermal equilibrium $\rho = \frac{1}{{\rm Tr} ( e^{-\beta H} )} e^{-\beta H} $, with $\beta = 1/T$ the inverse temperature. To make progress, one formally evaluates the trace in~\eqref{eq:def-expectation} by inserting a complete set of basis states
\begin{equation}
    {\rm Tr}\left[ \rho \O \right] = \sum_{\phi, \phi'} \bra{\phi'} \rho \ket{ \phi } \bra{\phi} \O \ket{ \phi' }\,,
\end{equation}
where, in field theory, we can think of $\ket{\phi}$ as an eigenstate of a field operator $\hat{\varphi}({\bf x})$, with a function-valued eigenvalue $\phi({\bf x})$, i.e., $\hat{\varphi}({\bf x}) \ket{\phi} = \phi({\bf x}) \ket{\phi}$. Then, if the operator $\ml{O}$ depends on $\hat{\varphi}$, we can replace the $\varphi$ operator by its corresponding eigenvalue $\phi$ when $\ml{O}$ acts on $|\phi\rangle$, i.e. $\ml{O}(\hat{\varphi})|\phi\rangle = \ml{O}(\phi)|\phi\rangle$. However, the set of basis states is complete at any given time slice, meaning that field configurations (i.e., \textit{states}) at different times can be expressed in terms of each other, which in general leads to the possibility that they may have complicated non-local expressions in terms of each other because the time-evolution operator will act non-trivially on $|\phi \rangle$.\footnote{Here we are thinking about having an initial basis of states $|\phi_i(t=0)\rangle$, which in the Schr\"odinger picture is evolved to $|\phi_i(t)\rangle = U(t) |\phi_i(t=0)\rangle$. In general, the overlap $\langle \phi_i(t) | \phi_j(0) \rangle $ is nonzero, meaning that the action of field operators acting on one basis will be different from the other.} This means that if the operator $\O$ depends on $\hat{\varphi}$ at different time slices, it is most convenient to insert a basis of field eigenstates around each field operator $\hat{\varphi}(x)$ at a given time, with the basis eigenstates being at the same time slice, so that the operator can be conveniently replaced by a function corresponding to an eigenvalue of $\hat{\varphi}(x)$ at that time. 

The standard way to insert the field operator eigenstates is the path integral formalism. All we have to do is to insert (infinitely) many bases of states that smoothly connect the starting point of the ``time-evolution'' operator $e^{-\beta H}$, which we can take to be anywhere in the $t$-complex plane, to the final point of the time evolution that is displaced by $-i\beta$ from the starting point. Two standard contours are shown in Figure~\ref{fig:keldysh-contour}, which correspond to the imaginary-time and real-time formalisms of thermal field theory respectively. Both contours are equally valid in the sense of connecting the starting and ending points. But they allow for a different set of possible insertions of operators. In particular, the real-time formalism explicitly allows for operators evaluated at any combination of real times to be inserted in the thermal expectation value. To allow for operator insertions at all real times, we choose this initial time to be $t_i=-\infty$.\footnote{In the sense that we take the starting point $t_i \to -\infty$ in every calculation.} The ordering of the operators is encoded in the path integral by the position along the time contour on which that operator is placed. For instance, operators on branch 2 are always behind operators on branch 1 in a correlation function in the sense of contour ordering. More generally, operators with a time coordinate ``closer'' to $t_f=-\infty -i\beta$ appear later than operators that are ``closer'' to $t_i=-\infty$ along the contour. The path integral representation of the generating functional of correlation functions in terms of the fields that are supported on each segment of the contour can be written as:
\begin{equation}
    Z[J_1,J_2] = \int \ml{D} \phi_E\, \ml{D}\phi_1\, \ml{D} \phi_2\, e^{-S_E[\phi_E] + iS[\phi_1] - iS[\phi_2] - \int_x \left[ J_1(x) \phi_1(x) - J_2(x) \phi_2(x) \right] }\,,
\end{equation}
where the $\phi_E$ field lives on the ``imaginary-time'' part of the contour, and the $\phi_1$ and $\phi_2$ fields live on the branches 1 and 2 respectively. $S_E[\phi]$ is the Euclidean action, and $S[\phi]$ is the real-time action. The symbol $\int_x$ is a short hand notation for $\int\diff^4x$. The negative sign associated with terms on the branch 2 is originated in the Hermitian conjugate of the time evolution operator.

Therefore, to enforce the explicit operator ordering in the correlator~\eqref{eq:elcorrelatorcalc}, we would place the fields at position ${\bs x}$ on the first branch of the Schwinger-Keldysh contour, and the fields at position ${\bs y}$ on the second branch. We note that since the first branch is time-ordered, and the second one is anti time-ordered, the path ordering of the operators in the Wilson lines is naturally implemented in each case. Then, we would expand in powers of the coupling constant and compute the $\ml{O}(g^2)$ correction to the correlator.

Conceptually, there is nothing preventing us from computing~\eqref{eq:elcorrelatorcalc} directly (without any reference to other operator orderings). However, in the light of certain issues (which we will discuss later) that are easier to address if we employ the KMS relations that the correlator satisfies at any definite temperature, we will formulate our calculation by starting from a more general object
\begin{align}\label{eq:elcorrelatorgeneral}
[g_E^{++}]^{da}_{ji,JI}(y,x) \equiv \Big\langle  \ml{T}_C \big[{E}_j(y) \ml{W}_{[( y^0, {\bs y}), (+\infty, {\bs y})]} \big]^d_J
\big[ \ml{W}_{[(+\infty, {\bs x}),(x^0, {\bs x})]} {E}_i(x) \big]^a_I \Big\rangle\,,
\end{align}
where $\ml{T}_C$ denotes the contour ordering (fields further along the contour are placed to the left) and $I,J$ are indices that indicate on which branch of the Schwinger-Keldysh contour the fields are located. This general correlator (\ref{eq:elcorrelatorgeneral}) reduces to the physically relevant correlator (\ref{eq:elcorrelatorcalc}) when we take $I=1$, $J=2$, and it also provides working definitions of other finite-temperature correlators with different contour orderings. For example, we can define the {\it time-ordered} version of this correlator to be the one where we take $I = J = 1$,\footnote{{ The correlation defined by $I=J=1$ is in general different from the following correlator
\begin{align}
\theta(y^0 - x^0) [g_E^{++}]^>(y,x)  + \theta(x^0 - y^0) [g_E^{++}]^<(y,x) \,, \nonumber 
\end{align}
See the discussions at the end of Section~\ref{sec:add-results}.}
} and also construct the {\it retarded} correlator by using that, in general, the time-ordered $G^{\ml{T}}$ and retarded $G^R$ 2-point functions in a thermal background satisfy
\begin{align}
G^{\ml{T}}(p) = G^R(p) + G^<(p) \,.
\end{align}

We stress that, since the physical object that determines the rates is given by the correlator~\eqref{eq:elcorrelatorcalc}, introducing these new correlators is, so far, more of a mathematical tool than an additional point of view revealing of some physical aspects of the calculation. However, we will comment on any physical aspect that becomes apparent from the new correlators as we explain the calculation. For now, given the KMS relations shown in Section~\ref{subsect:elcol}, we know that, mathematically, all we need to know to describe the rates is the following spectral function:
\begin{align}
\big[\rho_E^{++} \big]^{da}_{ji}(y,x) \equiv \big[g_E^{++} \big]^{>,da}_{ji}(y,x) - \big[g_E^{++} \big]^{<,da}_{ji}(y,x) \,,
\end{align}
and as such, we will focus on computing the spectral function. The usefulness of doing the calculation this way will become apparent when we discuss collinear finiteness of the result. 
{ We want to emphasize that the lesser correlation function $[g_E^{++}]^{<,da}_{ji}(y,x)$ is not equal to the correlator with $I=2,J=1$ in general, due to the different ordering of the gauge fields with respect to the electric field. Calculating $[g_E^{++}]^{<,da}_{ji}(y,x)$ requires going beyond the usual Schwinger-Keldysh contour, while calculating the correlator with $I=2,J=1$ can be done in the Schwinger-Keldysh formalism. In the following, when this subtlety becomes crucial in certain diagrams, we will use the KMS relation to replace  $[g_E^{++}]^{<,da}_{ji}(y,x)$ with $[g_E^{--}]^{>,da}_{ji}(-y,-x)$, which can be calculated in the conventional Schwinger-Keldysh formalism.
}

\subsubsection{Sign conventions and Feynman rules}

Before proceeding to any actual calculation, we must first establish some conventions regarding the flow of momenta through the diagrams, our working definitions of propagators, and the SU$(N_c)$ gauge theory Feynman rules. While these conventions are usually devoted to an appendix, we consider that, albeit somewhat technical, they highlight important aspects of the calculation that we will perform. Nonetheless, because we do not want to overload our development with technical details, in this section we will establish the mathematical machinery that we will use only for the purely gauge boson sector of the theory, leaving the details of the fermion and ghost propagators to Appendix~\ref{app:FeynmanRules}, where we will also repeat the definitions we now present.

As a starting point, let us introduce the \textit{free} gauge boson propagators of the theory in $R_\xi$ gauge. Depending on where the gauge boson fields $A_\mu^a$ are inserted on the Schwinger-Keldysh contour, we can have different types of propagators, with a general structure given by
\begin{align}
D^{Y,ab}_{\mu \nu}(k) = \delta^{ab} P_{\mu \nu}(k) D^Y(k)\,,
\end{align}
where $Y$ can be any of $>,<,\ml{T},\overline{\ml{T}}$, and
\begin{align}
P_{\mu \nu }(k) = - \left[ g_{\mu \nu} - (1 - \xi) \frac{k_\mu k_\nu}{k^2} \right]\,.
\end{align}
with the metric signature $(+,-,-,-)$. The different types of propagators in the free theory ($g=0$) are given by
\begin{align}
D^>(k) &= \left( \Theta(k_0) + n_B(|k_0|) \right) 2\pi \delta(k^2)\,, &  D^<(k) &= \left( \Theta(-k_0) + n_B(|k_0|) \right) 2\pi \delta(k^2)\,, \nonumber \\
D^{\ml{T}}(k) &= \frac{i}{k^2 + i0^+} + n_B(|k_0|) 2\pi \delta(k^2)\,, &  D^{\overline{\ml{T}}}(k) &= \frac{-i}{k^2 - i0^+} + n_B(|k_0|) 2\pi \delta(k^2)\,,
\end{align}
which are called Wightman functions (the two propagators on the first line), time-ordered propagator, and anti time-ordered propagator, respectively. As introduced earlier, $n_B(k_0) = (\exp(k_0/T)-1)^{-1}$ is the Bose-Einstein distribution. It is also useful to define
\begin{align}
    D^R(k) &= \frac{i}{k^2 + i0^+ {\rm sgn}(k^0) }\,, &  D^A(k) &= \frac{i}{k^2 - i0^+ {\rm sgn}(k^0) }\,, \nonumber \\ D^S(k) &= D^>(k) + D^<(k) = (1 + 2n_B(|k_0|)) 2 \pi \delta(k^2)\,, & &
\end{align}
as the \textit{free} retarded, advanced, and symmetric propagators respectively.

\begin{figure}
	\centering
	\begin{tabular}{  c  c  l  }
    \raisebox{-0.26in}{\includegraphics[height=0.6in]{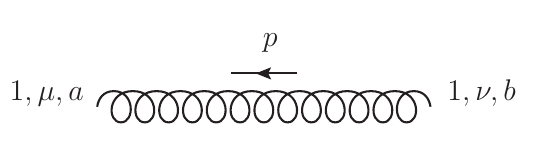}}  &=& $ \delta^{ab} P_{\mu\nu} D^{\ml{T}}(p)  $\\
	\raisebox{-0.26in}{\includegraphics[height=0.6in]{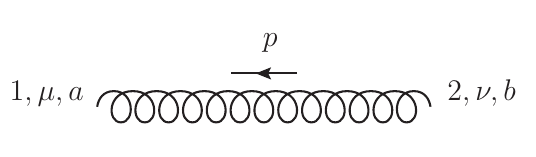}}  &=& $ \delta^{ab} P_{\mu\nu} D^{<}(p) $  \\
	\raisebox{-0.26in}{\includegraphics[height=0.6in]{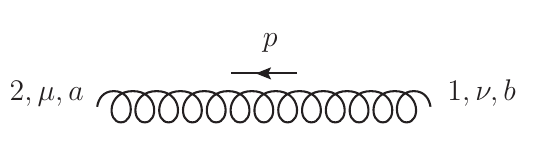}}  &=& $ \delta^{ab} P_{\mu\nu} D^{>}(p)  $\\
	\raisebox{-0.26in}{\includegraphics[height=0.6in]{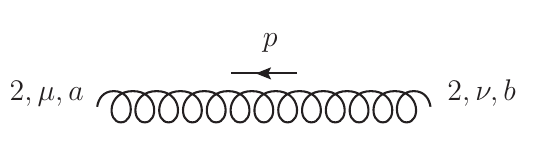}}  &=& $ \delta^{ab} P_{\mu\nu} D^{\overline{\ml{T}}}(p) $ 
	\end{tabular}
\caption{Feynman rules associated with different types of gauge boson propagators.}
\label{fig:rules-props-maintext}
\end{figure}

In terms of the indices of the Schwinger-Keldysh contour, we can compactly write our propagators as
\begin{align}
\mathbb{D}(p)_{JI} =  \begin{bmatrix} D^{\ml{T}}(p)  &  D^<(p) \\
D^>(p) & D^{\overline{\ml{T}}}(p)  \end{bmatrix} _{JI}\,,
\end{align}
for example, $\mathbb{D}(p)_{21} = D^>(p)$, and $\mathbb{D}(p)_{11} = D^{\ml{T}}(p)$. With
these definitions, correlations in position space of two fields in branches $I$ and $J$ of the Schwinger-Keldysh contour are given by
\begin{align}
\langle \phi_J(y) \phi_I(x) \rangle = \mathbb{D}(y-x)_{JI} = \int \frac{\dd^4 k}{(2\pi)^4} e^{-i k \cdot (y-x) } \, \mathbb{D}(p)_{JI}\,,
\end{align}
where $\phi$ can be thought of as a scalar field since we have taken out the non-Abelian indexes and the Lorentz structure.
Crucially, this convention of the Fourier transform defines the signs of the momenta appearing in the Feynman rules, to which we now turn. Pictorially, the momentum ``flow'' in a propagator $\mathbb{D}(p)_{JI}$ should be diagrammatically depicted as going from the field insertion of type $I$ towards the field insertion of type $J$. The Feynman rules that illustrate the relevance of having a consistent definition of momentum flow most clearly are those of the propagators themselves, in a manner consistent with our previous definitions. We list them in Figure~\ref{fig:rules-props-maintext}.

\begin{figure}
	\centering
	\begin{tabular}{  c  c  l  }
	\raisebox{-0.8in}{\includegraphics[height=1.6in]{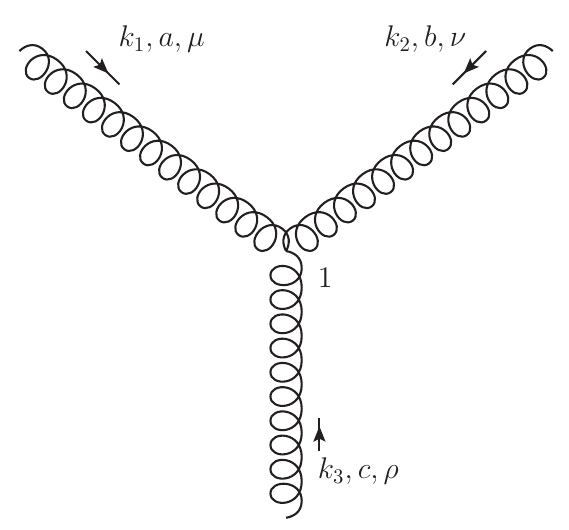}}  &=& $ gf^{abc}\Big( g_{\mu\nu} (k_1-k_2)_\rho + g_{\nu\rho} (k_2-k_3)_\mu + g_{\rho\mu} (k_3-k_1)_\nu \Big) $  \\
	\raisebox{-0.8in}{\includegraphics[height=1.6in]{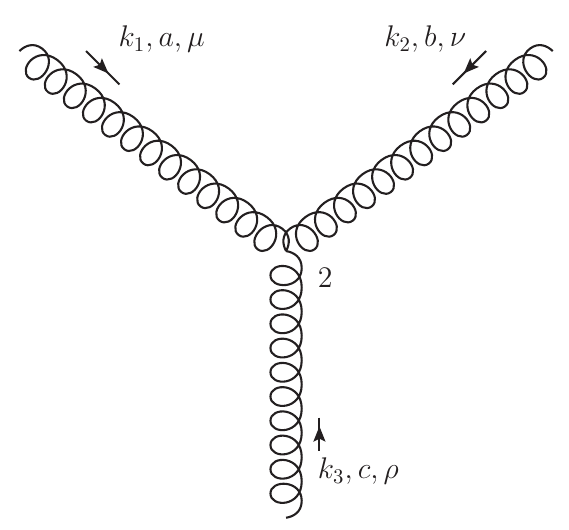}}  &=& $ -gf^{abc}\Big( g_{\mu\nu} (k_1-k_2)_\rho + g_{\nu\rho} (k_2-k_3)_\mu + g_{\rho\mu} (k_3-k_1)_\nu \Big) $
	\end{tabular}
\caption{Feynman rules associated to the 3-gauge boson vertex, for both types of fields on the branches of the Schwinger-Keldysh contour along the real-time directions.}
\label{fig:rules-3gluon-maintext}
\end{figure}

Two other ingredients in our diagrammatic calculations are particularly sensitive to the choice of signs related to the momentum flow. One ingredient is the 3-gauge boson vertex, which involves the incoming/outgoing momentum from each of its external legs explicitly in the corresponding Feynman rule. We show the 3-gauge boson vertex in Figure~\ref{fig:rules-3gluon-maintext}, for both fields of type 1 and 2 on the Schwinger-Keldysh contour, just to emphasize that we have doubled the field content of the theory from the start. From now on, we will write all possible vertex insertions that come from expanding the interacting pieces of the action with a $(-1)^{I+1}$ factor beside the vertex, where $I\in \{1,2\}$ is an index that tells us on which branch the vertex is to be evaluated, so that we can effectively write all possible combinations of indices more compactly. For example, by writing
\begin{equation}
\label{eqn:example_3vertex}
    \D(k)_{I'I} \D(p)_{I'J} \D(q)_{I'K} (-1)^{I'+1} \,,
\end{equation}
we indicate a 3-particle vertex with three incoming particles that have momenta $k,p,q$, and live on branches $I,J,K$, respectively. We implicitly sum over the repeated $I'$ indices in the last expression (\ref{eqn:example_3vertex}) (by taking values $I'\in\{1,2\}$).

\begin{figure}
	\centering
	\begin{tabular}{  c  c  l  }
	\raisebox{-0.8in}{\includegraphics[height=2in]{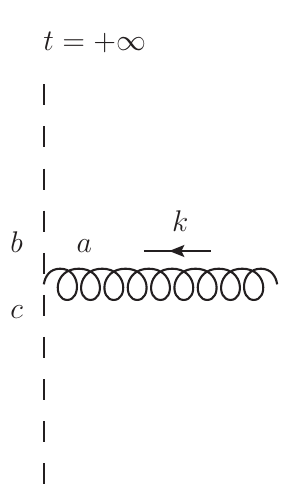}}  &=& $ \dfrac{gf^{abc}}{i k_0 + 0^+}$
	\end{tabular}
\caption{Feynman rule associated with gauge boson insertions of Wilson lines. The color index $b$ is to be contracted with the next operator along the Wilson line towards $t=+\infty$, while the index $c$ is contracted with the next operator in the direction towards the electric field insertion, which is at the lower end of the vertical dashed line and not shown here explicitly.}
    \label{fig:rules-momentum-Wilson}
\end{figure}

The other Feynman rule that is particularly sensitive to the sign convention of the momentum flowing through the diagram is the gauge boson insertion from the Wilson lines. There are two possibilities, depending on the sign convention of the incoming/outgoing momentum:
\begin{enumerate}
    \item Momentum flows away from the gauge boson insertion of the Wilson line at the spacetime point $z_s=(s,{\bs z})$ on the contour branch $K$ with color (index of the adjoint representation) given by $a$, towards the rest of the diagram (with vertex spacetime location, contour branch, color, and Lorentz index $(t',{\bs z}')$, $K'$, $a'$, and $\mu'$, respectively). In this case we have from the expansion of the Wilson line
    \begin{equation}
    \begin{split}
        \int_t^\infty \!\! \diff s \, e^{-\varepsilon s} \, \D(z'-z_s)_{\mu'0,K'K}^{a'a} &=  \delta^{a'a} \int_t^\infty \!\! \diff s \, e^{-\varepsilon s} \int \frac{\diff^d k}{(2\pi)^d} e^{-i k \cdot (z' - z_s)} P_{\mu' 0}(k) \D(k)_{K'K} \\
        &= \delta^{a'a} \int \frac{\diff^d k}{(2\pi)^d} e^{-i k \cdot (z' - z_t)} \frac{1}{-i k_0 + 0^+} P_{\mu' 0}(k) \D(k)_{K'K}\,,
    \end{split}
    \end{equation}
    where $z_t=(t,{\bs z})$ and we have introduced a positive infinitesimal $\varepsilon$ to make the Wilson line a well-defined operator as it approaches infinite time.
    \item Momentum flows towards the gauge boson insertion of the Wilson line at the spacetime point $z_s=(s,{\bs z})$ on the contour branch $K$ with color $a$, from the rest of the diagram (with vertex spacetime location, contour branch, color, and Lorentz index $(t',{\bs z}')$, $K'$, $a'$, and $\mu'$, respectively). In this case we have
    \begin{equation}
    \begin{split}
        \int_t^\infty \!\! \diff s \, e^{-\varepsilon s}\, \D(z_s-z')_{0\mu',KK'}^{aa'} &=  \delta^{aa'} \int_t^\infty \!\! \diff s \, e^{-\varepsilon s} \int \frac{\diff^d k}{(2\pi)^d} e^{-i k \cdot ( z_s-z')} P_{0\mu'}(k) \D(k)_{KK'} \\
        &= \delta^{aa'} \int \frac{\diff^d k}{(2\pi)^d} e^{-i k \cdot (z_t-z')} \frac{1}{i k_0 + 0^+} P_{0\mu'}(k) \D(k)_{KK'} \,.
    \end{split}
    \end{equation}
\end{enumerate}
We summarize these results diagrammatically in Figure~\ref{fig:rules-momentum-Wilson}, where we also include the relevant color factors $f^{abc}$, which come from expanding the Wilson line in the adjoint representation. We give the rest of the Feynman rules in Appendix~\ref{app:FeynmanRules}.

We will work in $d$ spacetime dimensions to regulate the potentially ultraviolet (UV) divergence, and take the limit $d\to 4$ at the end of the calculation after renormalization. In practice, we take the spacetime to be $(1,d-1)$ dimensional, i.e., the limit $d\to 4$ is taken by varying the number of spatial dimensions, while always holding the number of time-like dimensions fixed.

\subsubsection{Contributing Feynman diagrams}

Now we proceed to depict all the diagrams that contribute to the generalized thermal electric correlator~\eqref{eq:elcorrelatorgeneral} at next-to-leading order. As we will see momentarily, a natural way to group the terms coming from each Feynman diagram is to take a look at their propagator structures in terms of the different propagator combinations that appear. The first criterion to do this is to separate the different contributions by the number of propagators that appear in each diagram shown in Figure~\ref{fig:diagrams}.

\begin{figure}
    \centering
    \begin{subfigure}[b]{0.33\textwidth}
        \centering
        \includegraphics[height=1.4in]{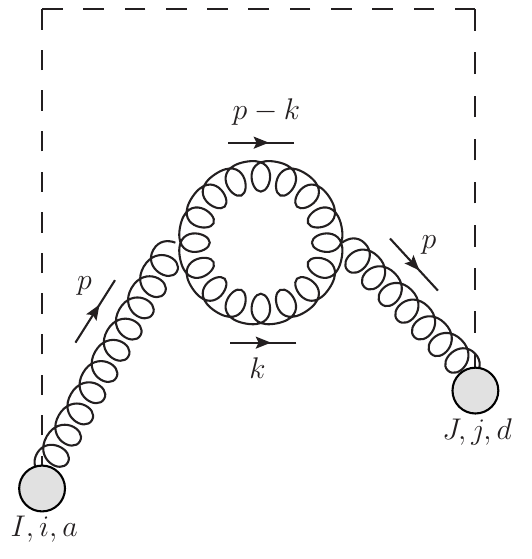}
        \caption*{$(1)$}\label{subfig:1}
    \end{subfigure}%
    ~ 
    \begin{subfigure}[b]{0.33\textwidth}
        \centering
        \includegraphics[height=1.4in]{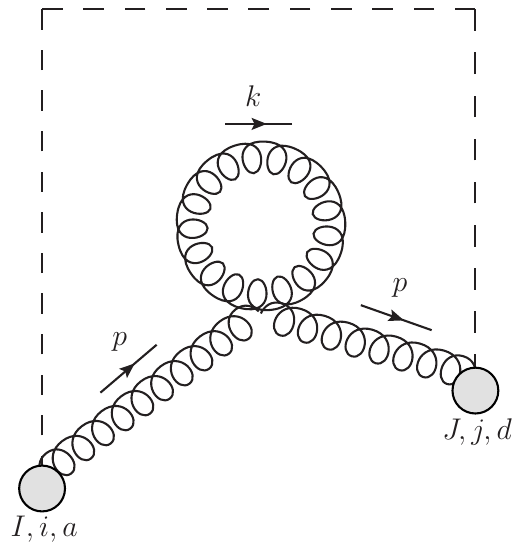}
        \caption*{$(2)$}\label{subfig:2}
    \end{subfigure}%
    ~ 
    \begin{subfigure}[b]{0.33\textwidth}
        \centering
        \includegraphics[height=1.4in]{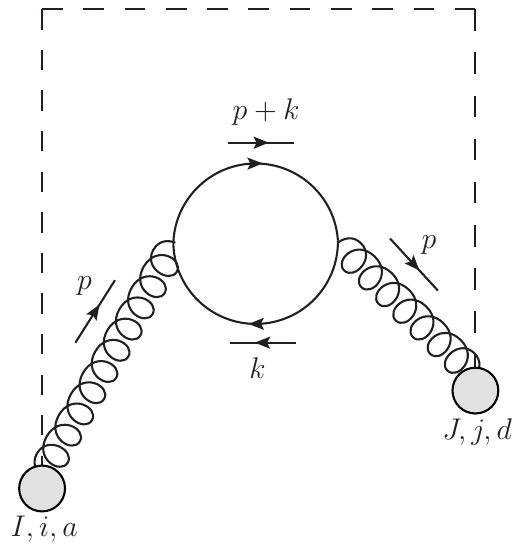}
        \caption*{$(f)$}\label{subfig:f}
    \end{subfigure}%

    \begin{subfigure}[b]{0.33\textwidth}
        \centering
        \includegraphics[height=1.4in]{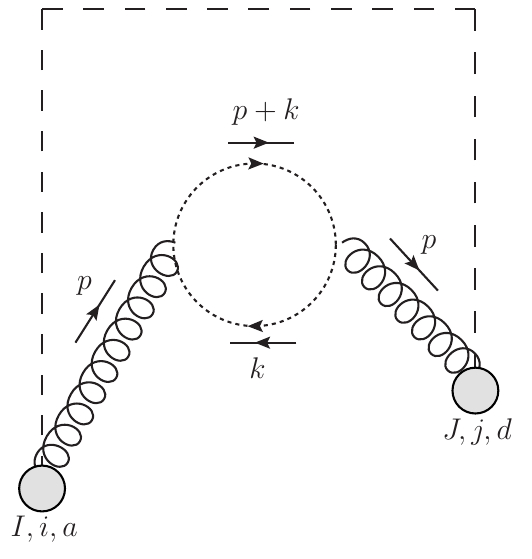}
        \caption*{$(g)$}\label{subfig:g}
    \end{subfigure}%
    ~ 
    \begin{subfigure}[b]{0.33\textwidth}
        \centering
        \includegraphics[height=1.4in]{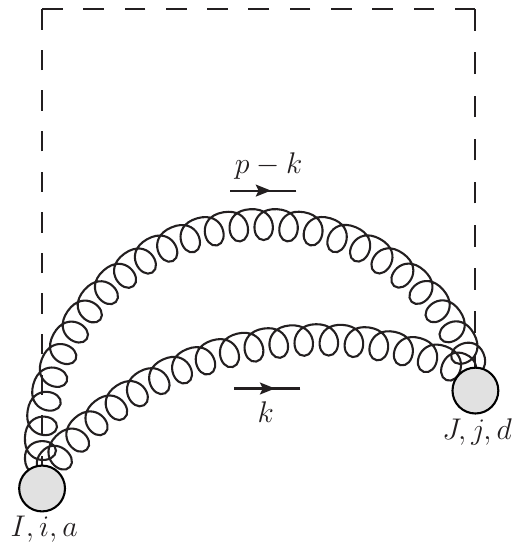}
        \caption*{$(3)$}\label{subfig:3}
    \end{subfigure}%
    ~ 
    \begin{subfigure}[b]{0.33\textwidth}
        \centering
        \includegraphics[height=1.4in]{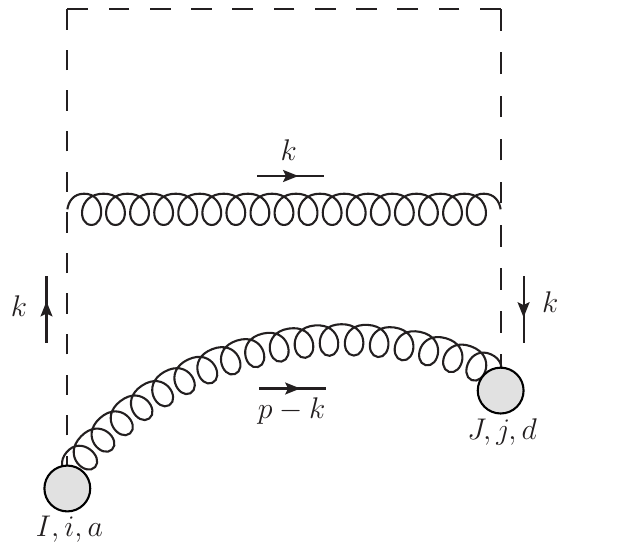}
        \caption*{$(4)$}\label{subfig:4}
    \end{subfigure}%
    
   \begin{subfigure}[b]{0.33\textwidth}
        \centering
        \includegraphics[height=1.4in]{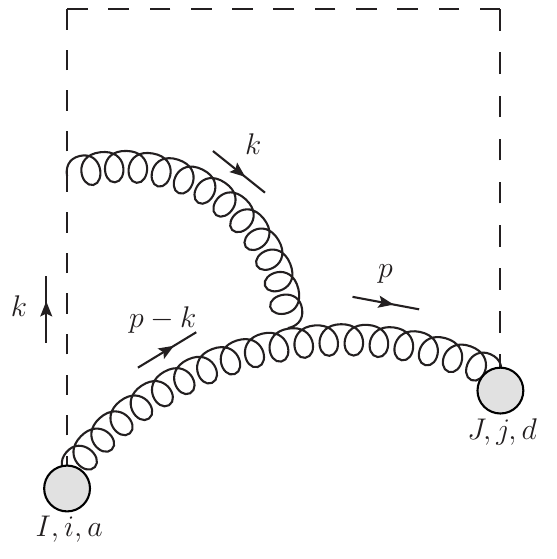}
        \caption*{$(5)$}\label{subfig:5}
    \end{subfigure}%
    ~ 
    \begin{subfigure}[b]{0.33\textwidth}
        \centering
        \includegraphics[height=1.4in]{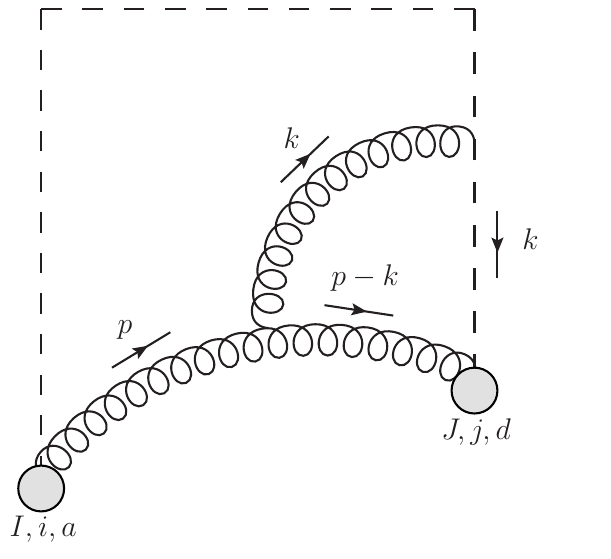}
        \caption*{$(5r)$}\label{subfig:5r}
    \end{subfigure}%
    ~  
    \begin{subfigure}[b]{0.33\textwidth}
        \centering
        \includegraphics[height=1.4in]{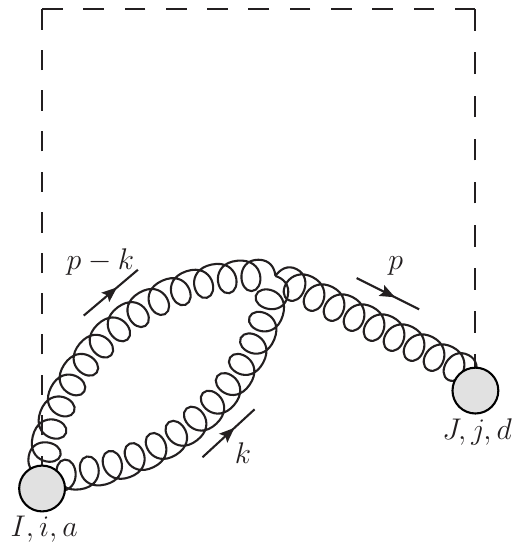}
        \caption*{$(6)$}\label{subfig:6}
    \end{subfigure}%
    
    \begin{subfigure}[b]{0.33\textwidth}
        \centering
        \includegraphics[height=1.4in]{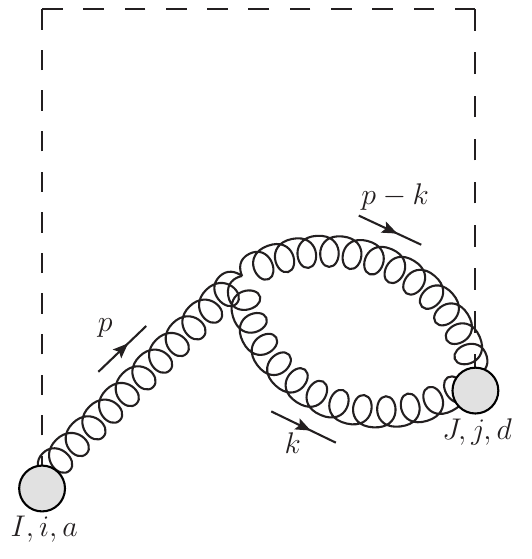}
        \caption*{$(6r)$}\label{subfig:6r}
    \end{subfigure}%
    ~
    \begin{subfigure}[b]{0.33\textwidth}
        \centering
        \includegraphics[height=1.4in]{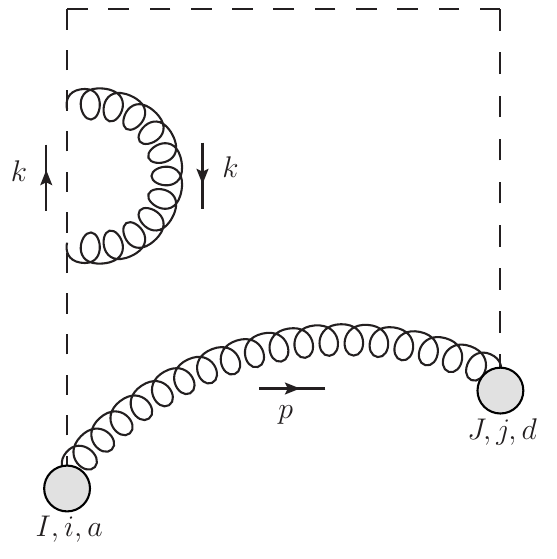}
        \caption*{$(7)$}\label{subfig:7}
    \end{subfigure}%
    ~      
   \begin{subfigure}[b]{0.33\textwidth}
        \centering
        \includegraphics[height=1.4in]{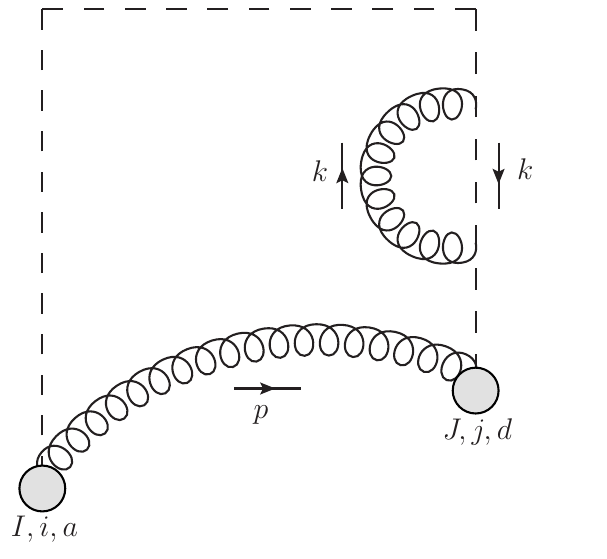}
        \caption*{$(7r)$}\label{subfig:7r}
    \end{subfigure}%
     
     \begin{subfigure}[b]{0.33\textwidth}
        \centering
        \includegraphics[height=1.4in]{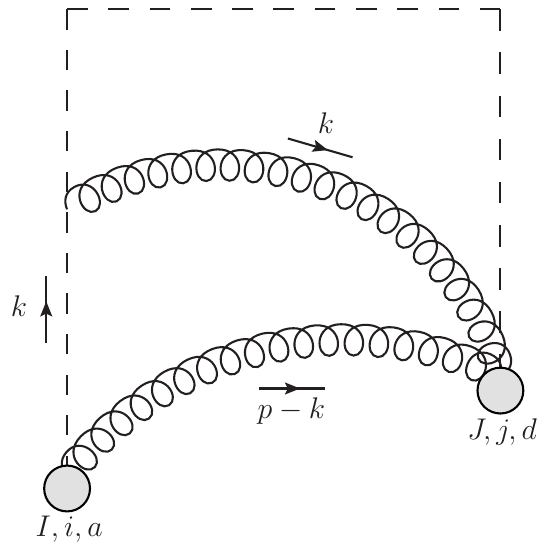}
        \caption*{$(8)$}\label{subfig:8}
    \end{subfigure}%
    ~ 
     \begin{subfigure}[b]{0.33\textwidth}
        \centering
        \includegraphics[height=1.4in]{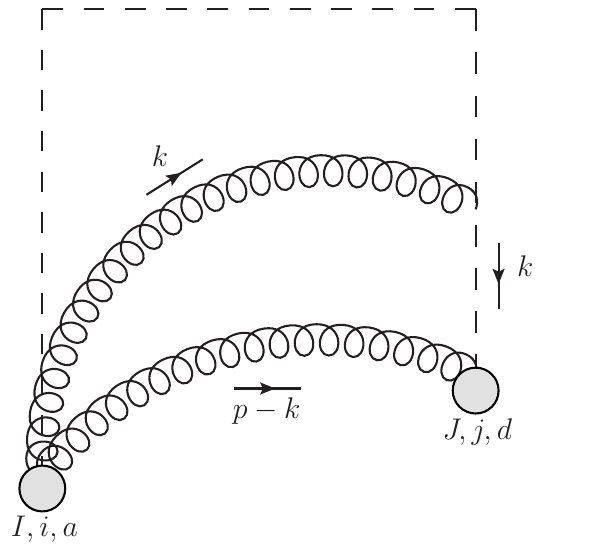}
        \caption*{$(8r)$}\label{subfig:8r}
    \end{subfigure}%
    ~
    \begin{subfigure}[b]{0.33\textwidth}
        \centering
        \includegraphics[height=1.4in]{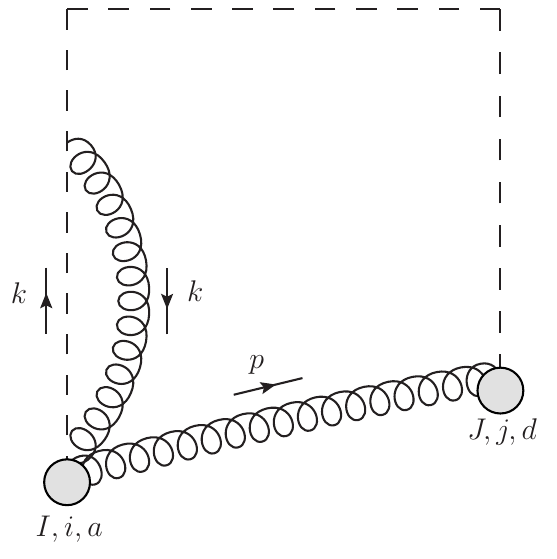}
        \caption*{$(9)$}\label{subfig:9}
    \end{subfigure}%
\end{figure}
\begin{figure}[htb]\ContinuedFloat
    \begin{subfigure}[b]{0.33\textwidth}
        \centering
        \includegraphics[height=1.4in]{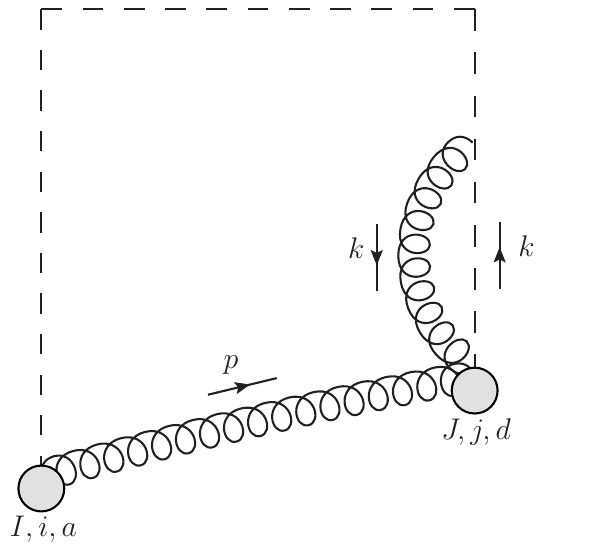}
        \caption*{$(9r)$}\label{subfig:9r}
    \end{subfigure}%
    ~   
    \begin{subfigure}[b]{0.33\textwidth}
        \centering
        \includegraphics[height=1.4in]{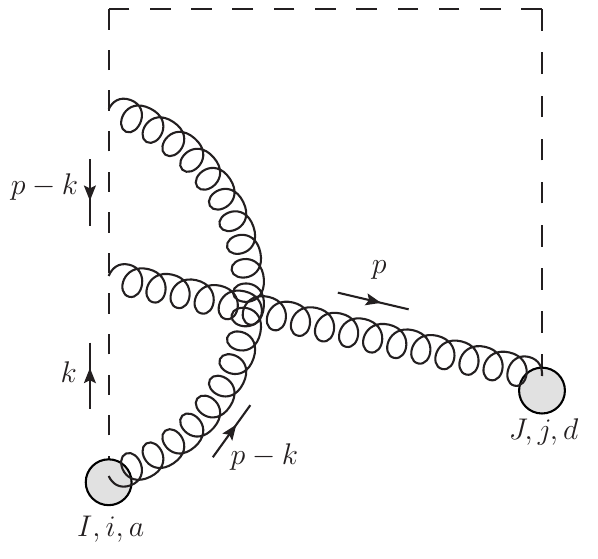}
        \caption*{$(10)$}\label{subfig:10}
    \end{subfigure}%
    ~    
    \begin{subfigure}[b]{0.33\textwidth}
        \centering
        \includegraphics[height=1.4in]{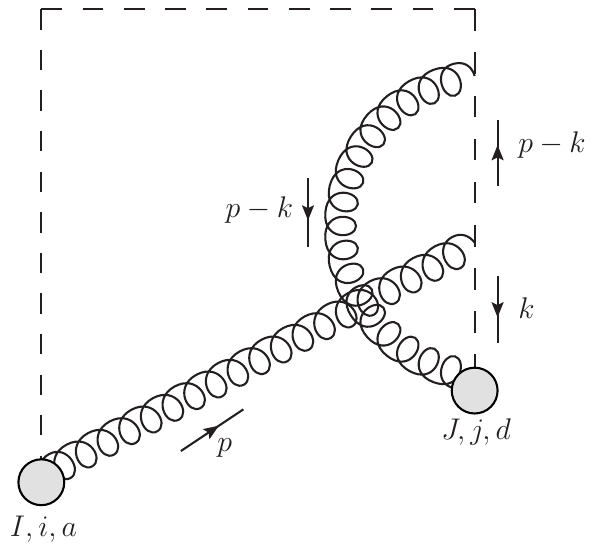}
        \caption*{$(10r)$}\label{subfig:10r}
    \end{subfigure}%
    
    \begin{subfigure}[b]{0.33\textwidth}
        \centering
        \includegraphics[height=1.4in]{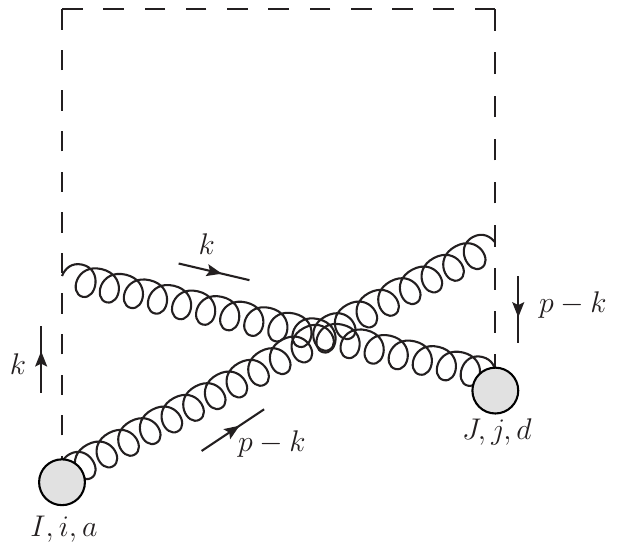}
        \caption*{$(11)$}\label{subfig:11}
    \end{subfigure}%
    \caption{List of all diagrams contributing to the electric field correlator~\eqref{eq:elcorrelatorgeneral}, with free indices $(I,i,a)$ and $(J,j,d)$ for the corresponding electric field insertions. The long dashed lines represent the Wilson lines while the short dashed lines label the ghost field. Solid lines indicate the fermion field. The two grey blobs are the electric fields.}
    \label{fig:diagrams}
\end{figure}

Each diagram in Figure~\ref{fig:diagrams}, labeled by $(X)$, can be written in the following form
\begin{align}
\big(X\big)^{da}_{ji,JI}(p) = \delta^{ad} g^2 \int \frac{\diff^d k}{(2\pi)^d} \ml{Q}^{(X)}(p,k)_{JI} V^{(X)}(p,k)_{ji}\,,
\end{align}
where $\ml{Q}^{(X)}(p,k)_{JI}$ is a sum of products of $\mathbb{D}$ propagators associated with the diagram $(X)$, with momenta flows that depend only on $p$ and $k$, and $V^{(X)}(p,k)_{ji}$ is a rational function of the momenta $p_\mu, k_\nu$ given by the appropriate vertex factors appearing in each diagram.

\begin{table*}
\begin{center}
\begin{tabular}{c||c}
\toprule
    Diagram $(X)$ & Propagator structure $\ml{Q}^{(X)}(p,k)_{JI}$ \\ \midrule
    $ (1), (g) $ & $\D(p)_{I'I}  \D(k)_{J'I'} \D(p-k)_{J'I'} \D(p)_{JJ'} (-1)^{I'+J'}$ \\ \midrule
    $ (f) $ & $ \D(p)_{I'I} {\rm Tr}[ \gamma^{\mu} \mathbb{S}(p-k)_{J'I'} \gamma^{\nu} \mathbb{S}(k)_{J'I'}  ] \D(p)_{JJ'} (-1)^{I'+J'}$ \\
 \bottomrule
\end{tabular}
\end{center}
\caption{Summary of propagator structures of diagrams with 4 propagators contributing to $\big[g_E^{++}\big]_{ji,JI}^{da}$. Summation over repeated $I',J'$ indices is implicit, even when there are three or more instances of such indices. Because fermionic propagators are intrinsically matrix-valued objects, we have made the choice to define the propagator structures with the gamma matrices included from the start. Fermionic propagators $\mathbb{S}_{IJ}$ are defined in Appendix~\ref{app:FeynmanRules}.}
\label{tab:4prop-structure}
\end{table*}

\begin{table*}
\begin{center}
\begin{tabular}{c||c}
\toprule
    Diagram $(X)$ & Vertex factors $V^{(X)}(p,k)_{ji}$  \\ \midrule \midrule
    $ (1) $ & $\dfrac{N_c}{2} (i p_0 g_{i \rho''}  - ip_i g_{0\rho''} ) P_{\mu \mu'}(k)  P_{\nu \nu'}(p-k)  (-i p_0 g_{j\rho'} + i p_j g_{0 \rho'})$ \\ & $\times \left[ g^{\rho'' \mu'} (p+k)^{\nu'} + g^{\mu' \nu'}(p - 2k)^{\rho''} + g^{\nu' \rho''} (k - 2p)^{\mu'} \right]$ \\ & $\times  \left[ g^{\rho' \mu} (-p-k)^{\nu} + g^{\mu \nu} (2k-p)^{\rho'} + g^{\nu \rho'}(2p-k)^{\mu} \right]$ \\ \midrule
    $ (g) $ & $(-1) N_c (i p_0 g_{i \mu}  - ip_i g_{0\mu} ) P^{\mu \mu'}(k) k_{\mu'} (p_{\nu'}-k_{\nu'}) P^{\nu' \nu}(p-k)  (-i p_0 g_{j\nu} + i p_j g_{0 \nu}) $\\ \midrule
    $ (f) $ & $ n_f C({\bs N}) (i p_0 g_{i \mu}  - ip_i g_{0\mu} ) (-i p_0 g_{j\nu} + i p_j g_{0 \nu})  $ \\
 \bottomrule
\end{tabular}
\end{center}
\caption{Summary of vertex factors of diagrams with 4 propagators contributing to $\big[g_E^{++}\big]_{ji,JI}^{da}$. The free $\mu,\nu$ indices in the fermion loop diagram $(f)$ are to be contracted with the free $\mu,\nu$ indices in the corresponding propagator structure.}
\label{tab:4prop-vertex}
\end{table*}

For diagrams with four propagators, 
we list their propagator structures in Table~\ref{tab:4prop-structure}, and their respective vertex factors in Table~\ref{tab:4prop-vertex}. Whenever two diagrams have the same propagator structure, we will list them together. As one might expect, it can be seen that the ghost diagram $(g)$ can be added to the gauge boson loop diagram $(1)$ without altering the propagator structure. This is not only convenient, but also necessary, because the Faddeev-Popov gauge-fixing procedure does not depend on whether the calculation is at finite temperature or not.

Next, we list the propagator structures of all diagrams with three propagators in Table~\ref{tab:3prop-structure}, and their respective vertex factors in Table~\ref{tab:3prop-vertex}. For the three-propagator structure, all contributions come purely from the Yang-Mills sector of the theory, which is,  in some sense, a result of the gauge invariance of the electric field correlator. As we will show in the next section, we will need all of these diagrams (plus those with two propagators) to compensate for the gauge dependence of diagram $(1)$.

\begin{table*}
\begin{center}
\begin{tabular}{c||c}
\toprule
    Diagrams $(X)$ & Propagator structure $\ml{Q}^{(X)}(p,k)_{JI}$ \\ \midrule
    $ (2) $ & $ \D(p)_{I'I}  \D(p)_{JI'}   \D(k)_{I'I'} (-1)^{I'+1} $ \\ \midrule
    $ (5),(6) $ & $ \D(p-k)_{I'I} \D(k)_{I'I} \D(p)_{JI'} (-1)^{I'+1} $ \\ \midrule
    $ (5r),(6r) $ & $\D(p-k)_{JI'} \D(k)_{JI'} \D(p)_{I'I} (-1)^{I'+1} $ \\ 
 \bottomrule
\end{tabular}
\end{center}
\caption{Summary of propagator structures of diagrams with 3 propagators contributing to $\big[g_E^{++}\big]_{ji,JI}^{da}$. Summation over repeated $I'$ indices is implicit, even when there are three or more instances of such indices.}
\label{tab:3prop-structure}
\end{table*}

\begin{table*}
\begin{center}
\begin{tabular}{c||c}
\toprule
    Diagram $(X)$ & Vertex factors $V^{(X)}(p,k)_{ji}$  \\ \midrule \midrule
    $ (2) $ & $ i N_c (i p_0 g_{i\mu} - ip_i g_{0\mu}) P_{\rho \rho'}(p) (-i p_0 g_j^{\rho} + i p_j g_{0}^{\rho}) \left[ g^{\mu \rho'} P_\nu^\nu(k) - P^{\mu \rho'}(k) \right]$ \\ \midrule
    $ (5) $ & $N_c \left[ g^{\rho \mu} (k-2p)^{\nu} + g^{\mu \nu}(p- 2k)^{\rho} + g^{\nu \rho} (k+p)^{\mu} \right]  P_{0\nu}(k) $  \\ 
    & $\quad \times  (p_0 g_{j\rho} - p_j g_{0 \rho}) ( (p-k)_0 g_{i\mu} - (p-k)_i g_{0\mu} )/(-ik_0 + 0^+)$\\ \midrule
    $ (5r) $ & $N_c \left[ g^{\rho \mu} (k-2p)^{\nu} + g^{\mu \nu}(p- 2k)^{\rho} + g^{\nu \rho} (k+p)^{\mu} \right]  P_{0\nu}(k)$ \\ &  $ \quad \times (p_0 g_{i\rho} - p_i g_{0 \rho}) ( (p-k)_0 g_{j\mu} - (p-k)_j g_{0\mu} )/({-ik_0 - 0^+})$ \\ \midrule
    $ (6) $ & $ N_c \left[ g^{\rho \mu} (-k-p)^{\nu} + g^{\mu \nu} ( 2k-p)^{\rho} + g^{\nu \rho} (2p-k)^{\mu} \right] $ \\ & $ \quad \times (i p_0 g_{j\rho} - i p_j g_{0 \rho})  P_{0\mu}(k) P_{i\nu}(p-k)$ \\ \midrule
    $ (6r) $ & $N_c \left[ g^{\rho \mu} (-k-p)^{\nu} + g^{\mu \nu} ( 2k-p)^{\rho} + g^{\nu \rho} (2p-k)^{\mu} \right] $ \\ &  $ \quad \times (i p_0 g_{i\rho} - i p_i g_{0 \rho})   P_{0\mu}(k) P_{j\nu}(p-k)$ \\
 \bottomrule
\end{tabular}
\end{center}
\caption{Summary of vertex factors of diagrams with 3 propagators contributing to $\big[g_E^{++}\big]_{ji,JI}^{da}$}
\label{tab:3prop-vertex}
\end{table*}

Finally, we discuss all remaining diagrams with two gauge boson propagators. In Figure~\ref{fig:diagrams}, we have presented the complete list of non-vanishing diagrams after performing the color indices contraction. From the diagrams with two gauge boson propagators, it is clear that the momenta flows in the propagators are decoupled (one of them carries $p$ and the other can be chosen to carry momentum $k$), and then we only need to analyze the vertex factors. It turns out that diagrams $(9),(9r),(10),(10r)$ do not contribute because of spacetime symmetries. Diagrams $(9)+(9r)$ vanish because one piece of the integrand is odd under $k_0 \to -k_0$, as can be verified explicitly using the Feynman rules, and the other piece of the sum is proportional to the integral of $\k$, which vanishes by rotational invariance of the plasma. It is necessary to take the sum so that the contribution from the pole in the $i0^+$ prescription from the Wilson line propagators cancels unambiguously. Similarly, $(10)$ and $(10r)$ vanish (separately) by rotational invariance because the integrand is proportional to $\k$. Therefore, regarding the diagrams with two gauge boson propagators, we only show the relevant propagator structures of non-vanishing diagrams in Table~\ref{tab:2prop-structure}, and their respective vertex factors in Table~\ref{tab:2prop-vertex}.

\begin{table*}
\begin{center}
\begin{tabular}{c||c}
\toprule
    Diagrams $(X)$ & Propagator structure $\ml{Q}^{(X)}(p,k)_{JI}$ \\ \midrule
    $ (3),(4),(8),(8r),(11) $ & $ \D(p-k)_{JI} \D(k)_{JI} $ \\ \midrule
    $ (7) $ & $ \D(k)_{II} \D(p)_{JI} $ \\ \midrule
    $ (7r) $ & $ \D(k)_{JJ} \D(p)_{JI} $ \\ \midrule
 \bottomrule
\end{tabular}
\end{center}
\caption{Summary of propagator structures of diagrams with 2 propagators contributing to $\big[g_E^{++}\big]_{ji,JI}^{da}$.}
\label{tab:2prop-structure}
\end{table*}

\begin{table*}
\begin{center}
\begin{tabular}{c||c}
\toprule
    Diagrams $(X)$ & Vertex factors $V^{(X)}(p,k)_{ji}$  \\ \midrule \midrule
    $ (3) $ & $ N_c \left( P_{00}(k) P_{ij}(p-k) - P_{0j}(k) P_{i0}(p-k) \right) $ \\ \midrule
    $ (4) $ & $ (-1) N_c \dfrac{(p_0-k_0)^2 g_{ij} + (p-k)_i (p-k)_j g_{00} }{k_0^2} P_{00}(k) $ \\ \midrule
    $ (7), (7r) $ & $ \dfrac{N_c}2 (p_0^2 g_{ij}  + p_i p_j g_{00}) \dfrac{P_{00}(k) }{k_0^2} $ \\ \midrule
    $ (8) $ & $N_c \dfrac{ i(p-k)_0 g_{ij} P_{00}(k) + i(p-k)_i  g_{00} P_{0j}(k) }{-ik_0 +0^+} $ \\ \midrule
    $ (8r) $ & $ N_c \dfrac{ i(p-k)_0 g_{ij} P_{00}(k) + i(p-k)_j  g_{00} P_{0i}(k) }{-ik_0 -0^+} $ \\ \midrule
    $ (11) $ & $ N_c \dfrac{-(p-k)_i k_j}{(-ik_0 + 0^+) (i (p-k)_0 + 0^+) }  $ \\
 \bottomrule
\end{tabular}
\end{center}
\caption{Summary of vertex factors of diagrams with 2 propagators contributing to $\big[g_E^{++}\big]_{ji,JI}^{da}$.}
\label{tab:2prop-vertex}
\end{table*}

Tables~\ref{tab:4prop-structure},~\ref{tab:4prop-vertex},~\ref{tab:3prop-structure},~\ref{tab:3prop-vertex},~\ref{tab:2prop-structure} and~\ref{tab:2prop-vertex}, describe the complete NLO calculations of the electric field correlator $\big[g_E^{++} \big]_{ji,JI}^{da}$ in full generality, for any type of correlation. As advertised, our objective is to compute the correlation with $J=2$ and $I=1$, but we will take a slightly less direct path and first compute the spectral function $\big[\rho_E^{++} \big]_{ji}^{da}$. A non-trivial check of our calculations is to verify that independently of our choice of $I$ and $J$ (i.e., on what branches of the Schwinger-Keldysh contour we evaluate the fields) the result is gauge invariant, since the electric field correlator is defined in a gauge invariant way. We now discuss the gauge invariance of the calculation in $R_\xi$ gauge.

\subsection{Gauge invariance in \texorpdfstring{$R_\xi$}{R_x} gauge}
As a consistency check of our calculation, we verify our result is gauge invariant. To recapitulate, the purpose of doing this is twofold: i) we can make sure that we have included and accounted for all diagrams contributing to the correlator at NLO, which also partially justifies the neglect of the Wilson lines at infinite time, whose contributions vanish in our calculations, as discussed in section~\ref{sect:scope}, and ii) we will verify that the way gauge-dependent parts cancel is independent of what type of correlation function we are calculating. That is to say, our proof will not rely on whether we choose to calculate a Wightman or a time-ordered correlator: we will prove gauge invariance for the full matrix-valued correlation function with the Schwinger-Keldysh contour indices.

We must show all the $\xi$ gauge parameter dependent parts cancel when summing over all diagrams. Depending on the number of gauge boson propagators in a diagram, we may have $\xi^2$, $\xi^3$ and $\xi^4$ terms showing up in the calculations. First, we note that due to the nature of the electric field, a single gauge boson propagator connected with the electric field automatically has its $\xi$ dependent part in the propagator removed. This may reduce the naive number of powers of $\xi$ in a diagram by 2. It turns out that at NLO in $R_\xi$ gauge, there are two independent cancellations that must take place: one for the terms proportional to $(1-\xi)^2$, and the other with the pieces proportional to $1-\xi$.

An important issue is how to connect the $\xi$ coefficients of diagrams with different numbers of propagators. For instance, diagram $(1)$ is the only diagram with four propagators that has a term proportional to $\xi^2$, so the $\xi^2$ term must be cancelled by diagram(s) with fewer gauge boson propagators. The crucial property that allows us to prove these cancellations without actually performing the integrals, (which is apparent from the free gauge boson equation of motion) is:
\begin{align}\label{eq:propcancellation}
-i p^2 \D(p)_{JI} (-1)^{I+1} = \mathbbm{1}_{JI} \,,
\end{align}
where no summation is implied by repeated indices, and $I=1$ is assumed to be a time-ordered branch of the Schwinger-Keldysh contour. With this, one can reduce the number of propagators in any diagram if a factor of the corresponding momentum squared appears.

We will verify the cancellation of the terms proportional to $(1-\xi)^2$ explicitly first, and then outline the technically more involved cancellation of the terms proportional to $(1-\xi)$, which is explained in Appendix~\ref{app:gauge-invariance} in detail.

\subsubsection{Cancellation at \texorpdfstring{$O((1-\xi)^2)$}{O_{(1-x)}}}

There are four diagrams that contribute at order $\xi^2$ when we plug
\begin{align}
P_{\mu \nu }(k) = - \left[ g_{\mu \nu} - (1 - \xi) \frac{k_\mu k_\nu}{k^2} \right] \,,
\end{align}
into our vertex factors, which are diagrams $(1)$, $(3)$, $(6)$, and $(6r)$.

\begin{figure}[h]
    \begin{subfigure}[b]{0.49\textwidth}
        \centering
        \includegraphics[height=2.0in]{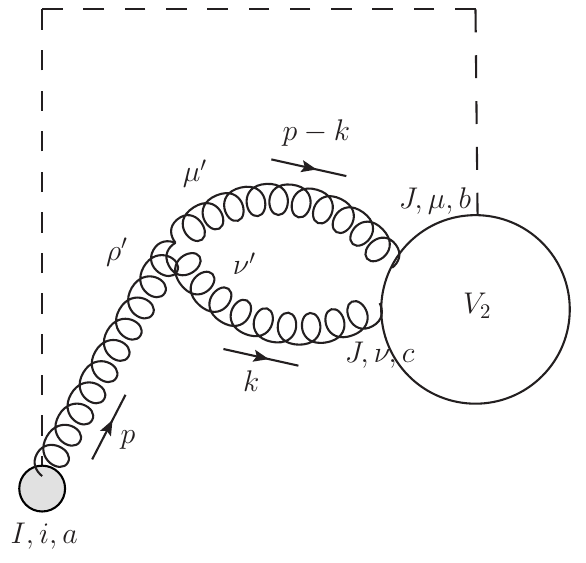}
        \caption{}\label{subfig:V4_1}
    \end{subfigure}%
    ~
    \begin{subfigure}[b]{0.49\textwidth}
        \centering
        \includegraphics[height=2.0in]{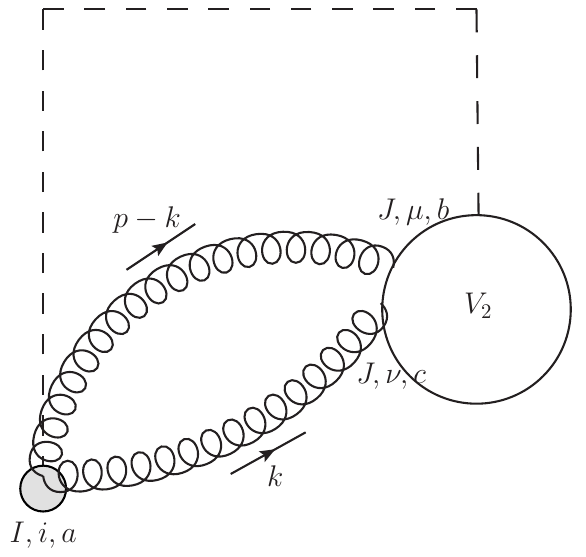}
        \caption{}\label{subfig:V4_2}
    \end{subfigure}%
    \caption{Diagrams with $O((1-\xi)^2)$ gauge dependence, with $V_2$ an arbitrary 2-gauge boson vertex to which the electric field, represented by the grey blob on the left, is connected in either of the two ways (a) and (b) shown.}
    \label{fig:gaugeProofDiagsxi2}
\end{figure}

It turns out that the cancellation can be proven from the common sub-diagram structures of these four. 
In particular, we consider the sub-diagrams of Figure~\ref{fig:gaugeProofDiagsxi2}. As we already emphasized, any diagram containing a term proportional to $(1-\xi)^2$ contributes in this way: the only diagrams that can contain two factors of $(1-\xi)$ are those that have two propagators that are not directly connected to an electric field via a single gauge boson line. 
Moreover, if we group our diagrams as (1)+(6r) and (3)+(6), we find each combination corresponds exactly to one of the two structures shown in  Figure~\ref{fig:gaugeProofDiagsxi2}. If we explicitly add the diagrams in Figure~\ref{fig:gaugeProofDiagsxi2}, we obtain\footnote{For the purposes of our calculation, we only need to consider the situation where the gauge boson propagators that are attached to $V_2$ have the same Schwinger-Keldysh indices $J$. In general, the $J$ indices can be different.}
\be
    && \big(ip_0 \D_{i\rho'}^{aa'}(p)_{I'I} - ip_i \D_{0 \rho'}^{aa'}(p)_{I'I} \big) \D_{ \mu' \mu}^{b' b}(p-k)_{JI'} \D_{\nu' \nu}^{c'c}(k)_{JI'} (-1)^{I'+1} \nonumber \\
&\times& g f^{a'b'c'} \left[ g^{\rho' \mu'} (2p-k)^{\nu'} + g^{\mu' \nu'} (2k-p)^{\rho'} + g^{\nu' \rho'} (-k-p)^{\mu'} \right] \nonumber \\
& +& gf^{ab'c'} \left[ \D_{i\mu}^{c'b}(p-k)_{JI} \D_{0\nu}^{b'c}(k)_{JI} + \D_{i\nu}^{c'c}(k)_{JI} \D_{0\mu}^{b'b}(p-k)_{JI} \right] \,.
\ee
Focusing on the $(1-\xi)^2$ piece, we can replace the propagators $\D_{\mu \nu}^{ab}(p-k)$ and $\D_{\mu \nu}^{ab}(k)$ with
\begin{align}
\D_{\mu \nu}^{ab}(p)_{JI} \to \delta^{ab} (1-\xi) \frac{p_\mu p_\nu}{p^2} \D(p)_{JI} \,,
\end{align}
where the last propagator $\D(p)_{JI}$ is the scalar thermal propagator. The replacement yields
\begin{align}
    & (-ip_0 g_{i \rho'} + ip_i g_{0 \rho'}) (1 - \xi)^2 \frac{(p-k)_{\mu} (p-k)_{\mu'} }{(p-k)^2} \frac{k_\nu k_{\nu'}}{k^2} \D(p)_{II'} \D(p-k)_{I'J} \D(k)_{I'J} (-1)^{I'+1} \nonumber \\
    & \quad \quad \quad \times g f^{abc} \left[ g^{\rho' \mu'} (2p-k)^{\nu'} + g^{\mu' \nu'} (2k-p)^{\rho'} + g^{\nu' \rho'} (-k-p)^{\mu'} \right] \nonumber \\
    & + gf^{abc} (1-\xi)^2 \left[  - \frac{(p-k)_i (p-k)_\mu}{(p-k)^2} \frac{k_0 k_\nu}{k^2} + \frac{k_i k_\nu}{k^2} \frac{(p-k)_0 (p-k)_\mu}{(p-k)^2} \right] \D(p-k)_{IJ} \D(k)_{IJ} \nonumber \\
    &= g f^{abc} (1-\xi)^2 \frac{k_\nu (p-k)_\mu}{k^2 (p-k)^2} \bigg[ \D(p)_{II'} \D(p-k)_{I'J} \D(k)_{I'J} (-1)^{I'+1} i p^2 (p_0 k_i - p_i k_0) \nonumber \\
    & \quad \quad \quad\quad\quad\quad\quad\quad\quad\quad\quad\quad\quad\quad\quad\quad\quad\quad + \D(p-k)_{IJ} \D(k)_{IJ} (p_0 k_i - k_0 p_i) \bigg] \nonumber \\
    &= g f^{abc} (1-\xi)^2 \frac{k_\nu (p-k)_\mu}{k^2 (p-k)^2} (p_0 k_i - p_i k_0) \bigg[ - \mathbbm{1}_{II'} \D(p-k)_{I'J} \D(k)_{I'J} + \D(p-k)_{IJ} \D(k)_{IJ}  \bigg] \nonumber \\
    &= 0 \,.
\end{align}
This proves that all the terms proportional to $(1-\xi)^2$ in Figure~\ref{fig:gaugeProofDiagsxi2} cancel out. Since all the diagrams with the contributions proportional to $(1-\xi)^2$ appear in this way, we have shown that the full NLO result has no $(1-\xi)^2$ terms. Now we have to verify that the remaining $(1-\xi)$ dependence also cancels.

\subsubsection{Cancellation at \texorpdfstring{$O(1-\xi)$}{O_{1-x}}}

As illustrated by the previous cancellation, the general strategy we will use to prove gauge invariance relies on identifying the factor of a momentum flowing through the diagram ($p$, $k$, or $p-k$) squared, and using them to ``cancel'' some propagator insertions so that the diagrams with a greater number of propagators, such as $(1)$, can be added seamlessly with other diagrams with fewer gauge boson propagators.

Because of the reflection symmetry relating diagrams $(X)$ in Figure~\ref{fig:diagrams} with diagrams $(Xr)$ (implemented by reversing the flow of momentum and exchanging the external color, spatial, and contour indices), we actually only need to verify that the following cancellation takes place:
\begin{align}\label{eq:gauge-cancellation}
\frac{\partial}{\partial (1-\xi)} \left( \frac{1}{2} \big[ (1) + (2) + (3) + (4) \big] + (5) + (6) + (7) + (8) \right) = 0 \,,
\end{align}
for any value of $\xi$. We have not included diagram $(11)$ in this list because it is automatically $R_\xi$ gauge-invariant: all of its gauge boson propagators are connected to a electric field via a single attachment, which gives a momentum structure that automatically removes all $\xi$ dependence in the propagator:
\begin{align}
\big[ i p_0 g_{i\mu} - i p_i g_{0\mu} \big] P^{\mu \nu}(p) =  (-1) \big[ i p_0 g_{i}^{\nu} - i p_i g_{0}^{\nu} \big] \,.
\end{align}

We show that~\eqref{eq:gauge-cancellation} holds identically in Appendix~\ref{app:gauge-invariance} by explicitly working out each diagram. Specifically, we show that the contributions linear in $(1-\xi)$ cancel. It turns out that all of the gauge-dependent diagrams need to be taken into account: the cancellation does not hold for any separate subset of diagrams. To illustrate schematically how the cancellations take place, we summarize them as follows:
\begin{align}
    (1)_{(1-\xi), \, \not\propto p^2} + (2)_{(1-\xi)} &= 0\,, \\
    \Big(\frac{1}{2} (1)_{(1-\xi), \, \propto p^2} + \big((5)+(6)\big)_{(1-\xi)} \Big)_{\propto (p-k)^2} + (7)_{(1-\xi)} &= 0\,, \\
    \Big(\frac{1}{2} (1)_{(1-\xi), \, \propto p^2} + \big((5)+(6)\big)_{(1-\xi)} \Big)_{\propto p^2} + \frac{1}{2} \left( (3)_{(1-\xi)} + (4)_{(1-\xi)} \right) + (8)_{(1-\xi)} &= 0\,,
\end{align}
where the notation ${(X)}_{(1-\xi)}$ represents the piece of diagram $(X)$ that is linear in $(1-\xi)$. Also, with the subscripts ${}_{\propto p^2}$ we only keep the terms in a given vertex factor that are proportional to the quantity in the subscript while with the subscripts ${}_{\not\propto p^2}$ we keep all the remaining terms that are not proportional to the quantity in the subscript. These subscripts essentially label which propagator in the diagram is cancelled via~\eqref{eq:propcancellation}.\footnote{However, we warn the reader that they depend on the convention taken for the momentum flow in the diagram, and so this structure of cancellations is strictly true only if the calculation is carried out consistently with the conventions we have taken so far and in Appendix~\ref{app:gauge-invariance}.}

In summary, after adding up all the diagrams, the result is independent of $\xi$. This verifies the expectation from the construction of the electric field correlation function, where the Wilson line insertions exactly guarantee the gauge invariance of the correlator. This means that we can confidently choose any (non-singular) gauge to perform the calculation. Throughout the rest of this work, we will choose the Feynman gauge, $\xi=1$.

\subsection{Calculations in Feynman gauge}

We now proceed to evaluate each diagram that contributes to the chormoelectric correlator. However, to simplify the analysis, we will solely focus on calculating the object that contributes to the inclusive bound state formation/dissociation rates, which is the integrated spectral function
\begin{align}
\label{eqn:rho_p_integrated}
\varrho^{++}_E(p_0) = \frac{1}{2} \int \frac{\diff^{d-1} \p}{(2\pi)^{d-1}} \delta^{ad} \delta_{ij} \big[ \rho_E^{++} \big]_{ji}^{da}(p_0,\p) \,.
\end{align}
The overall factor of $1/2$ is a choice of normalization, just to cancel the factor of $2$ when we write the spectral function as the real part of the retarded propagator, see Eq.~(\ref{eq:spectral-retarded}) from below. 

As a baseline, we first show, explicitly in $d=4-\epsilon$, how one can obtain the spectral function at LO in a way that is also applicable to higher order contributions. Then, we will consider the diagrams contributing at NLO. Since a naive treatment of the integrals in $d=4$ leads to UV divergences, we use will dimensional regularization with $d = 4 - \epsilon$ and the $\overline{\rm MS}$ renormalization scheme throughout the calculation of these diagrams. We will only take the limit $\epsilon \to 0$ at the very end for diagrams that are UV divergent while for those that are UV finite, we take $d=4$ as early as possible. We will split the calculation into three blocks: First, we start with the calculation of the textbook diagrams that contribute to the gauge boson self-energy in Section~\ref{sec:gluon-self-energy}, quoting the calculation of the $T=0$ quantum field theory result in the $\overline{\rm MS}$ scheme from textbooks as an input to our calculation, and also explicitly calculating the finite-temperature pieces, verifying that the results are consistent. Then, we devote a section to the diagrams with three propagators $(5)$, $(5r)$, $(6)$, $(6r)$ in~\ref{sec:3-propagator-evaluation}, because they share similar properties in the pole structure of their propagators. We calculate these four diagrams for both the vacuum ($T=0$) contribution and the finite temperature contribution. Finally, in section~\ref{sec:2-propagator-evaluation} we discuss the calculation of the remaining diagrams, which are all made up of only two propagators.

In the first two forthcoming subsections, we will make direct use of the relation
\begin{align}\label{eq:spectral-retarded}
\rho_G(p) = G^>(p) - G^<(p) = 2{\rm Re} \left\{ G^R(p) \right\} = 2 {\rm Re} \left\{ G^T(p) - G^<(p) \right\} \,,
\end{align}
which is a well-known relation between the spectral function and the retarded correlation function. The relation between the spectral function and the retarded propagator generally holds for two-point functions of operators that are local in time. This is directly applicable to the LO result, as well as to the NLO contributions from the traditional gauge boson self-energy (diagrams $(1),(2),(f),(g)$), because they only involve correlation functions of local operators. In the case of diagrams involving Wilson lines, it is not obvious that the relation between the spectral function and the retarded propagator is true. Therefore we feel compelled to first present the calculation in a way that does not use this property explicitly, and thus we will resort to using the usual definition $\rho_G(p) = G^>(p) - G^<(p)$ for these cases. 

\subsubsection{Leading order result: a single gauge boson propagator}

As an introductory step, we consider the LO calculation of the momentum-integrated spectral function:
\begin{equation}
    \left. \varrho_E^{++}(p_0) \right|_{\rm{LO}} = \delta_{ij} \delta^{ad}  \int \frac{\diff^{d-1}\p}{(2\pi)^{d-1}} \delta^{ad}\, {\rm Re} \left\{  \frac{(i p_0 g_{i\mu} - i p_i g_{0 \mu} ) i P^{\mu \nu}(p) (-i p_0 g_{j\nu} + i p_j g_{0 \nu} ) }{p_0^2 - \p^2 + i0^+ {\rm sgn}(p_0)}  \right\}\,,
\end{equation}
which we have calculated by taking the real part of the corresponding retarded correlation. The retarded correlation for the electric fields is constructed by applying derivatives on the retarded correlation function of two gauge fields.\footnote{At this point, there is a subtlety that needs to be addressed in the definition of the retarded correlation function, because derivatives do not commute with the time-ordering operator that defines the correlator (in this case, a retarded correlation). The usual choice when defining time-ordered correlation functions in QFT textbooks is to take the derivatives outside the time-ordering symbol (i.e., $\mathcal{T}^\ast$-product), because it is this choice that respects Lorentz covariance. For our present purposes this is inconsequential, because when we take the real part of the retarded correlator the ambiguity is removed as $p_0$ can be replaced by $|\p|$. But in any calculation where retarded objects are involved, one needs to have a consistent definition. We follow standard practice, and define retarded/time-ordered correlations with the time-ordering operator acting first on the fundamental fields, and then calculate the action of derivatives on them so that the time-ordered correlator is compatible with Lorentz covariance.} Evaluating directly in $d$ dimensions, we obtain
\begin{align}
\left. \varrho_E^{++}(p_0) \right|_{\rm{LO}} = (N_c^2-1) \frac{(d-2) \pi \Omega_{d-1}}{2(2\pi)^{d-1}}  |p_0|^{d-1} {\rm sgn}(p_0) \,.
\end{align}
We will keep these $d$-dependent prefactors wherever possible in our calculation, as they will simplify the analysis when we later introduce the coupling constant renormalization to compensate for the UV divergences.

\subsubsection{Traditional gauge boson self-energy: diagrams \texorpdfstring{$(1)$}{(1)}, \texorpdfstring{$(2)$}{(2)}, \texorpdfstring{$(g)$}{(g)}, \texorpdfstring{$(f)$}{(f)}} \label{sec:gluon-self-energy}

Now we turn to the calculation of the the gauge boson self-energy contributions to the integrated spectral function. We start by quoting the textbook result~\cite{Schwartz:2013pla} for the $T=0$ case, where the time-ordered self-energy is given in the $\overline{\rm MS}$ scheme
\begin{equation}
\begin{split}
        i \mathcal{M}_{\ml{T}}^{ad, \mu \nu}(p) = i \delta^{ad} \frac{g^2}{16 \pi^2} (p^2 g^{\mu \nu} - p^\mu p^\nu) & \bigg[ N_c \left( \frac{10}{3\epsilon} + \frac{31}{9} + \frac{5}{3} \ln \left( \frac{\mu^2}{-p^2-i0^+} \right) \right) \\ & - n_f C({\bs N}) \left( \frac{8}{3\epsilon} + \frac{20}{9}  + \frac{4}{3} \ln \left( \frac{\mu^2}{-p^2-i0^+} \right) \right) + \O(\epsilon) \bigg] \,,
\end{split}
\end{equation}
where ${\ml{T}}$ indicates the quantity is time-ordered.
The retarded self-energy can be obtained by replacing $(p^2 + i0^+)$ with $(p^2 + i0^+{\rm sgn}(p_0))$. Since the spectral function is an odd function of $p_0$, we can actually focus only on $p_0>0$ and therefore ignore the distinction between retarded and time-ordered self-energies in this case. Then, as long as we take $p_0>0$, we can compute the vacuum contribution to the momentum-integrated spectral function as
\begin{equation}
\begin{split}
    & \left. \varrho_E^{++}(p_0) \right|_{{\rm NLO}, \, T=0}^{\rm self-energy} \\ &= \delta^{ad} \int \frac{\diff^{d-1} \p}{(2\pi)^{d-1}} {\rm Re} \left\{ \frac{i (p_0 g_i^{\mu'} - p_i g_0^{\mu'} ) (i P_{\mu \mu'}(p)) i \mathcal{M}_{\ml{T}}^{ad, \mu \nu}(p) (i P_{\nu \nu'}(p)) (-i p_0 g_j^{\nu'} + i p_j g_0^{\nu'} ) }{(p^2 + i0^+)^2}  \right\}\,,
\end{split}
\end{equation}
where we have included the two ``external'' propagators that are present in our original formulation in the Schwinger-Keldysh contour. One can directly check from the path integral that the complete retarded correlation has the structure $D^R(p) i \ml{M}_R D^R(p) $ in terms of the retarded self-energy $\ml{M}_R$. It then follows that
\begin{equation}
\begin{split}
    \left. \varrho_E^{++}(p_0) \right|_{{\rm NLO}, \, T=0 }^{\rm self-energy} &= (N_c^2 - 1) \frac{(d-2) \pi \Omega_{d-1}}{2 (2\pi)^{d-1}} \frac{g^2}{4\pi^2} p_0^{d-1} \bigg[ N_c \left( \frac{5}{6\epsilon} + \frac{14}{9} + \frac{5}{12} \ln \left( \frac{\mu^2}{4 p_0^2} \right)  \right)   \\ & \quad \quad \quad \quad \quad \quad \quad \quad \quad \quad \quad \quad \quad  - n_f C({\bs N}) \left( \frac{2}{3\epsilon} + \frac{10}{9} + \frac{1}{3}  \ln \left( \frac{\mu^2}{4 p_0^2} \right) \right) \bigg]\,,
\end{split}
\end{equation}
is the full contribution from the gauge boson self-energy for $p_0>0$. The extension to $p_0<0$ is given by naturally continuing the function with the rule $ \varrho_E^{++}(-p_0) = -\varrho_E^{++}(p_0) $. It is also straightforward to check that the fermion sector matches the result of the Abelian case~\cite{Binder:2020efn}.

Now we explain the calculation of the finite temperature pieces. Since the fermion loop contribution has already been discussed in previous work by some of us, we will simply quote the result at the end of this section, and focus only on the gauge boson and ghost diagrams for the purpose of this discussion. We commence by analyzing the propagator structure of these diagrams in the retarded case, i.e., (looking at the first row of Table~\ref{tab:4prop-structure})
\begin{align}
\ml{Q}^{(1)}_{11} - \ml{Q}^{(1)}_{12} = [D^R(p)]^2 \left[ D^{\ml{T}}(k) D^{\ml{T}}(p-k) - D^<(k) D^<(p-k) \right]\,,
\end{align}
where the equality follows after some algebra. Here we plan to compute the spectral function by taking the real part of the retarded propagator since the diagram (1) only consists of local operators, as discussed below Eq.~(\ref{eq:spectral-retarded}). This also verifies that the correspondence between retarded and time-ordered self-energies we used earlier is consistent: if we set $p_0>0$ in the $T=0$ case, the combination of propagators $D^<(k) D^<(p-k)$ vanishes identically because $D^<(p-k)$ is proportional to $\theta(-p_0+k_0)$ and is nonzero only if $p_0-k_0 < 0$, implying $k_0>0$, but then $D^<(k)$ vanishes due to its own theta function. Then, the retarded self-energy reduces to the structure $D^{\ml{T}}(k) D^{\ml{T}}(p-k)$, which is exactly the time-ordered self-energy, meaning that, as expected, $G^R(p) = G^T(p)$ in vacuum when $p_0>0$.

For the full calculation of both vacuum and finite temperature contributions, what we need to calculate is
\begin{equation}
\begin{split}
    & \left. \varrho_E^{++}(p_0) \right|_{{\rm NLO} }^{\rm self-energy} \\ &= (N_c^2 - 1) g^2 \int_\p \int_k N(p,k) [D^R(p)]^2 \left[ D^{\ml{T}}(k) D^{\ml{T}}(p-k) - D^<(k) D^<(p-k) \right]\,,
\end{split}
\end{equation}
where, in terms of Table~\ref{tab:4prop-vertex}, $N(p,k) = \delta_{ij} \left[V^{(1)}(p,k)_{ji} + V^{(g)}(p,k)_{ji} \right]$, and we have introduced the shorthands $\int_\p = \int \frac{\diff^{d-1}\p}{(2\pi)^{d-1}}$, $\int_k = \int \frac{\diff^d k}{(2\pi)^d}$. One can then manipulate the propagators that depend on $k$ to show that\footnote{The procedure that leads to this equality is only valid if the numerator $N(p,k)$ does not have any poles as a function of $k_0$. This is not strictly true in the case where we have Wilson lines, of which their nonlocality in time introduces a pole at $k_0 = 0$, and a more careful treatment would be needed.}
\begin{equation}
\begin{split}
    & \left. \varrho_E^{++}(p_0) \right|_{{\rm NLO} }^{\rm self-energy} \\ &= (N_c^2 - 1) g^2 \int_\p \int_k N(p,-k) [D^R(p)]^2 \left[ D^<(k) D^R(p+k) +  D^<(p+k) D^A(k) \right]\,,
\end{split}
\end{equation}
which has been written in a form that is more appropriate for the finite temperature calculation. To see why this is so, note that we can write the advanced and retarded propagators as
\begin{align}
D^R(k) &= i \int \frac{\diff k_0'}{2\pi} \frac{D^>(k_0',\k) - D^<(k_0',\k)}{k_0 - k_0' + i0^+} \\
D^A(k) &= i \int \frac{\diff k_0'}{2\pi} \frac{D^>(k_0',\k) - D^<(k_0',\k)}{k_0 - k_0' - i0^+}\,,
\end{align}
which, after some algebra, leads to (we define $\q = \p + \k$)
\begin{equation} \label{eq:integrated-gluon-self-energy-before-integration}
\begin{split}
\left. \varrho_E^{++}(p_0) \right|_{{\rm NLO} }^{\rm self-energy} &=  (N_c^2 - 1) g^2 \int_{\k, \q} \frac{1+2n_B(|\k|) }{ 2|\k| 2|\q|} \\ & \quad \times {\rm Re} \left\{ \frac{-i}{( (p_0+ i0^+)^2 - (\q - \k)^2 )^2 } \sum_{\sigma_1,\sigma_2} \frac{\sigma_2 N((p_0,\q-\k),-k)_{k_0 = \sigma_1 |\k|} }{p_0 + \sigma_1 |\k| - \sigma_2 |\q| + i0^+ } \right\} \,.
\end{split}
\end{equation}
This is the expression for the complete contribution (vacuum + finite temperature) in an arbitrary number of dimensions $d$. Since we already evaluated the vacuum contribution, we can drop the $1$ in $1 + 2n_B(k)$ and set $d=4$ right away to evaluate the remaining piece, because the presence of the Bose-Einstein distribution guarantees that the integrals are convergent in the UV limit.

Naively, however, an issue arises in the collinear limit, where $\k$ becomes parallel to $\p$ (or, equivalently, to $\q$). This is because when one takes the real part of this expression, as there is a factor of $i$ in the numerator, one is effectively evaluating the sum of the residues of the poles at the locations where the denominator becomes zero. This is equivalent to what one would do, in a standard QFT textbook, by using ``cutting rules''. The problem appears when one tries to evaluate these ``cuts'' separately, because the integrals over the remaining momentum degrees of freedom (after taking the residue) separately diverge for each pole, and only their sum gives a finite, meaningful result. There are various ways to deal with this, both with and without introducing extra regulators, and, importantly, the result is independent of the choice of the methods we use to deal with this divergence. We discuss two of these methods in detail in Appendix~\ref{app:collinear-integrals}.

Regardless of the choice of integration ordering, one arrives at
\begin{equation}
\begin{split}
&\left. \varrho_E^{++}(p_0) \right|_{{\rm NLO}}^{\rm gauge boson+ghost} - \left. \varrho_E^{++}(p_0) \right|_{{\rm NLO}, \, T=0}^{\rm gauge boson+ghost} \\ &=  \frac{(N_c^2 - 1) g^2}{(2\pi)^3} N_c \int_{0}^\infty \diff k \,2 n_B(k) \left[ -2 k p_0 + (k^2+p_0^2) \ln \left| \frac{k+p_0}{k-p_0} \right| + k p_0 \ln \left| \frac{k^2-p_0^2}{p_0^2} \right| \right] \,,
\end{split}
\end{equation}
which has no known (to us) closed form. The interested reader can note that, asymptotically, the $\O(1/k)$ piece in the series expansion of the term in square brackets is $\frac{5}{3} \frac{p_0^3}{k}$, which is closely related to the fact that the dimensionally-regularized $T=0$ UV divergence in these diagrams goes as $\frac56 \frac{1}{\epsilon}$. To get the factor of $1/2$ correctly, one has to do the computation completely in dimensional regularization, which requires a level of involvement that is unnecessary for our present purposes, since here we focus on the finite temperature piece which has no UV divergence.

Combining self-energy contributions from the fermion loop, the ghost loop and the gauge boson loop, we obtain
\begin{equation}
\begin{split}
&\left. \varrho_E^{++}(p_0) \right|_{{\rm NLO}}^{\rm self-energy} \\
&= g^2 (N_c^2 - 1) \bigg\{ \frac{(d-2) \pi \Omega_{d-1}}{2 (2\pi)^{d-1} (4\pi^2)} p_0^{d-1} \bigg[ N_c \left( \frac{5}{6\epsilon} + \frac{14}{9} + \frac{5}{12} \ln \left( \frac{\mu^2}{4 p_0^2} \right)  \right)   \\ & \quad \quad \quad \quad \quad \quad \quad \quad \quad \quad \quad \quad \quad \quad \quad - n_f C({\bs N}) \left( \frac{2}{3\epsilon} + \frac{10}{9} + \frac{1}{3}  \ln \left( \frac{\mu^2}{4 p_0^2} \right) \right) \bigg] \\ 
& +  \int_{0}^\infty \!\! \diff k \frac{2 N_c n_B(k)}{(2\pi)^3}   \bigg[ -2 k p_0 + (k^2+p_0^2) \ln \left| \frac{k+p_0}{k-p_0} \right|  + k p_0 \ln \left| \frac{k^2-p_0^2}{p_0^2} \right| \bigg] \\
&  + \int_0^\infty \!\! \diff k \frac{2 N_f n_F(k)}{(2\pi)^3} \bigg[ -2k p_0 + (2k^2 + p_0^2) \ln \left| \frac{k+p_0}{k-p_0} \right| + 2 k p_0 \ln \left| \frac{k^2 - p_0^2}{p_0^2} \right| \bigg] \bigg\}\,,
\end{split}
\end{equation}
up to $\O(\epsilon)$ terms that we need not keep track of in the limit $\epsilon \to 0$. We do keep $\O(\epsilon)$ terms in the prefactor of the $1/\epsilon$ pole because they will also appear in front of other UV poles, and thus contributing at $\O(\epsilon^0)$ in the end.

\subsubsection{Diagrams \texorpdfstring{$(5)$}{(5)}, \texorpdfstring{$(5r)$}{(5r)}, \texorpdfstring{$(6)$}{(6)}, and \texorpdfstring{$(6r)$}{(6r)}} \label{sec:3-propagator-evaluation}

We will compute both the vacuum and finite temperature contributions of these diagrams. Given the presence of non-local operators in the correlation, to avoid having to deal with the $1/k_0$ poles coming from the Wilson lines using contour integration, we will not calculate the spectral function by taking the real part of the corresponding retarded correlator, but rather, we will write down the spectral function directly in terms of the Wightman functions. As before, we work with $p_0>0$.

{ It is instructive to note where the difficulty with the $1/k_0$ poles comes from, and what is necessary to do in order to properly deal with them. Our approach allows us to evaluate the difference between the $(I=1, J=2)$ and $(I=2, J=1)$ correlation functions, but for these diagrams, the $[g_E^{++}]_{12}$ does not agree exactly with the definition of the correlator $[g_E^{++}]^{<}$ that is the KMS conjugate~\eqref{eqn:KMS_for_g++} of the physical correlator $[g_E^{++}]^{>}$. This is so because the time-ordering implicit in $[g_E^{++}]_{12}$ does not follow the corresponding matrix product ordering of the Wilson lines.\footnote{ Similar features of the time-ordering of operators explain the difference between the heavy-quark diffusion coefficient and the quarkonium transport coefficients at NLO.} To formulate the spectral function of physical interest without extending the Schwinger-Keldysh contour (so as to accommodate non-trivial operator orderings), it is more efficient to use the relation $[g_E^{++}]^{<}(t,{\bs x}) = [g_E^{--}]^{>}(-t,-{\bs x})$. The only new ingredient to do this calculation is the Feynman rule of a gauge boson insertion in a Wilson line going towards $t=-\infty$, which, after some algebra, on can show that only amounts to flipping the sign of the $i0^+$ prescription in our Feynman rule for the Wilson lines going towards $t=+\infty$.
}

{ We denote the sum of the vertex factors from diagrams $(5)$ and $(6)$ as
\begin{equation}
\begin{split}
    N^{(5),(6)}(p,k) = \delta_{ij} & \big[ V^{(5)}(p,k)_{ji} + V^{(6)}(p,k)_{ji}  \big]\,,
\end{split}
\end{equation}
where the superscript refers to the corresponding diagrams. After some algebra, using the relations between $[g_E^{--}]^>$ and $[g_E^{++}]^<$, and}
using the propagator structures in Table~\ref{tab:3prop-structure}, one finds that
{ \begin{equation}
\begin{split}
   \frac{1}{2} \delta^{ad} \delta_{ij} \big[ \rho_E^{++} \big]^{da}_{ji} &= g^2 (N_c^2-1) \int_k {\rm Re} \bigg\{ N^{(5),(6)}(p,k) \big[ \big( D^>(p) - D^<(p) \big) D^{\ml{T}}(k) D^{\ml{T}}(p-k)  \\
   & \quad \quad \quad \quad \quad \quad \quad \quad   - D^{\overline{\ml{T}}}(p) \big( D^>(k) D^>(p-k) - D^<(k) D^<(p-k) \big) \big] \bigg\} \, .
\end{split}
\end{equation}
This can be decomposed into the contributions that we can obtain from the $(I=1, J=2)$ and $(I=2, J=1)$ correlation functions, and an extra piece, as we will show in a moment.}

{ It is helpful to denote the sum of all of the vertex factors associated to these diagrams}, symmetrized under $k \to p-k$, by
\begin{equation}
\begin{split}
    N_{3p}(p,k) = \frac{\delta_{ij}}{2} & \big[ V^{(5)}(p,k)_{ji} + V^{(5r)}(p,k)_{ji} + V^{(6)}(p,k)_{ji} + V^{(6r)}(p,k)_{ji} \\ & \, +  V^{(5)}(p,p-k)_{ji} + V^{(5r)}(p,p-k)_{ji} + V^{(6)}(p,p-k)_{ji} + V^{(6r)}(p,p-k)_{ji} \big]\,,
\end{split}
\end{equation}
where the subscript $3p$ indicates that the vertex factor originates from the diagrams with three propagators. { This quantity, as opposed to $N^{(5),(6)}$ (which has both real and imaginary parts), is manifestly a purely imaginary number. }

{ Using these definitions, it follows that
\begin{equation}
\begin{split}
 \left. \varrho_E^{++}(p_0) \right|_{\rm NLO}^{5-6} &= g^2 (N_c^2-1) \! \int_{k,{\bs p}} \!\!\! {\rm Re} \bigg\{ i \left[ {\rm Im} \, N^{(5),(6)}(p,k)\right] \big[ \big( D^>(p) - D^<(p) \big) D^{\ml{T}}(k) D^{\ml{T}}(p-k)  \\
   & \quad \quad \quad \quad \quad \quad \quad \quad \quad \quad \quad - D^{\overline{\ml{T}}}(p) \big( D^>(k) D^>(p-k) - D^<(k) D^<(p-k) \big) \big] \bigg\} \\
   & \quad + g^2 (N_c^2-1) \! \int_{k,{\bs p}} \!\!\! {\rm Re} \bigg\{\left[  {\rm Re} \, N^{(5),(6)}(p,k)\right] \big[ \big( D^>(p) - D^<(p) \big) D^{\ml{T}}(k) D^{\ml{T}}(p-k)  \\
   & \quad \quad \quad \quad \quad \quad \quad \quad \quad \quad \quad - D^{\overline{\ml{T}}}(p) \big( D^>(k) D^>(p-k) - D^<(k) D^<(p-k) \big) \big] \bigg\} \\
   &= i g^2  (N_c^2-1) \tilde{\mu}^\epsilon \int_{\k,\p} \frac{1+2n_B(k)}{2k} \sum_{\sigma_1}  N_{3p}((p_0,\p),(\sigma_1 k, \k)) \\
   & \quad \quad \quad \quad \quad \quad \quad \quad \quad \times {\rm Re}\left\{ \frac{i}{((p_0+i\epsilon)^2 -\p^2 )((p_0- \sigma_1 k + i\epsilon)^2 - (\p-\k)^2 )} \right\} \\
   & \quad + g^2  (N_c^2-1) \pi \int_{\k,\p}   \frac{ \left[ k_0 N^{(5),(6)}(p,k) \right]_{k_0=0} }{\k^2} \ml{P} \left( \frac{1}{p_0^2 - (\p - \k)^2} \right) \delta(p^2)   \,,
\end{split}
\end{equation}
where the first set of integrals, containing the Bose-Einstein distribution explicitly, is to be identified with the contribution from the $(I=1, J=2)$ and $(I=2, J=1)$ correlation functions, whereas the last line is a consequence of carefully handling the $k_0=0$ pole using the physical KMS conjugate $[g_E^{++}]^{<}$ of $[g_E^{++}]^{>}$.}


We have kept everything in $d$ dimensions explicitly. { The last line can be done explicitly, and $d=4$ may be set right away for this piece in the final result}. To compute the remaining 6-dimension integral, our strategy is to first carry out the $\p$ integral by introducing Feynman parameters to merge the denominators using the standard formulae for loop integrals in $d=4-\epsilon$ dimensions, and then perform the integrals over the remaining solid angle of $\k$ and the Feynman parameters, leaving only the integral over $k$ to be dealt with (details of the calculation can be found in Appendix~\ref{app:3-prop-detail}). We then get a result of the form
\begin{equation}
 \left. \varrho_E^{++}(p_0) \right|_{\rm NLO}^{5-6} = g^2 p_0^3  N_c (N_c^2-1) \bigg[ { \frac{\pi^2/2}{(2\pi)^3}} -  \frac{\Omega_{3-\epsilon} \tilde{\mu}^\epsilon}{(4\pi)^{(3-\epsilon)/2}} \Gamma \! \left( \frac{-1+\epsilon}{2} \right) \int_0^\infty \!\!\! \diff k \frac{ 1+2n_B(k)}{(2\pi)^{d-1}} K(k;\epsilon) \bigg]
 \,,
\end{equation}
with only the final integral { over $k$} to be evaluated. The expression of $K(k;\varepsilon)$ can be found in Appendix~\ref{app:3-prop-detail}, and is obtained by carrying out the steps we just outlined. An important feature of $K(k;\epsilon)$ is that if one naively takes the limit $\epsilon \to 0$ before integrating over $k$, the integrand becomes UV divergent. However, as we show in Appendix~\ref{app:3-prop-detail}, it turns out that the sum of these diagrams in dimensional regularization is UV finite in vacuum as long as we take $\epsilon \to 0$ after performing the $k$ integration.
After dealing with the potentially divergent pieces in a careful way, i.e., extracting the terms that become UV divergent in the $\epsilon \to 0$ limit and integrating over them before taking $\epsilon \to 0$, one obtains
\begin{equation}
\begin{split}
    \left. \varrho_E^{++}(p_0) \right|_{{\rm NLO}, \, T=0 }^{5-6}
    = \frac{g^2 N_c }{(2\pi)^3} (N_c^2-1) p_0^3 \left[ { 1 + \frac{\pi^2}{3} } \right] + \mathcal{O}(\epsilon)\,.
\end{split}
\end{equation}

The $T>0$ contribution is free of UV divergences because of the presence of the Bose-Einstein distribution, and so one can simply take $\epsilon=0$ to get the physical result. One then obtains the full NLO contribution from these diagrams
\begin{equation}
\begin{split}
    &\left. \varrho_E^{++}(p_0) \right|_{{\rm NLO}}^{5-6} = \frac{g^2 N_c  (N_c^2-1)}{(2\pi)^3} p_0^3 \bigg[ { 1 + \frac{\pi^2}{3} }  \\ 
    & \quad \quad \quad \quad \quad \quad \quad \quad \quad + \int_0^\infty \diff k \, \frac{2 n_B(k)}{p_0} \frac{k^2 p_0 + (k^3 + p_0^3) \ln \big| \frac{k-p_0}{p_0} \big| + (p_0^3 - k^3) \ln \big| \frac{k + p_0}{p_0} \big| }{k p_0^2 - k^3} \bigg] \,,
\end{split}
\end{equation}
plus $\O(\epsilon)$ terms which are irrelevant as the result is already finite. The denominator $k p_0^2 - k^3$ is actually a Cauchy principal value distribution, which gives a finite result after integration.

\subsubsection{Diagrams \texorpdfstring{$(3)$}{(3)}, \texorpdfstring{$(4)$}{(4)}, \texorpdfstring{$(7)$}{(7)}, \texorpdfstring{$(7r)$}{(7)}, \texorpdfstring{$(8)$}{(8)}, \texorpdfstring{$(8r)$}{(8r)}, \texorpdfstring{$(11)$}{(11)} } \label{sec:2-propagator-evaluation}

Finally, we evaluate the diagrams that have no 3-gauge boson vertices, and are purely contractions of fields through propagators. { There is no issue with operator ordering here, because only diagrams where the gauge boson insertions at the Wilson lines are contracted with fields on the other side of the correlator contribute. The diagrams in which the gauge bosons from the Wilson lines are contracted with fields on the same side, i.e., diagrams $(9)$, $(9r)$, $(10)$, and $(10r)$ give vanishing contributions.} There are two types of propagator structures here, as can be seen from Table~\ref{tab:2prop-structure}, but it turns out it is necessary to calculate them together in order to cancel possible IR divergences, specifically between diagrams $(4)$, $(7)$, and $(7r)$. Calculating the momentum-integrated spectral function from the difference of the corresponding Wightman functions gives, after performing all integrals that do not involve the temperature dependent pieces,
\begin{align}
\left. \varrho_E^{++}(p_0) \right|_{{\rm NLO}}^{3-11} &= g^2 (N_c^2 - 1) N_c p_0^{-\epsilon} \frac{\pi \Omega_{3-\epsilon}^2}{4(2\pi)^{6-2\epsilon}} {\tilde{\mu}}^\epsilon \int_0^\infty \frac{\diff k }{k^{1+\epsilon}} \nonumber  \\
& \quad \times \bigg( (n_B(k) -n_B(k+p_0) ) (-(3-\epsilon)p_0^2 + (p_0+k)^2 )\Theta(p_0  + k) (p_0 + k) \nonumber \\
& \quad - (n_B(k) -n_B(k-p_0) ) (-(3-\epsilon)p_0^2 + (p_0-k)^2 ) \Theta( k - p_0) (-p_0 + k) \nonumber \\
& \quad + (1 + n_B(k) + n_B(p_0-k)  ) (-(3-\epsilon)p_0^2 + (p_0 - k)^2 ) \Theta(p_0 - k) (p_0 - k) \nonumber \\
& \quad - (1 + n_B(k) + n_B(-p_0-k)  ) (-(3-\epsilon)p_0^2 + (p_0 + k)^2) \Theta(-p_0 - k) (-p_0 - k) \nonumber \\
& \quad- (1 + 2n_B(k)) (-(2-\epsilon) p_0^2) p_0 \bigg) \,,
\end{align}
which contains both the finite-temperature contributions as well as the vacuum parts, in arbitrary $d = 4 - \epsilon$ dimensions.

It is instructive to evaluate the vacuum part separately from the rest. We obtain (assuming $p_0>0$ as before)
\be
\left.\varrho_E^{++}(p_0) \right|_{{\rm NLO},\, T=0}^{3,4,7,7r,8,8r,11} &=& (N_c^2 -1) \frac{(d-2) \pi \Omega_{d-1}}{2 (4\pi^2) (2\pi)^{d-1}} g^2 p_0^{d-1}  N_c \nn\\
&\times& \bigg[ \frac{1}{\epsilon} + \frac{7}{12}  + \frac{1}{2}  \ln \left( \frac{\mu^2}{p_0^2} \right) + \frac{\gamma_E + \psi(3/2)}{2} \bigg] \,,
\ee
where we have kept the same prefactors that appeared in the gauge boson self-energy and in the LO result in arbitrary dimensions, so that the UV divergent pieces can be added straightforwardly.

The contribution at finite temperature can be written in $d=4$, or equivalently $\epsilon=0$, without any issue of possible divergences. Doing so, we get
\be
& & \varrho_E^{++}(p_0) \big|_{{\rm NLO}}^{3,4,7,7r,8,8r,11} \nn\\
&=& g^2 (N_c^2 -1) N_c \bigg\{ \frac{(d-2) \pi \Omega_{d-1}}{2 (4\pi^2) (2\pi)^{d-1}}  p_0^{d-1}   \bigg[ \frac{1}{\epsilon} + \frac{7}{12}  + \frac{1}{2}  \ln \left( \frac{\mu^2}{p_0^2} \right) + { 1 - \ln(2)} \bigg] \nn\\[4pt]
&+& \frac{1}{2(2\pi)^3} \int_0^\infty \frac{\diff k}{k} \bigg[  6k^2 p_0 n_B(k) 
+ |k-p_0| (-3p_0^2 + (p_0 - k)^2) n_B(|p_0-k|) \nn \\[4pt]
&-& |k+p_0| (-3p_0^2 + (p_0+k)^2 ) n_B(|p_0+k|)  \bigg] \bigg\}\,, \label{eq:2-prop-fin-result}
\ee
which completes the calculation of all the NLO pieces. We will add all results in the subsequent section~\ref{sec:add-results}{, where we will further rearrange~\eqref{eq:2-prop-fin-result} such that $n_B(k)$ becomes an overall factor in the integrand}.

\subsection{Remarks on infrared and collinear safety}

Up to the UV renormalization that we will discuss in the next subsection, we have obtained explicitly finite results of the spectral function $\varrho_E^{++}(p_0)$ at NLO. That is to say, all potentially infrared and collinear divergent diagrams and subdiagrams have been added up to a physical result. In this subsection, we will discuss the infrared and collinear structures of the diagrams in detail, highlighting the aspects that require particular care when performing these calculations.

\subsubsection{IR aspects}

We first discuss the cancellation of the divergences in the IR limit. Specifically, in bosonic thermal field theory the Bose-Einstein distribution introduces an extra $1/|k_0|$ factor as we let $k_0 \to 0$ in $n_B(|k_0|)$, which means that Feynman diagrams at 1-loop and beyond are potentially more singular than their vacuum counterparts in the low-energy limit $|k_0| \ll T$. Moreover, if we look at an individual Feynman diagram, e.g., diagram $(4)$ (see Table~\ref{tab:2prop-vertex}), one can explicitly see that one encounters an IR divergence because of the extra factors of $1/k_0$ coming from the Wilson line, in addition to the one that originates from the thermal distribution.

Operationally, however, individual Feynman diagrams with a single gauge boson insertion from a Wilson line are not IR divergent because the corresponding $1/k_0$ factor flips sign when approaching zero from either side (positive or negative), while the $1/|k_0|$ factor originated in the thermal distribution appearing in the free thermal propagator does not change sign, meaning that the integrand is an odd function of $k_0$ near the origin $k_0=0$. Since the real parts of the $1/k_0$ factors coming from Wilson lines must be interpreted as principal values, the integral around $k_0=0$ gives a finite contribution for these diagrams.

Diagrams with two gauge boson insertions from Wilson lines, with those gauge bosons connected through a propagator, are nonetheless individually divergent, because now we have a factor of $1/k_0^2$ coming from the Wilson lines that does not cancel with its reflected version $k_0 \to -k_0$ as we take $k_0 \to 0$. There are two such types of diagrams in our calculation: diagram $(4)$, which has a net momentum flow $p$ through the loop, and diagrams $(7)$, $(7r)$, which have their momentum flow $p$ disconnected from the loop momentum $k$ (see Figure~\ref{fig:diagrams}). To be explicit, if we take $(7)$ and $(7r)$ in Feynman gauge at $T=0$, we have
\begin{align}
    (7) &= \frac{N_c}{2} (p_0^2 \delta_{ij} - p_i p_j) \D(p)_{JI} \int_k \frac{1}{k_0^2} \D(k)_{II} \,, \\
    (7r) &= \frac{N_c}{2} (p_0^2 \delta_{ij} - p_i p_j) \D(p)_{JI} \int_k \frac{1}{k_0^2} \D(k)_{JJ} \,.
\end{align}

The problem is then clear: As we have written them, $(7)$ and $(7r)$ are unambiguously IR divergent for any $\p$, as is their sum, so it appears we have an ill-defined intermediate result. Nonetheless, in the case of our spectral function, we can get extra guidance on the IR divergence cancellation because the IR divergence in the contributions of $(7)$ and $(7r)$ to the spectral function becomes localized at $p_0^2 = \p^2$. It turns our that the integrals over $k$ are independent of whether the propagator indices are $I$ or $J$. So when we build the spectral function by taking $(J=2,I=1) - (J=1,I=2)$, we get a distribution with support only on $p_0^2 = \p^2$. Since distributional functions need to be integrated if one is interested in calculating an experimental observable, the cancellation of this IR divergence, if it happens, must be obtained by integrating over $\p$ in the remaining diagrams near $\p^2 = p_0^2$. Indeed, that is what actually happens. To illustrate this point, we point out that the contribution to the spectral function coming from diagrams $(3)$, $(4)$, $(8)$, $(8r)$, $(11)$, can be calculated explicitly in vacuum (for $p_0>0$) by integrating over $k$ in $d=4$:
\begin{equation}
\left. \frac{\delta_{ij} \delta^{ad} [\rho_E^{++}]^{da}_{ji}(p_0,\p)}{g^2 N_c (N_c^2 -1)} \right|^{(3),(4),(8),(8r),(11)}_{{\rm NLO}, \, T=0} = \frac{-1}{4\pi} \Theta (p_0 - |\p|) \frac{4 |\p| p_0^3 + (p_0^4 - \p^4) {\rm arctanh}(|\p|/p_0) }{ p_0 |\p| (p_0^2 - \p^2)}\,,
\end{equation}
which is manifestly divergent as $|\p| \to p_0$. Crucially, if one integrates this over a region including $|\p|=p_0$, then a divergence appears, with opposite sign to the one originated from diagrams $(7)$ and $(7r)$. This is one way to see the cancellation of this IR divergence between diagrams $(7)$, $(7r)$ and diagrams $(3)$, $(4)$, $(8)$, $(8r)$, $(11)$, in which we first integrate over $k$ and then integrate ${\bs p}$ over a region that contains $|{\bs p}|=p_0$. Both ${\bs p}$ integrals (for $(7)$, $(7r)$ and for $(3)$, $(4)$, $(8)$, $(8r)$, $(11)$) give divergent results, but with opposite signs, and these divergent pieces will cancel each other explicitly to give a finite result if we regulate the Wilson line $1/k_0$ contributions in the same way in all diagrams and then remove the regulator after the sum is performed.

On the other hand, if one first integrates over ${\bs p}$ on an arbitrary region, holding off on the IR divergent integrals over $k$, adds the results, and then performs the integration over $k$ for all diagrams ($(3)$, $(4)$, $(7)$, $(7r)$, $(8)$, $(8r)$, $(11)$) simultaneously, one does not encounter any IR divergences at all. This can be interpreted as an explicit verification that the result of performing the loop momentum $k$ integrals defines a distribution as a function of ${\bs p}$ that gives finite results over any integration region.
While at finite temperature it is no longer possible to get explicit expressions in terms of ordinary functions after integrating over $k$, the cancellation between the IR singularities of the different diagrams happens in the same way.

In vacuum, one could have tried to deal with these divergent diagrams by using dimensional regularization to regulate both the UV and IR divergences, and since diagrams $(7)$ and $(7r)$ give scaleless integrals, one can take them to be zero in dimensional regularization. But then, when one calculates diagram $(4)$ and integrates over some range of $\p$ that includes $|\p|=p_0$, one will encounter an uncompensated IR divergence. Since the theory has been already regulated in dimensional regularization with the above choice ($\epsilon_{\rm UV} = \epsilon_{\rm IR} = \epsilon$), one will find that to calculate the correct renormalization group running of the theory (which is in principle a UV effect), we would need to calculate the $1/\epsilon$ pole contribution by inspecting the IR divergence of diagram $(4)$ instead (whereas this $1/\epsilon$ pole would have come from the UV divergence of diagrams $(7)$ and $(7r)$ if we had not taken these diagrams to be zero). Of course, the simplest way to get the renormalization group equations is to just keep track of the UV divergences separately from the IR divergences. This is even more natural when we go to finite temperature, where the finite temperature contributions in the loop integrals of $(7)$ and $(7r)$ are manifestly UV finite, but clearly IR divergent. For that reason, we decided to first compute the integral over $\p$ in our actual calculation, because with that approach we render the IR sector finite from the start and thus we can identify all divergences as UV divergences, up to unphysical collinear singularities that we now discuss.

\subsubsection{The collinear limit}

Another nontrivial aspect of the calculation, at least from the point of view of a naive diagrammatic computation of the formation/dissociation rates, is how the result becomes finite in the limit where the momentum flow through the propagators in our diagrams becomes collinear, i.e., $\k$ becomes parallel to $\p$.

\begin{figure}
    \centering
    \includegraphics[width=\textwidth]{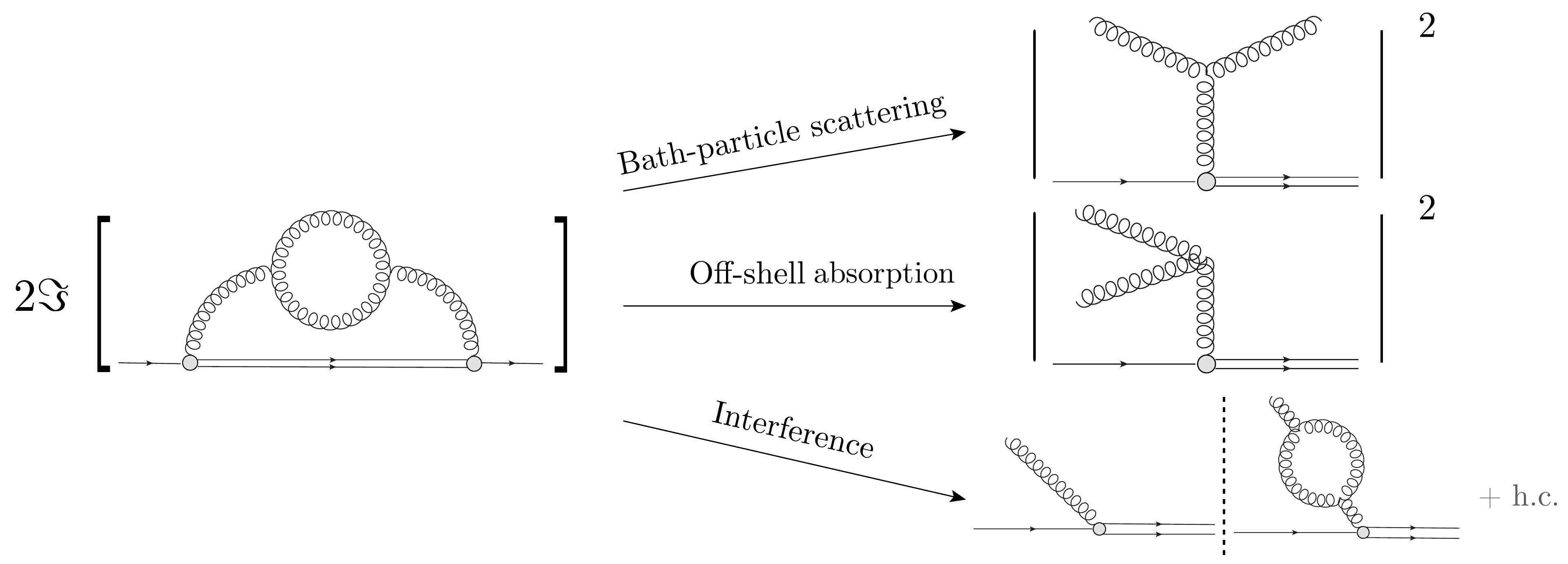}
    \caption{Decomposition of the imaginary part of the forward scattering amplitude of a singlet field with a gauge boson self-energy in terms of ``cut'' diagrams. When taken individually, these diagrams present singularities in the collinear limit, which only cancel after all terms are added.}
    \label{fig:collinear-gluon-self-energy}
\end{figure}

The way in which how potential collinear divergences are cancelled is most clearly highlighted in Figure~\ref{fig:collinear-gluon-self-energy}, where the imaginary part of the forward scattering amplitude of a singlet field, associated with the gauge boson self-energy is decomposed in terms of products of tree-level amplitudes by means of the optical theorem.\footnote{Note that the cutting rules at finite temperature can involve loop momentum flow in either direction, i.e., the cuts do not determine that momentum flows in only one direction as in vacuum.} If treated individually, the ``cut'' diagrams are each separately divergent in the collinear limit $\p \parallel \k $, and therefore when one integrates over momenta to obtain a cross-section for each diagram the result appears to be divergent as well. This also means that, in general, one cannot get a reliable estimate of the size of the cross-section by simply looking at one diagram, e.g., scattering by particles in the thermal bath (upper right subdiagram in Figure~\ref{fig:collinear-gluon-self-energy}).

In practice, this means that when we evaluate the imaginary part of an amplitude, we must sum over all ``cuts'', which manifest as poles of propagators before carrying out the integrals. In our approach to the calculation, this is relevant when we take the real part of a retarded correlation, as shown in section~\ref{sec:gluon-self-energy}. When we write the contributions to the spectral function from diagrams $(5)$, $(5r)$, $(6)$, $(6r)$ in section~\ref{sec:3-propagator-evaluation}, we cannot evaluate the contributions from each pole of the propagators separately. Broadly speaking, one has to either regulate the divergence, calculate each contribution, add them up, and then remove the regulator, or, calculate the contribution of all poles simultaneously from the start. In obtaining our results, we follow two different integration orders, one that relies on the introduction of a regulator, and the other that does not, both of which are discussed in Appendix~\ref{app:collinear-integrals}. Both approaches agree and give the same final result, even though the resulting algebraic expressions, while equivalent, may have significant differences in the way they look (i.e., in terms of what functions they are expressed) when obtained from one method or the other.

A rigorous justification of why the result is finite is as follows: When integrating over $\p$, the numerator is a holomorphic function of the integration momentum $|\p|$ and the angle between $\p$ and $\k$. For definiteness, let us work under the assumption that we first perform the integral over $|\p|$. Given the propagator structure of the diagrams, all of the integrands we consider can be written in the form
\begin{equation}
    \ml{I}(z) = \frac{H(z)}{(z-z_1)^n (z-z_2)^m} \,,
\end{equation}
where $z$ is the complexified $|\p|$ integration variable, $z_1$ and $z_2$ are the positions of the poles (which are functions of the variables over which we are not integrating at this point, and are infinitesimally displaced off the real line by $\pm i0^+$, as usual in a propagator), and $n,m$ give the order of the poles, assuming that $H$ has no zeros at $z_1$ or $z_2$. Then, because of the concrete expressions we have for the vertex factors, $H(z)$ can be taken to be real on the real line, and then we can obtain the contribution to the retarded correlation functions by taking the imaginary part of the integral of $\ml{I}(z)$. This means that all we have to do is to evaluate the residues at their corresponding poles, because only they can yield an imaginary part. This leads to
\begin{equation}
\begin{split}
    {\rm Im} \left\{ \int \ml{I}(z) \diff z \right\} &=  \frac{\pi}{(n-1)!} \left. \frac{\diff^{n-1}}{\diff z^{n-1}} \left( \frac{H(z)}{(z-z_2)^m} \right) \right|_{z=z_1} + \frac{\pi}{(m-1)!} \left. \frac{\diff^{m-1}}{\diff z^{m-1}} \left( \frac{H(z)}{(z-z_1)^n} \right)  \right|_{z=z_2} \,.
\end{split}
\end{equation}
In this language, the collinear limit is realized when the two poles merge, i.e., $z_2 \to z_1$. If we naively evaluate each pole separately, this will generically lead to divergences because we will have contributions of the form $\lim_{z_2\to z_1} H(z_1)/(z_1-z_2)^k$ for some positive power $k$. However, if we take $z_1=z_2$ from the start, we see that the result actually must be finite and equal to
\begin{equation}
    {\rm Im} \left\{ \int \ml{I}(z) \diff z \right\} =  \frac{\pi}{(m+n-1)!} \left. \frac{\diff^{m+n-1}}{\diff z^{m+n-1}} \big( H(z)\big) \right|_{z=z_1=z_2} \,,
\end{equation}
which is simply a matter of evaluating the residue of a pole of higher order. We can also see that the limit is well defined from the point of view of $z_1 \neq z_2$, for which we show the two cases that are relevant for our purposes:
\begin{enumerate}
    \item $n=m=1$ gives
    \begin{equation}
        {\rm Im} \left\{ \int \ml{I}(z) \diff z \right\} = \pi \frac{H(z_1) - H(z_2)}{z_1-z_2} \xrightarrow{z_2 \to z_1} \pi H'(z_1) \,,
    \end{equation}
    which is the definition of the derivative of a function.
    \item $n=2$, $m=1$ case was already discussed in Ref.~\cite{Binder:2020efn} and gives
    \begin{equation}
    \begin{split}
        {\rm Im} \left\{ \int \ml{I}(z) \diff z \right\} &= \pi \left[ \frac{H'(z_1)}{(z_1-z_2)} - \frac{H(z_1)}{(z_1-z_2)^2} + \frac{H(z_2)}{(z_2-z_1)^2} \right] \\
        &= \pi \frac{H(z_2) - H(z_1) - (z_2 - z_1) H'(z_1) }{(z_2-z_1)^2} \xrightarrow{z_2 \to z_1} \frac{\pi}{2} H''(z_1) \,,
    \end{split}
    \end{equation}
    as can be verified, e.g., by a Taylor expansion.
\end{enumerate}

This conclusively shows that the apparent collinear singularities, which arise because setting one propagator on-shell induces a divergent behavior on other propagators in the collinear limit where $\p \parallel \k$, are all compensated within each uncut diagram, and therefore that any consistent method of calculating the integrals will give the same result.

\subsection{The complete NLO integrated spectral function}

\subsubsection{Coupling constant renormalization} \label{sec:coupling-renorm}

Before proceeding to give the final result at NLO, we discuss including the effects of the renormalization of the coupling constant. This is most easily understood if we recall that the quantity appearing in the transition rates is actually
\begin{equation}
    g^2 [g_E^{++}]^{da}_{ji}(y,x) = g^2 \Big\langle \big[E_j(y) \ml{W}_{[( y^0, {\bs y}), (+\infty, {\bs y})]} \big]^d
\big[ \ml{W}_{[(+\infty, {\bs x}),(x^0, {\bs x})]} E_i(x) \big]^a \Big\rangle,
\end{equation}
i.e., with an extra factor of $g^2$ that will be renormalized due to its own loop effects. Since the above expression comes directly from vertices in the Lagrangian, we can interpret every appearance of $g^2$ as a factor of the bare coupling constant, and then the appropriate substitution to include the corrections due to coupling constant renormalization is to substitute
\begin{align}
g^2 \to Z_{g^2} g^2.
\end{align}
Since $Z_{g^2} = 1 + \O(g^2)$, including modifications due to the renormalization of $g^2$ only affects our NLO calculation by changing the prefactor of the LO calculation, because any substitution of $g^2 \to Z_{g^2} g^2$ in our NLO results would lead to NNLO effects, of which we are not keeping track here.

At NLO, we can write
\be
\big( g^2 \varrho_E^{++}(p_0) \big) \big|_{\rm NLO} =  g^2(Z_{g^2}-1)\varrho_E^{++}(p_0)\big|_{\rm LO} + g^2 \varrho_E^{++}(p_0)\big|_{\rm NLO}\,.
\ee
So we need to combine the NLO result of the spectral function with the LO result multiplied by the factor $(Z_{g^2}-1)$. It is well known that
\begin{equation}
    Z_{g^2} =  1 - \frac{g^2}{8\pi^2 \epsilon} \left( \frac{11 N_c}{3} - \frac{4}{3} n_f C({\bs N}) \right) \,.
\end{equation}
Therefore, we obtain that
\begin{align}
g^2(Z_{g^2}-1)\varrho_E^{++}(p_0)\big|_{\rm LO} =  - g^4 (N_c^2-1) \frac{(d-2) \pi \Omega_{d-1}}{2 (4\pi^2) (2\pi)^{d-1}}  p_0^{d-1} \left( \frac{11 N_c}{6 \epsilon} - \frac{2}{3 \epsilon} n_f C({\bs N}) \right),
\end{align}
which we can now add to our previous results.

\subsubsection{Adding all results} \label{sec:add-results}

We are now in a position to present the final result for the momentum-integrated spectral function. Importantly, the $1/\epsilon$ pole cancels altogether in $g^2\varrho_E^{++}(p_0)$, which means this quantity and $g^2[g_E^{++}]$ are scale independent. Conveniently, we have kept all of the prefactors of the $1/\epsilon$ poles in the same form, and so the cancellation is straightforward. All that remains is to add up the finite pieces, where the limit $d \to 4$ (or $\epsilon \to 0$) is taken explicitly because there is no obstacle to it at this point. Therefore, adding up our results from sections~\ref{sec:gluon-self-energy},~\ref{sec:3-propagator-evaluation},~\ref{sec:2-propagator-evaluation}, and~\ref{sec:coupling-renorm}, we obtain
\begin{equation} \label{eq:spectral-full-result}
\begin{split} 
    &\left. g^2 \varrho_E^{++}(p_0) \right|_{{\rm LO} + {\rm NLO}} \\
    &= \frac{ g^2 ( N_c^2 -1 ) p_0^3}{(2\pi)^3} \bigg\{ 4\pi^2 + g^2  \bigg[ \left( \frac{11}{12} N_c - \frac13 N_f \right) \ln \left( \frac{\mu^2}{ p_0^2}\right) + a_g N_c  - a_F N_f   \bigg] \bigg\} \\
    & \,\, + \frac{g^4 (N_c^2-1)}{(2\pi)^3} \bigg\{ \int_0^\infty \!\! \diff k \, 2 N_f n_F(k) \bigg[ -2k p_0 + (2k^2 + p_0^2) \ln \left| \frac{k+p_0}{k-p_0} \right|  + 2 k p_0 \ln \left| \frac{k^2 - p_0^2}{p_0^2} \right| \bigg] \\
    & \quad \quad \quad \quad \quad \quad \, + \int_{0}^\infty \!\! \diff k \, 2 N_c n_B(k) \bigg[ -2 k p_0 + (k^2+p_0^2) \ln \left| \frac{k+p_0}{k-p_0} \right|  + k p_0 \ln \left| \frac{k^2-p_0^2}{p_0^2} \right| \bigg]  \\ 
    & \quad \quad \quad \quad \quad \quad \, + \int_0^\infty \diff k \, \frac{2 N_c n_B(k)}{k} \ml{P} \left( \frac{p_0^2}{ p_0^2 - k^2} \right) \bigg[ k^2 p_0 + (k^3 + p_0^3) \ln \left| \frac{k-p_0}{p_0} \right| \\ 
    & \quad \quad \quad \quad \quad \quad \quad \quad \quad \quad \quad \quad \quad \quad \quad \quad \quad \quad \quad \quad \quad \quad \quad  + ( -k^3 + p_0^3) \ln \left| \frac{k + p_0}{p_0} \right| \bigg]  \\
    & \quad \quad \quad \quad \quad \quad \, + \int_0^\infty \!\! \diff k \, 2N_c n_B(k) \mathcal{P} \left( \frac{k^3 p_0}{k^2 - p_0^2} \right) \bigg\}\,,
\end{split}
\end{equation}
{ where we have rearranged the integrand of~\eqref{eq:2-prop-fin-result} so that $n_B(k)$ appears as an overall factor.} Also, we have denoted $N_f = n_f C({\bs N})$, and the numbers $a_g$ and $a_F$ are given by
\begin{align}
    a_g &= { \frac{149}{36} + \frac{\pi^2}3 - \frac{11}{12}\ln(4)  \approx 6.157987}\,, \nonumber \\
    a_F &= \frac{10}{9} - \frac{1}{3} \ln(4) \approx 0.649013 \,, \label{eq:consts}
\end{align}
and the coupling constant is given in the $\overline{\rm MS}$ scheme, with all finite pieces accounted for. Since in light of section~\ref{sec:coupling-renorm}, $g^2 \varrho_E^{++}$ does not run after the coupling constant renormalization is included, the RHS of~\eqref{eq:spectral-full-result} multiplied by $g^2$ is independent of $\mu$. So we have
\begin{equation}
\begin{split}
    &0 = 4\pi^2 \frac{\diff g^2}{\diff\ln \mu} \left(1 + \O(g^2) \right) + g^4  \left( \frac{11}{6} N_c - \frac23 N_f \right) + \O(g^6) \\
    &\implies \beta(\alpha) = \frac{\diff \alpha}{\diff\ln \mu} = -\frac{\alpha^2}{2\pi} \left( \frac{11}{3} N_c - \frac43 N_f \right) + \O(\alpha^3)\,,
\end{split}
\end{equation}
where $\alpha = \frac{g^2}{4\pi}$, verifying the expected running of the coupling for non-Abelian gauge theory.

Although somewhat lengthy, it is remarkable that we can write an explicit expression (even though there is one integral left to be done, with which we deal numerically) for the full momentum-integrated spectral function at NLO, which involves $O(20)$ diagrams in non-Abelian gauge theory with Wilson line insertions. 

{ At this point, let us summarize the similarities and differences with the previous results in the literature. The first thing to note is that, up to an overall normalization factor, the contribution of fermions to the spectral function is the same as that obtained in the U$(1)$ case~\cite{Binder:2020efn}, as expected. Secondly, by assembling together all terms that have the same logarithm factors, the finite temperature part of our NLO result of the spectral function can be shown to be equal to that obtained in~\cite{Burnier:2010rp}, which calculates the correlator~\eqref{eq:correlator-intro-Eller} by means of the imaginary time formalism. However, we believe this is just a coincidence at NLO. Since the zero temperature piece obtained here differs from that of~\cite{Burnier:2010rp}, we expect that at NNLO in perturbation theory the results of these two correlators will be different in the temperature-dependent terms as well. This is consistent with our initial observation in the introduction that they are, in fact, different correlators. 

At this point, it is worth to expound on the fact that in the quantum theory there is a clear difference between the two correlators
\begin{align}
    \left\langle {E}^a_i(t) \ml{W}^{ab}(t,0)  {E}^b_i(0) \right\rangle_T \neq \left\langle {\rm Tr}_{\rm color} \left[ U(-\infty,t) E_i(t) U(t,0) E_i(0) U(0,-\infty) \right] \right\rangle_T \label{eq:quarkonia-neq-hq}
\end{align}
given by the fact that operators in quantum mechanics at different times, in general, do not commute. In the correlator on the left of~\eqref{eq:quarkonia-neq-hq}, all of the gauge field $A_0$ operator insertions from Wilson lines occur between the two electric fields (relative to the thermal density matrix $e^{-\beta H}$), as opposed to the correlator of the heavy quark diffusion coefficient, which has explicit gauge field insertions that are in between the electric fields as well as adjacent to the thermal density matrix. This means that, even if one could conceivably relate the SU$(N_c)$ color structures of both correlators, going from one correlator to the other in the full quantum theory involves evaluating several non-trivial quantum-mechanical commutators, and there is therefore no reason to expect that the two correlation functions be equal.

Finally, since we computed the zero temperature result explicitly as well, we can compare it with the appropriate limit of the field strength correlator considered in~\cite{Eidemuller:1997bb}:
\begin{align} \label{Eidemuller-Jamin-correlator}
\left\langle 0 \left| \ml{T} \left( F_{\mu \nu}^a(t)  \ml{W}_{[t,0]}^{ab} F_{\rho \sigma}^b(0) \right)  \right| 0 \right\rangle \,.
\end{align}
After rearranging their results, we find that the $\overline{\rm MS}$ finite constant in this time-ordered correlator and the one in our present work are the same:
\begin{align}
a_g^{{\rm ref.} \, [113]} = a_g^{\rm this \, work}  \,.
\end{align}

While this is a strong check of our results, we want to stress that depending on how one implements the interplay between time-ordering of the operators acting on the Fock space of the theory and the SU$(N_c)$ matrix products in the Wilson lines, one could be defining mathematically different objects that are relevant to different physics. For instance, due to the non-local nature of the electric fields dressed with Wilson lines, we have in general
\begin{align}
\Big\langle \ml{T} & \big[{E}_i(y) \ml{W}_{[( y^0, {\bs y}), (+\infty, {\bs y})]} \big]^a
\big[ \ml{W}_{[(+\infty, {\bs x}),(x^0, {\bs x})]} {E}_i(x) \big]^a \Big\rangle_T \nonumber \\ 
&\neq \theta(y^0 - x^0) \Big\langle  \big[{E}_i(y) \ml{W}_{[( y^0, {\bs y}), (+\infty, {\bs y})]} \big]^a
\big[ \ml{W}_{[(+\infty, {\bs x}),(x^0, {\bs x})]} {E}_i(x) \big]^a \Big\rangle_T \nonumber \\ 
& \quad + \theta(x^0 - y^0) \Big\langle  
\big[ \ml{W}_{[(+\infty, {\bs x}),(x^0, {\bs x})]} {E}_i(x) \big]^a \big[{E}_i(y) \ml{W}_{[( y^0, {\bs y}), (+\infty, {\bs y})]} \big]^a \Big\rangle_T \,, \label{eq:discrepancy-T-ordered}
\end{align}
where we have omitted the Wilson lines at infinite time since they do not contribute in our calculations. As the notation suggests, the difference between them is due to operator ordering. In our notation, the correlator on the right hand side corresponds to the time-ordered version of the object we are physically interested in, whereas the one on the left hand side corresponds to the choice $I=J=1$ on the Schwinger-Keldysh contour because of the explicit time-ordering symbol. Concretely, the prescription ordering the operators along the Schwinger-Keldysh contour implies that the disposition of the vector potential $A_\mu$ operators in the Wilson lines, while remaining the same in terms of SU$(N_c)$ indices, will be different in terms of the quantum-mechanical operator ordering for the two correlators. However, as we show in Appendix~\ref{app:T-ordered-vacuum-5-5r} by direct computation, this prescription ($I=J=1$) gives the same result at NLO for the vacuum finite piece result after we set ${\bs y}\to {\bs x}$, as the one given in~\cite{Eidemuller:1997bb}, which is equal to that of the physical answer for heavy particles bound state formation/dissociation given in the present work. Therefore, while we have shown that at NLO one cannot distinguish between the correlator in~\eqref{Eidemuller-Jamin-correlator}, and the ones in~\eqref{eq:discrepancy-T-ordered} with ${\bs y}\to{\bs x}$, we do not expect such an equality to hold exactly, nor at higher orders in perturbation theory. Hence, we stress that it is of utmost importance to rigorously define the correlator that is of physical interest to each situation when comparing similar-looking correlation functions.}


This completes the technical section of this work, and now we apply this to the calculation of transition rates of heavy particle states in the thermal environment that we have discussed so far.

\section{Results}
\label{sec:results}
\subsection{Electric field correlator}
The NLO result of the non-Abelian electric field correlator, as computed in the previous section and cumulatively given in Eq.~(\ref{eq:spectral-full-result}), enters the thermally averaged bound state formation cross section in Eq.~(\ref{eqn:reco2}) and the dissociation rates in Eq.~(\ref{eqn:disso}) as:
\be
G^{>}_{ii}(\Delta E) = 2\big( 1+n_B(\Delta E)\big) \rho_E^{++}(\Delta E)\,,
\ee
where $\Delta E= {\bs p}_{\rm rel}^2/M - E_{\ml{B}}$ for bound state formation and $\Delta E= -{\bs p}_{\rm rel}^2/M + E_{\ml{B}}$ for bound state dissociation. The factor of 2 comes from the factor of $1/2$ in the definition of the momentum integrated spectral function~(\ref{eqn:rho_p_integrated}), which is just a convention. In either case, the NLO result can be written as
\begin{align}
\label{eqn:nlo_G}
&\big(G^{>}_{ii}(\Delta E)\big)^{\rm{LO+NLO}} = \big(G^{>}_{ii}(\Delta E) \big)^{\rm LO} \times \\ &\left[1+ \alpha N_c R_g^{T=0}(\mu/\Delta E)+ \alpha N_c R_g^{T\neq 0}(\Delta E/T) + \alpha N_f R_f^{T=0}(\mu/\Delta E) + \alpha N_f R_f^{T\neq 0}(\Delta E/T)  \right] \,.\nonumber
\end{align}
Here $N_f = n_f C({\bs N})$, with $n_f$ the number of fermion ``flavors'' in the plasma background and $C({\bs N})$ the normalization of the generator matrices corresponding to the group representation in which these fermions are ($C({\bs N}) = 1/2$ for the fundamental representation). $N_c$ is the number of ``colors'' of the gauge group SU($N_c$).
These NLO results are valid for any singlet-adjoint or adjoint-singlet transition for both quarkonium and DM. The translation of our general SU($N_c$) result to the U($1$) case is just the replacement $N_c \rightarrow 0$ and $N_f=n_f$ in Eq.~(\ref{eqn:nlo_G}). 

\begin{figure}
    \begin{subfigure}[t]{1\textwidth}
	    \centering
	    \includegraphics[height=3in]{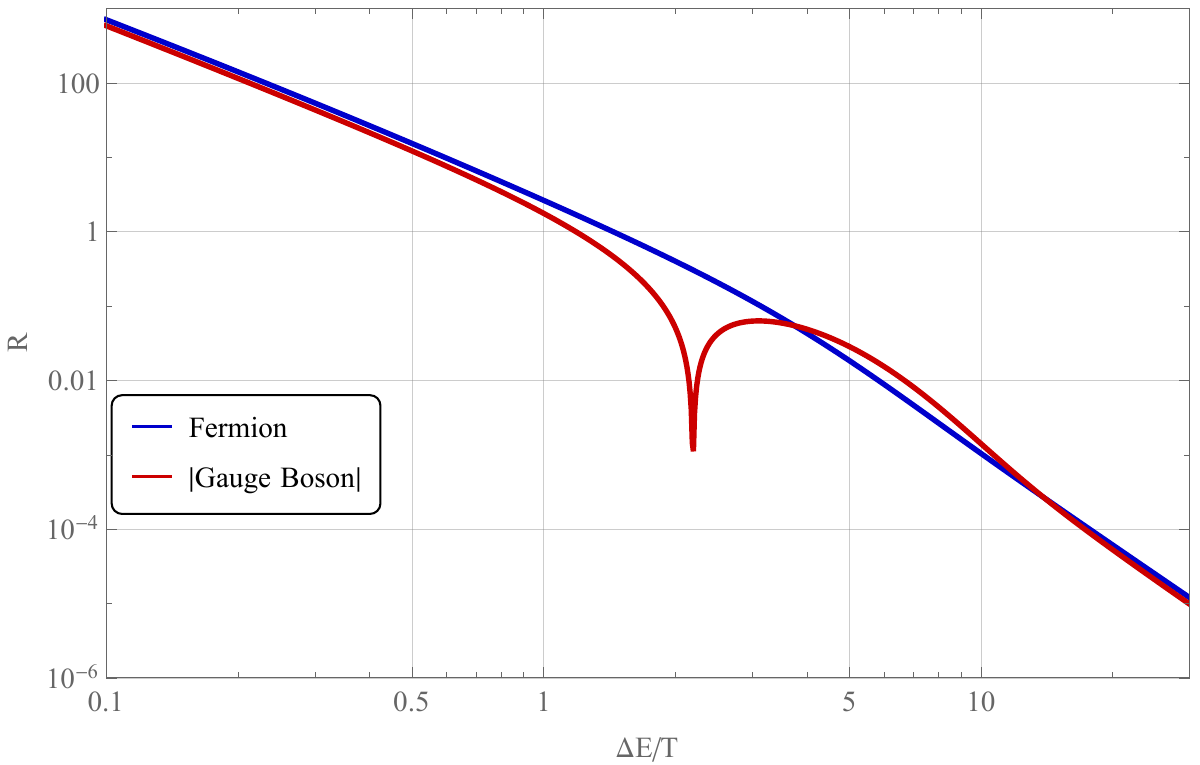}
	    \caption{}
	    \label{fig:8.1}
	\end{subfigure}%
    
    \begin{subfigure}[t]{1\textwidth}
	    \centering
	    \includegraphics[height=3in]{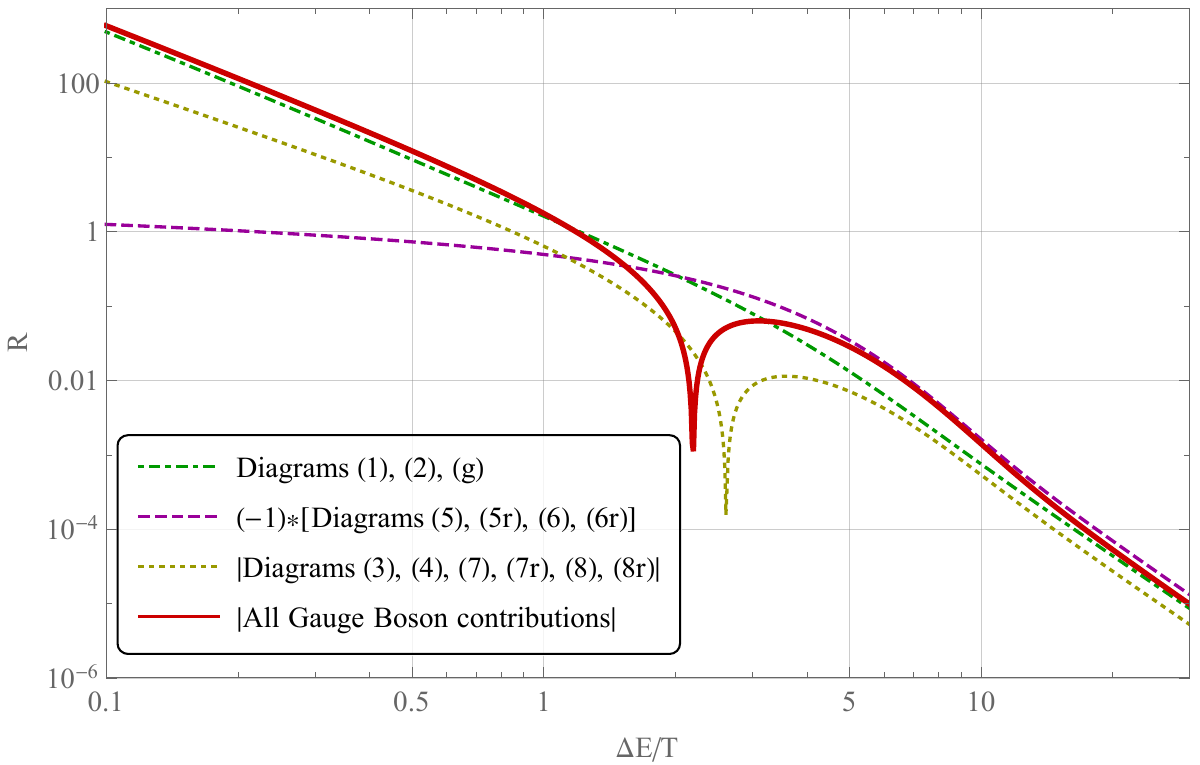}
	    \caption{}
	    \label{fig:8.2}
	\end{subfigure}%
\caption{Finite temperature parts of the $R$ functions. In (a), the blue curve corresponds to the fermion loop contribution Eq.~(\ref{eq:fermion}), in agreement with Ref.~\cite{Binder:2020efn}, and the red solid line depicts the gauge sector contribution Eq.~(\ref{eq:gauge}), which includes contributions from all diagrams except diagram $(f)$. $R_{g}^{T\neq0}$ is positive in the high temperature regime, changes the sign at the dip and becomes negative in the low temperature regime, whereas $R_{f}^{T \neq 0}$ is always positive. However, we note that the full LO $+$ NLO contribution to the rates, including the vacuum contributions, stays positive. In (b), the non-vanishing finite temperature contributions from different diagrams are shown in dashed lines, each of which can be gauge-dependent. Only the sum, plotted as the red solid curve, is gauge invariant.}
\label{fig:R}
\end{figure}

These new four $R$ functions incorporate the NLO vacuum and finite temperature corrections. They are dimensionless, independent of the value of the coupling constant, and independent of the SU($N_c$) group theory factors. Concretely, they are given by
\begin{align}
& R_f^{T=0}(\mu/\Delta E) =- \frac{1}{\pi} \left[ \frac{1}{3} \ln\left( \frac{\mu^2}{\Delta E^2} \right) + a_f\right] \,, \\
& R_g^{T=0}(\mu/\Delta E) = \frac{1}{\pi} \left[ \frac{11}{12} \ln\left( \frac{\mu^2}{\Delta E^2} \right)+ a_g \right] \,, \\
& R_f^{T\neq0}(x=\Delta E/T) = \frac{1}{\pi} x^{-3} \times \nn\\
\label{eq:fermion}
&\int_0^\infty \!\! \diff y \, 2 n_F(y) \bigg[ -2y x + (2y^2 + x^2) \ln \left| \frac{y+x}{y-x} \right|  + 2 y x \ln \left| \frac{y^2 - x^2}{x^2} \right| \bigg] \,, \\
& R_g^{T\neq0}(x=\Delta E/T) =  \frac{1}{\pi} x^{-3} \times \nn\\
&\bigg\{ \int_{0}^\infty \!\! \diff y \, 2 n_B(y) \bigg[ -2 y x + (y^2+x^2) \ln \left| \frac{y+x}{y-x} \right|  + y x \ln \left| \frac{y^2-x^2}{x^2} \right| \bigg] \nonumber\\
    &  \, + \int_0^\infty \diff y \, \frac{2 n_B(y)}{y} \ml{P} \left( \frac{x^2}{ x^2 - y^2} \right) \bigg[ y^2 x + (y^3 + x^3) \ln \left| \frac{y-x}{x} \right| + ( -y^3 + x^3) \ln \left| \frac{y + x}{x} \right| \bigg]  \nonumber\\
    & \, + \int_0^\infty \diff y 2 n_B(y) \ml{P} \left( \frac{y^3x}{ y^2 - x^2} \right)\bigg\} ,\label{eq:gauge}
\end{align}
where $\Delta E = p^2_\ma{rel}/M-E_{\ml{B}}$, the constants $a_f$ and $a_g$ are given in Eq.~(\ref{eq:consts}), and the equilibrium distributions in these dimensionless variables are $n_B(y) = (e^{y}-1)^{-1}$ and $n_F(y) = (e^{y}+1)^{-1}$. The temperature dependent $R$ functions are plotted in Fig.~\ref{fig:R} for different contributions. For $T \gg \Delta E$, the gauge part (marked as red in Fig.~\ref{fig:8.1}) is of comparable size with the fermion contribution (marked as blue in Fig.~\ref{fig:8.1}).\footnote{By looking at the plot one sees that these results grow at high temperatures, which will eventually force a breakdown of the fixed-order perturbative calculation we have adopted here. To have quantitative control over the regime $T \gg \Delta E$, one needs to include higher-order contributions as well, e.g., using HTL resummation. See Section~\ref{sect:highT}.} While the fermion part stays always positive, the gauge part flips the sign and becomes negative when $T\lesssim \Delta E$. However, the full LO $+$ NLO contribution to the rates, including the vacuum contributions, stays positive. In Fig.~\ref{fig:8.2}, we plot contributions from different diagrams for the gauge part,\footnote{We only include diagrams that give a nonvanishing contribution to the $R$ ratios discussed in this section. As discussed earlier, this means that we do not need to include $(9)$, $(9r)$, $(10)$, $(10r)$, and we also do not include $(11)$ as it gives a vanishing contribution once we integrate over the whole ${\bs p}$ domain.} which can be individually gauge dependent and thus can contain unphysical information. Only the sum of all contributions in the gauge part is gauge independent.

\subsection{Rates}
For practical applications, we need to calculate the rates for dynamical processes in the context of quarkonia and DM. To this end, while it is the main focus of the present work, it does not suffice to have just the knowledge of the environment plasma: one also needs to know the properties of the two-body states that propagate through the thermal environment. In particular, we need to compute the averaged dipole overlap integral $|\langle \psi_{\ml{B}} | {\bs r} | \Psi_{{\bs p}_\ma{rel}} \rangle |^2$ that appears in the dissociation rate~\eqref{eqn:disso} and recombination cross section~\eqref{eqn:reco}, by using the wavefunctions of the scattering and bound states. 

Firstly, the wavefunctions can be obtained by solving the corresponding Schr\"odinger equations:
\begin{align}
\left[ -\frac{\nabla^2}{2\mu} + V_{\text{s}}({\bs r})\right]\Psi_{{\bs p}_{\text{rel}}}({\bs r})&= E_{{\bs p}_{\text{rel}}} \Psi_{{\bs p}_{\text{rel}}}({\bs r}) \,,\\
\left[ -\frac{\nabla^2}{2\mu} + V_{\text{b}}({\bs r}) \right]\psi_{nlm}({\bs r})&= E_{nlm} \psi_{nlm}({\bs r})\,,
\end{align}
with the standard normalization conditions given by, respectively:
\begin{align}
\int \text{d}^3 r\, \Psi^{*}_{{\bs p}_{\text{rel}}}({\bs r}) \Psi_{{\bs p}_{\text{rel}}^{\prime}}({\bs r})&=(2\pi)^3 \delta^3({\bs p}_{\text{rel}}-{\bs p}_{\text{rel}}^{\prime}) \,,\\
\int \text{d}^3 r \, \psi_{nlm}^{*}({\bs r}) \psi_{n^{\prime}l^{\prime}m^{\prime}}({\bs r}) &= \delta_{n n^{\prime}}\delta_{l l^{\prime}}\delta_{m m^{\prime}} \,.
\end{align}
Secondly, the dipole overlap integral in terms of these spatial wavefunctions, after the third component of the orbital angular momentum has been averaged, is given by (c.f. Eq.~(\ref{eqn:average_ml})):
\be
\big| \langle \psi_{\ml{B}} | {\bs r} | \Psi_{{\bs p}_\ma{rel}} \rangle \big|^2 = \frac{1}{2l+1}\sum_{m_l=-l}^l 
\big| \langle \psi_{nlm_l} | {\bs r} | \Psi_{{\bs p}_\ma{rel}} \rangle \big|^2 \,.
\ee

We now review the analytic results of the dipole overlap integrals for general Coulomb potentials, which have been worked out in Refs~\cite{Brambilla:2011sg,Yao:2020xzw} for the 1S and 2S states and in Ref~\cite{Yao:2020xzw} for the 2P state. We consider the more general group decomposition (c.f., Eq.~(\ref{eq:group})):
\be
\bs{R} \otimes \bs{R^{\prime}} = \bigoplus_{\bs{\hat{R}}} \bs{\hat{R}} \,,
\ee
which results in the following Coulomb potential entering the Schr\"odinger equations (see, e.g., Ref.~\cite{Harz:2018csl}):
\begin{align}
V(r)= - \frac{\alpha}{2r}\left[C_2(\bs{R}) + C_2(\bs{R^{\prime}})-C_2(\bs{\hat{R}})\right] \,.
\end{align}
We absorb the group dependent part into an effective coupling for bound states and scattering states as $V_{s/b}=- \alpha^{\text{eff}}_{s/b}/r$.
Then, the dipole overlap integral for the 1S state is:
\begin{align}
 \Delta E^3 \big| \langle \psi_{10} | {\bs r} | \Psi_{{\bs p}_\ma{rel}} \rangle \big|^2  =  \frac{2^8\pi^2 \alpha_b^{\text{eff}}}{\mu^2} \frac{\zeta_s \zeta_b^4(1+\zeta_s^2)(1-\frac{\zeta_s}{2 \zeta_b})^2}{(1+\zeta_b^2)^3(1-e^{-2\pi \zeta_s})} e^{-4 \zeta_s \text{arccot}(\zeta_b)} \,,
\end{align}
where we introduced $\zeta_i = \alpha_i^{\text{eff}}/v_{\text{rel}}$. For 2S and 2P states (the Coulombic 2P state corresponds to the 1P state in the quarkonium spectra), we have\footnote{Which we obtain from Ref.~\cite{Yao:2020xzw} by replacing $\eta \rightarrow - \zeta_s$, $a_B \rightarrow (\mu \alpha^{\text{eff}}_b)^{-1}$ and $a_B {p}_{\text{rel}} \rightarrow \zeta_b^{-1}$. Note that $\Delta E^3=\mu^3 v_{\text{rel}}^6(1+\zeta_b^2)^3/2^3$ for the 1S state while $\Delta E^3=\mu^3 v_{\text{rel}}^6(1+(\zeta_b/2)^2)^3/2^3$ for the 2S and 2P states.} 
\begin{align}
    \Delta E^3 \big| \langle \psi_{20} | {\bs r} | \Psi_{{\bs p}_\ma{rel}} \rangle \big|^2 &= \frac{2^{-1}\pi^2 \alpha_b^{\text{eff}}}{\mu^2} \frac{\zeta_s \zeta_b^4(1+\zeta_s^2)}{[1+(\zeta_b/2)^2]^5(1-e^{-2\pi \zeta_s})} e^{-4 \zeta_s \text{arccot}(\zeta_b/2)} \nonumber \\
    &\times \left[\zeta_s \zeta_b - 4 (\zeta_b - \zeta_s)^2 +8 \left(1-\frac{\zeta_s}{2 \zeta_b}\right)\right]^2 \,,
\end{align}
and
\begin{align}
      \Delta E^3 \big| \langle \psi_{21} | {\bs r} | \Psi_{{\bs p}_\ma{rel}} \rangle \big|^2 &= \frac{2^{-3}\pi^2 \alpha_b^{\text{eff}}}{3 \mu^2} \frac{\zeta_s \zeta^{4}_b }{[1+(\zeta_b/2)^2]^5(1-e^{-2\pi \zeta_s})} e^{-4 \zeta_s \text{arccot}(\zeta_b/2)} \nonumber \\
    &\times \bigg[ { \frac{1}{3}} \left( -8 \zeta_s(\zeta_s^2-2) + 12 (2 \zeta_s^2-1)\zeta_b - 18 \zeta_s \zeta_b^2 + 3 \zeta_b^3 \right)^2 \nonumber \\
    &+{ \frac{32}{3}} \left(3\zeta_b-2\zeta_s\right)^2(\zeta_s^4+5 \zeta_s^2 + 4) \bigg] \,.\label{eq:2p}
\end{align}

\begin{figure}
    \centering
    \includegraphics[scale=0.6]{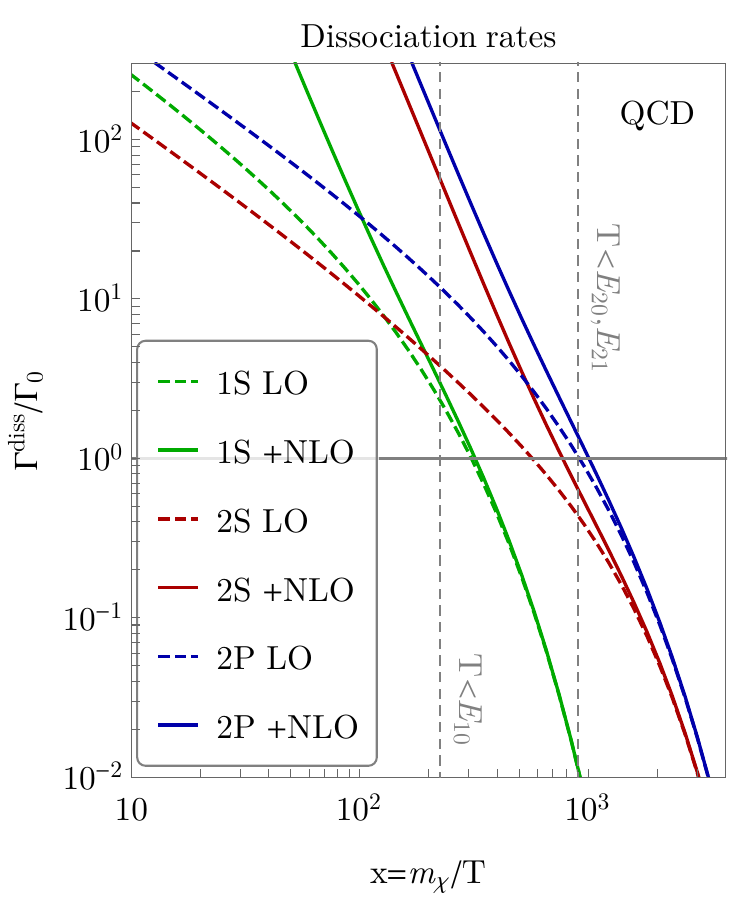}
    \includegraphics[scale=0.6]{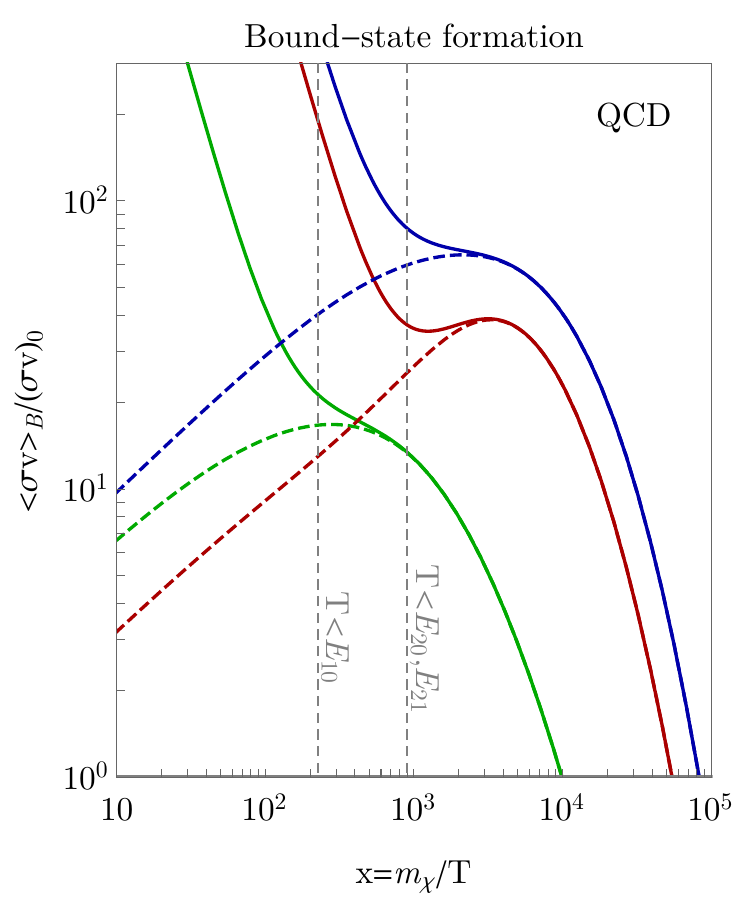}
        \includegraphics[scale=0.6]{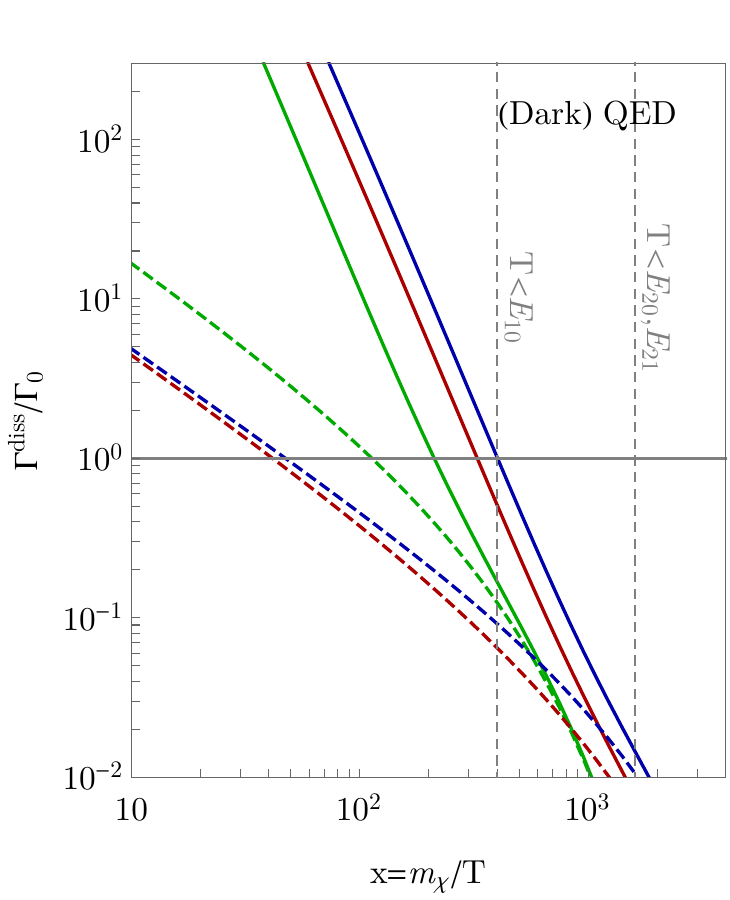}
    \includegraphics[scale=0.6]{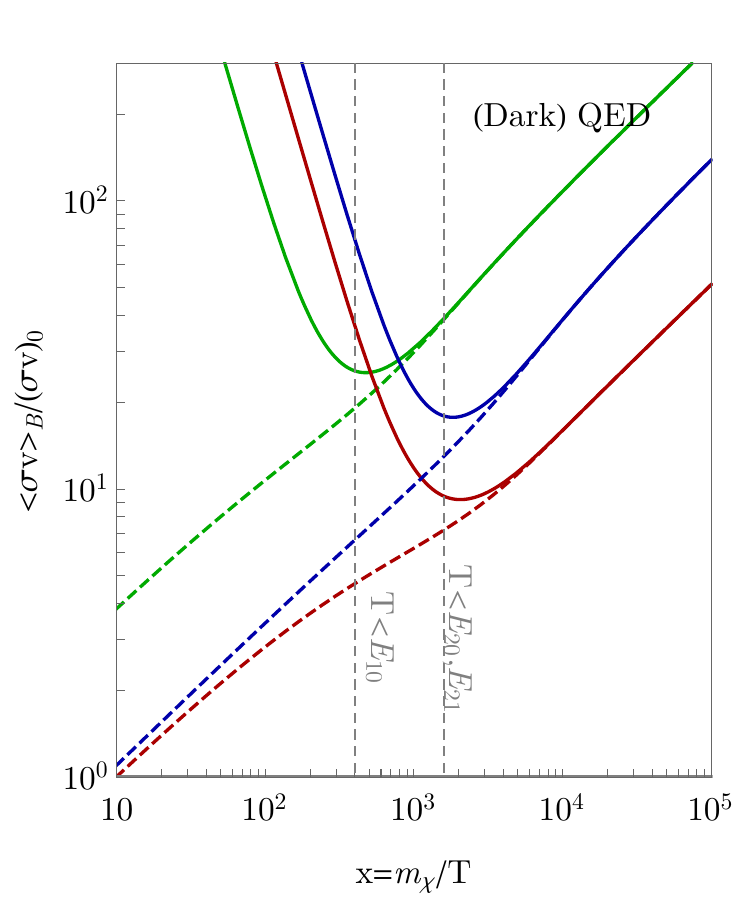}
    \caption{Bound state dissociation rates and thermally averaged recombination cross sections for two cases: (1) QCD with $\alpha=0.1$, $N_c =3$, $n_f C({\bs N})= 3$ (upper row) and (2) QED with $n_f C({\bs N})= 3$ (lower row). Dissociation rates are normalized by  $\Gamma_0=\alpha^5m_{\chi}/2$, while BSF cross sections are normalized by $(\sigma v)_0= \pi \alpha^2/m_{\chi}^2$.
    Since the potential for $\mathcal{S}(\chi \bar{\chi})_{\rm{adj}}$ is repulsive, the BSF cross section for QCD drops at a lower temperature.}
    \label{fig:flippingrates}
\end{figure}

To compare with previous work in the DM field, we consider the LO result of the ground state formation cross section~(\ref{eq:BSFcrossx}):
\begin{align}
(\sigma v_{\text{rel}})^{\text{LO}}_{1S}&= \frac{\pi \alpha \alpha^{\text{eff}}_b}{\mu^2} \frac{C_2({\bs R})s_{\chi} s_{\bar{\chi}}}{3 g_{\chi} g_{\bar{\chi}}} 2^9 \left(1- \frac{\alpha_s^{\text{eff}}}{2 \alpha_b^{\text{eff}}} \right)^2 S(\zeta_s,\zeta_b)(1+n_B(\Delta E))\,,\label{eq:eqLO1s} \\
S(\zeta_s,\zeta_b)&= \frac{2 \pi \zeta_s}{1-e^{-2\pi \zeta_s}} \frac{ \zeta_b^4(1+\zeta_s^2)}{(1+\zeta_b^2)^3} e^{-4 \zeta_s \text{arccot}(\zeta_b)} \,.
\end{align}
In particular, we consider charged scalars in a representation ${\bs R}$ and its conjugate $\bs{R'}=\bs{\overline{R}}$ of SU($N_c$), with $g_{\chi} =g_{\bar{\chi}}= d_{\bs R}$ and  $s_{\chi}=s_{\bar{\chi}}= 1$. 
For $\mathcal{S}(\chi \bar{\chi})_{\rm{adj}} \rightleftharpoons  \mathcal{B}(\chi \bar{\chi})_{\rm{1}}$ transitions, we have either $\bs{\hat{R}}={\bs 1}$ or $\bs{\hat{R}}=\bs{adj}$. For the adjoint-to-singlet transition $\mathcal{S}(\chi \bar{\chi})_{\rm{adj}} \rightarrow  \mathcal{B}(\chi \bar{\chi})_{\rm{1}}$, we have $\alpha_s^{\text{eff}} = \alpha [2 C_2(\bs{R})- C_2(\bs{adj}) ]/2$ and $\alpha_b^{\text{eff}} = \alpha [2 C_2(\bs{R})]/2$. Inserting these into Eq.~(\ref{eq:eqLO1s}) reproduces the first case in Eq.~(2.42a) of Ref.~\cite{Harz:2018csl} derived from the Bethe-Salpeter equation. For the singlet-to-adjoint transitions $ \mathcal{S}(\chi \bar{\chi})_{\rm{1}} \rightarrow \mathcal{B}(\chi \bar{\chi})_{\rm{adj}}$ where: $\alpha_s^{\text{eff}} = \alpha [2 C_2(\bs{R}) ]/2$ and $\alpha_s^{\text{eff}} = \alpha [2 C_2(\bs{R})- C_2(\bs{adj})]/2$, inserting these into Eq.~(\ref{eq:eqLO1s}) reproduces the second case in Eq.~(2.42a) of Ref.~\cite{Harz:2018csl}.

On the quarkonium side, we have fermions in the fundamental and antifundamental representations of SU($3$) with $g_{\chi} =g_{\bar{\chi}}= 3 s_{\chi} $ and $s_{\chi}=s_{\bar{\chi}}= 2$. Here, for octet-to-singlet transitions $\mathcal{S}(\chi \bar{\chi})_{\rm{8}} \rightarrow  \mathcal{B}(\chi \bar{\chi})_{\rm{1}}$: $\alpha_s^{\text{eff}} = \alpha [2 C_2(\bs{3}) - C_2(\bs{8}) ]/2$ and $\alpha_b^{\text{eff}} = \alpha C_2(\bs{3}) $ with $C_2(\bs{3})=4/3$ and $C_2(\bs{8})=3$. This reproduces the traditional leading order result that has been obtained from potential nonrelativistic QCD (pNRQCD) in Ref.~\cite{Yao:2017fuc,Yao:2018nmy} for quarkonium recombination. The LO result of quarkonium dissociation has been worked out from pNRQCD in Ref.~\cite{Brambilla:2011sg}.

To obtain the NLO rates, one can simply plug Eq.~(\ref{eqn:nlo_G}) and the results of the overlap integrals into Eqs.~(\ref{eqn:disso},~\ref{eqn:reco2}). For the bound state formation cross section (which is not yet thermally averaged), we have
\begin{align}
&(\sigma v_{\text{rel}})_\mathcal{B}^{\text{LO}+\text{NLO}}=(\sigma v_{\text{rel}})^{\text{LO}}_\mathcal{B}\times \\ &\left[1+ \alpha N_c R_g^{T=0}(\mu/\Delta E)+ \alpha N_c R_g^{T\neq 0}(\Delta E/T) + \alpha N_f R_f^{T=0}(\mu/\Delta E) + \alpha N_f R_f^{T\neq 0}(\Delta E/T)  \right] \,.\nonumber
\end{align}
Using this, and an analogous expression for the dissociation rate, we calculate relevant rates and show in  Fig.~\ref{fig:flippingrates} our finite temperature LO and NLO results for bound state formation and dissociation. We consider two cases: (1) QCD with $\alpha=0.1$, $N_c =3$, $n_f C({\bs N})= 3$ and (2) QED with $n_f C({\bs N})= 3$. One can recognize that inside a thermal plasma, the NLO effects flip the hierarchy among the rates of different bound states with different quantum numbers. This is because excited states have smaller $\Delta E/T$ compared to the ground state, such that the enhancement from NLO effects impacts excited states more than the ground state. This is a first indication that excited states for Dark Matter could matter in the relic abundance computation. This deserves a dedicated study. It is also interesting to note that transitions among bound states could have even smaller values of $\Delta E/T$, implying that these rates are enhanced as well, and that our NLO results can also be used to study, e.g., the feed down of excited states into the ground state, which eventually decays (annihilates) fastest.

\subsection{Comparison to effective treatment}
\label{sect:highT}

\begin{figure}
    \centering
    \includegraphics[scale=0.6]{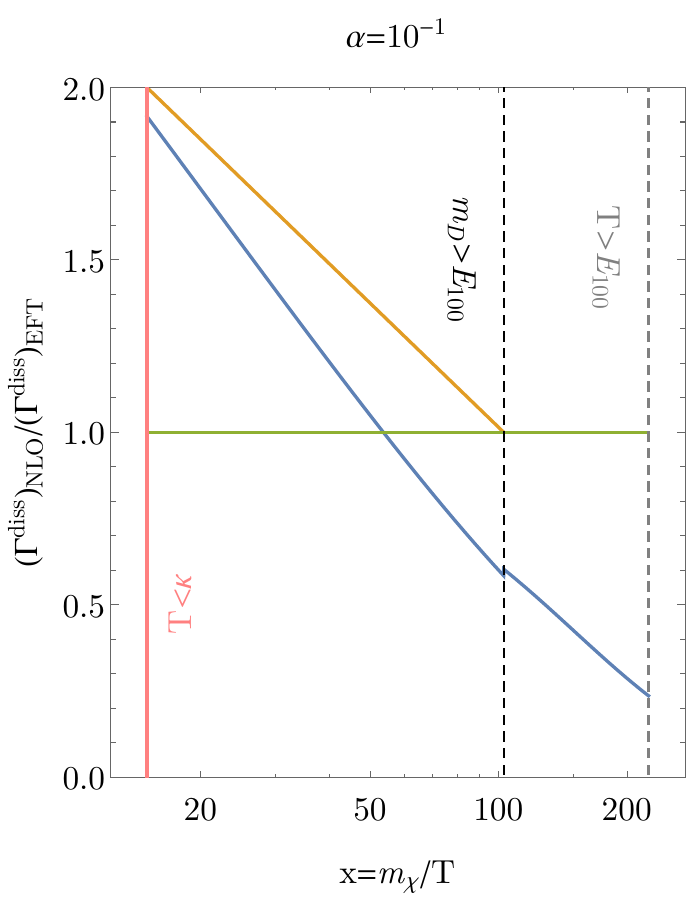}
    \includegraphics[scale=0.6]{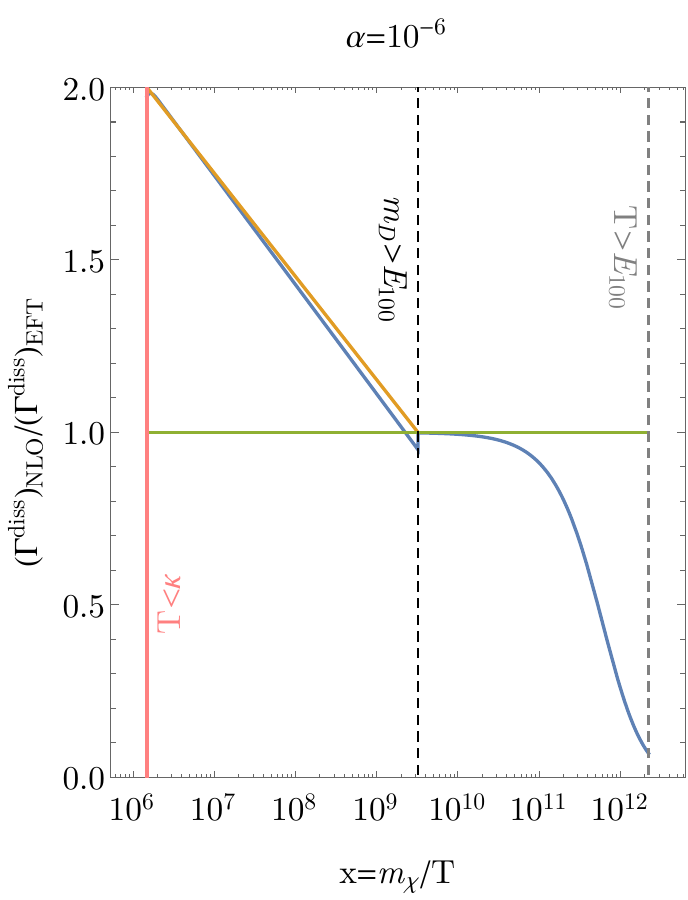}
    \caption{Ratios (blue lines) of the 1S state dissociation rate in our calculation and that in Ref.~\cite{Brambilla:2013dpa} for two values of the coupling constant $\alpha$. Only the fermion loop contribution (diagram $(f)$) is included in the comparison, since the gauge sector of Ref.~\cite{Brambilla:2013dpa} is not gauge invariant. In the hierarchy $\kappa \gg T \gg E \gg m_D$ where HTL resummation is not important, the two calculations agree well when the scales are largely separated. In the hierarchy $\kappa \gg T  \gg m_D \gg E$ where HTL resummation is important, our fixed order result overestimates that in Ref.~\cite{Brambilla:2013dpa} by a logarithmic factor , $C(T)$, depicted by the yellow lines.}
    \label{fig:my_label2}
\end{figure}

We compare our NLO dissociation rate to effective results for the two hierarchies $\kappa \gg T \gg E \gg m_D$ and $\kappa \gg T  \gg m_D \gg E$, where { $\kappa$} is the Bohr momentum { satisfying $\kappa \sim Mv$,} and $m_D$ is the Debye screening mass. These two hierarchies have been studied in Ref.~\cite{Brambilla:2013dpa} (see also references therein) by using pNRQCD and expanding the relevant in-medium gauge boson propagators under the aforementioned hierarchies of scales. The ratio of the 1S state dissociation rate calculated in our framework to the effective treatment in Ref.~\cite{Brambilla:2013dpa} is shown as blue lines in Fig.~\ref{fig:my_label2} for the fermion loop diagram $(f)$. Only the fermion loop contribution is included in the comparison, since the gauge sector of Ref.~\cite{Brambilla:2013dpa} is not gauge invariant. We consider cases where the scales entering the two hierarchies are rather well separated ($\alpha=10^{-6}$, right panel) and less separated ($\alpha=10^{-1}$, left panel).

For $\kappa \gg T \gg E \gg m_D$,  it can be seen by comparing the two ratios with different $\alpha$'s that as the coupling $\alpha$ becomes smaller, the scales  are more separated, and consequently the effective treatment agrees well with our fixed order calculation.
Technically, the agreement we see here is a result of $E \gg m_D$, where the HTL approximation of the NLO electric field correlator terms \emph{can} be resummed (but not necessarily needed). In this sense, both our fixed NLO computation and the HTL resummed results in the effective treatment are valid for this regime.
That the electric field correlator does not require resummation even up to $E\gtrsim m_D$ was mentioned in Ref.~\cite{Burnier:2010rp}. Notice that our result is also valid for $E\gtrsim T$.

For $\kappa \gg T  \gg m_D \gg E$, HTL resummation is needed and our fixed order calculation does not apply in this case. Nevertheless, we empirically find, if scales are well separated, that our unresummed NLO result in this regime overestimates the dissociation rate compared to the effective treatment only by a logarithmic term
\begin{align}
C(T) \equiv \frac{\ln(\kappa T/ T_{\star}^2)}{\ln(\kappa/T_{\star})} \,,
\end{align}
where $T_{\star}$ is the temperature which solves $m_D = E_{100}$. The logarithmic overestimation can be seen in Fig.~\ref{fig:my_label2} by comparing the blue lines with the yellow lines, in which the yellow lines depict the factor $C(T)$. A logarithmic error for $m_D \gg E$ was also mentioned in Ref.~\cite{Burnier:2010rp}.

The insights that we obtained from these two comparisons are useful for DM relic abundance calculations with metastable bound states. In the thermal production case we generically expect that ionization equilibrium is maintained until a temperature where $m_D\sim E$. In ionization equilibrium, the collision term is independent of the actual value of the dissociation and recombination rates. Therefore the logarithmic enhancement may not play a significant role in such a situation, and our fixed order calculation is sufficient for describing the decoupling from ionization equilibrium where the actual size matters. This is in particular expected for the most often studied case of capture and dissociation of the ground state. We caution however that $E\gtrsim m_D$ may be satisfied for the ground state for a particular temperature while not for some of the (higher) excited states with much lower $E$. Thus, the less explored case with many excited states may require resummation at relevant times.

\section{Conclusions}
\label{sec:conclusion}
In this paper, we studied the electric field correlator for medium-induced bound state formation and dissociation in a non-Abelian plasma. The non-Abelian electric field correlator is constructed in a gauge invariant way by appropriately taking into account the Wilson lines. In the hierarchy $Mv\gg T$, it encodes all essential information of the non-Alelian plasma that determines the heavy particle bound state dissociation and formation rates that appear in the Boltzmann and rate equations for quarkonium transport in the QGP and DM bound state formation in the early universe.

We carried out NLO calculations in $R_\xi$ gauge and showed that all the $\xi$ dependence cancels when summing over all contributing diagrams. The finite temperature part of our fixed order calculation is valid for $E\gtrsim m_D$, while resummation is needed otherwise. We also systematically investigated the UV and IR behaviors of the electric field correlator. On the UV side, we found that the renormalization group running of the electric field correlator is fully determined by the running of the coupling constant, which is similar to the case of the correlator defining the heavy quark diffusion coefficient (see e.g.~\cite{CaronHuot:2009uh,Burnier:2010rp,Laine:2021uzs}). More specifically, we found that the multiplication of the electric field correlator and the coupling constant $g^2$ is scale independent. On the IR side, we explicitly demonstrated that all the collinear and infrared divergences cancel. Using the NLO result of the electric field correlator, we calculated bound state dissociation rates and formation cross sections for the ground and the first two excited states in a Coulomb potential. Finally, we discussed the applicability range of our calculations.

 The non-Abelian electric field correlator studied here is for bound state formation and dissociation and different from the one for the heavy quark diffusion coefficient~\cite{Casalderrey-Solana:2006fio,Eller:2019spw,CaronHuot:2007gq,Burnier:2010rp}, though coincidentally the finite temperature parts agree at NLO (the vacuum parts differ). Our NLO perturbative calculation is the first systematic and rigorous study of this non-Abelian electric field correlator at finite frequency. Since the correlator is defined at finite frequency, it may be difficult to calculate it on a Euclidean lattice. But one may be able to calculate it by using simulations of classical Yang-Mills theories~\cite{Laine:2009dd,Boguslavski:2020tqz,Boguslavski:2021baf}. Furthermore, in the strong coupling limit, the AdS/CFT correspondence (see Ref.~\cite{Casalderrey-Solana:2011dxg} for a review) may be useful to calculate this correlator. It can also be extracted from experimental data on quarkonium suppression in heavy ion collisions, by using Bayesian analysis~\cite{Bernhard:2019bmu,Nijs:2020roc}. Anisotropic effects on the electric field correlator can also be studied in a similar way to Refs.~\cite{Dumitru:2007hy,Dumitru:2009ni,Dumitru:2009fy}. Our NLO calculation serves as a baseline to compare with in future studies, which can tell us more information about the properties of the QGP. 

The framework we used to derive the Boltzmann and rate equations is based on (1) vacuum Schr\"odinger wave functions and (2) factorization of the subsystem and environmental density matrices, both of which are valid in the hierarchy $Mv \gg T$. In this hierarchy, thermal effects on the heavy particle potentials are subleading and the interaction between the bound state and the environment is suppressed by the multipole expansion. Beyond this hierarchy, i.e., in the high temperature regime $ T \gtrsim Mv$, one may need to resort to a different framework, especially for highly excited states. Finally, for $m_D\gg Mv^2$, the HTL resummation will become crucial in the calculations of bound state dissociation rates and formation cross sections, although the physical impacts for thermally produced DM may be small.

\acknowledgments

We would like to thank Feng Luo and other participants of the online workshop `Quarkonia meet Dark Matter' for discussion.
BSH and XY are supported by the U.S. Department of Energy, Office of Science, Office of Nuclear Physics under grant Contract Numbers DE-SC0011090. TB was supported by World Premier International Research Center Initiative (WPI), MEXT, Japan, by the JSPS Core-to-Core Program Grant Number JPJSCCA20200002 and by JSPS KAKENHI Grant Number 20H01895.
KM was supported by MEXT Leading Initiative for Excellent Young Researchers Grant Number JPMXS0320200430.

{ During the final completion of this work, we became aware of Ref.~\cite{Garny:2021qsr} that studied similar dipole transitions.
This allowed us to cross-check and correct two prefactors in the 2P transition probability in Eq.~(\ref{eq:2p}).}

\appendix
\section{KMS relation for electric field correlator}
\label{app:kms}
We assume the environment is invariant under spacetime translation. Then we can write the greater (``$>$") electric field correlation as
\be
[g_E^{++}]^{>}(t,{\bs x}) = \frac{1}{Z}\Tr_E \Big( \big[{E}(t,{\bs x}) \ml{W}_{[( t, {\bs x}), (+\infty, {\bs \infty})]} \big]
\big[ \ml{W}_{[(+\infty, {\bs \infty}),(0, {\bs 0})]} {E}(0,{\bs 0}) \big] e^{-\beta H_E}\Big)\,,\nn\\
\ee
where we omitted the color and spatial indexes for simplicity. Here $Z=\Tr_E e^{-\beta H_E}$, $H_E$ is the environment Hamiltonian and $\ml{W}_{[( t, {\bs x}), (+\infty, {\bs \infty})]}$ denotes the two segments of the Wilson lines from the spacetime point $( t, {\bs x})$ to $(+\infty, {\bs \infty})$, one along the time axis and the other along the spatial direction. Using $\phi(t) = e^{iHt} \phi(t=0) e^{-iHt}$, we find
\be
&&[g_E^{++}]^{>}(t,{\bs x}) \nn\\
&=& \frac{1}{Z}\Tr_E \Big( \big[{E}(t,{\bs x}) \ml{W}_{[( t, {\bs x}), (+\infty, {\bs \infty})]} \big] e^{-\beta H_E}
\big[ \ml{W}_{[(+\infty-i\beta, {\bs \infty}),(-i\beta, {\bs 0})]} {E}(-i\beta,{\bs 0}) \big] \Big)\nn\\
&=& \frac{1}{Z}\Tr_E \Big( \big[ \ml{W}_{[(+\infty-i\beta, {\bs \infty}),(-i\beta, {\bs 0})]} {E}(-i\beta,{\bs 0}) \big] \big[{E}(t,{\bs x}) \ml{W}_{[( t, {\bs x}), (+\infty, {\bs \infty})]} \big] e^{-\beta H_E}\Big) \nn\\
&=& [g_E^{++}]^{<}(t+i\beta,{\bs x}) \label{eq:g++KMS} \,.
\ee
Similarly, one can show
\be
[g_E^{--}]^{>}(t,{\bs x}) = [g_E^{--}]^{<}(t+i\beta,{\bs x}) \,.
\ee

There is also a relation between $[g_E^{++}]^{>}$ and $[g_E^{--}]^{<}$. To see this, we need to apply the parity and time reversal transformations. The parity transformation $\ml{P}$ acting on the gauge field is given by
\begin{align}
\ml{P} A^\mu(t, {\bs x}) \ml{P}^{-1} &= A_\mu(t, -{\bs x}) \\
\ml{P} F^{\mu\nu}(t, {\bs x}) \ml{P}^{-1} &= F_{\mu\nu}(t, -{\bs x}) \,,
\end{align}
while the time reversal transformation is given by
\begin{align}
\ml{T} A^\mu(t, {\bs x}) \ml{T}^{-1} &= A_\mu(-t, {\bs x}) \\
\ml{T} F^{\mu\nu}(t, {\bs x}) \ml{T}^{-1} &= - F_{\mu\nu}(-t, {\bs x}) \,.
\end{align}
Under the parity and time reversal transformations, the Wilson line
\begin{align}
\ml{W}_{[a,b]} = {\rm P} \exp\Big( ig \int_b^a \diff x^\mu A_\mu(t, {\bs x}) \Big) \,,
\end{align}
where ${\rm P}$ is path ordering, changes according to
\begin{align}
\ml{P}\ml{T}  \ml{W}_{[a,b]}  \ml{T}^{-1} \ml{P}^{-1} &= \overline{{\rm P}} \exp\Big( -ig \int_{b}^{a} \diff x^\mu A_\mu(-t, -{\bs x}) \Big) =
\ml{W}_{[-b,-a]} \,.
\end{align}
Also, the time reversal operator is anti-unitary, which means under the time reversal operation
\be
\langle n | O(t) | m \rangle = \langle n | \ml{T}^{-1} \ml{T} O(t)   | m \rangle =  \langle  O(-t) m | n\rangle =  \langle m |  O^\dagger(-t)  | n \rangle \,,
\ee
where $O$ is an arbitrary operator.
This implies under a time reversal operation
\be
&&\Tr(O_1(t_1) O_2(t_2) e^{-\beta H}) = \sum_{n,m} e^{-\beta E_n} \langle n| O_1(t_1) |m\rangle \langle m | O_2(t_2) |n\rangle \nn \\
&\xrightarrow{\ml{T}} & \sum_{n,m} e^{-\beta E_n} \langle m | O_1^\dagger (-t_1)| n \rangle \langle n | O_2^\dagger(-t_2) | m \rangle = \Tr(O_2^\dagger(-t_2) O_1^\dagger(-t_1) e^{-\beta H}) \,.
\ee
Applying the parity and time reversal transformations leads to
\be
[g_E^{++}]^>(t, {\bs x})
&=& \Tr_E \Big( E(t,{\bs x}) \ml{W}_{[(t,{\bs x}),(+\infty,{\bs \infty})]} 
\ml{W}_{[(+\infty, {\bs \infty}),(0, {\bs 0})]} E(0, {\bs 0})
\frac{e^{-\beta H_E}}{Z}\Big) \nn \\
&\xrightarrow{\ml{P}\ml{T}} & \Tr_E\Big(  E(0,{\bs 0})
\ml{W}_{[(0,{\bs 0}), (-\infty, {\bs \infty})]}
\ml{W}_{[(-\infty, {\bs \infty}),(-t, -{\bs x})]} E(-t, -{\bs x})
\frac{e^{-\beta H_E}}{Z}
\Big) \nn \\
&=& [g_E^{--}]^<(-t, -{\bs x}) \,.
\ee

\section{Feynman rules} \label{app:FeynmanRules}

Two-point functions on the Schwinger-Keldysh contour are defined as:
\begin{align}
S(x-y)&= \langle \ml{T}_C \psi(x) \bar{\psi}(y) \rangle,\\
C(x-y)&= \langle \ml{T}_C c(x)\bar{c}(y) \rangle,\\
D_{\mu \nu}(x-y)&= \langle \ml{T}_C A_{\mu}(x)A_{\nu}(y) \rangle,
\end{align}
where $\psi$ is a fermion, $c$ an anti-commuting scalar (ghost) and $A$ a gauge boson field. $\ml{T}_C$ is the time-ordering operator along the contour, placing fields inserted at later sections of the contour (closer to $t=-i\beta$) to the left of the field insertions closer to $t=0$ in the correlation function. For an arbitrary covariant gauge, the propagators are given by
\begin{align}
D^{Y,ab}_{\mu \nu}(k) = \delta^{ab} P_{\mu \nu}(k) D^Y(k),
\end{align}
where $Y$ can be any of $>,<,\ml{T},\overline{\ml{T}}$, and
\begin{align}
P_{\mu \nu }(k) = - \left[ g_{\mu \nu} - (1 - \xi) \frac{k_\mu k_\nu}{k^2} \right].
\end{align}
with metric signature $(+,-,-,-)$. The free propagators in Fourier space are given by
\begin{align}
D^>(k) &= \left( \Theta(k_0) + n_B(|k_0|) \right) 2\pi \delta(k^2)\,, \quad\quad\ \,  D^<(k) = \left( \Theta(-k_0) + n_B(|k_0|) \right) 2\pi \delta(k^2) \nonumber \\
D^\ml{T}(k) &= \frac{i}{k^2 + i0^+} + n_B(|k_0|) 2\pi \delta(k^2) \,, \quad\quad D^{\overline{\ml{T}}}(k) = \frac{-i}{k^2 - i0^+} + n_B(|k_0|) 2\pi \delta(k^2) \nonumber \\ D^S(k) &= D^>(k) + D^<(k) = (1 + 2n_B(|k_0|)) 2 \pi \delta(k^2) \,.
\end{align}
We also need to state what are the Fermionic propagators $\mathbb{S}_{IJ}$. 
We will only explicitly use the Wightman functions $\mathbb{S}_{21}$ and $\mathbb{S}_{12}$, which are given by
\begin{align}
\mathbb{S}_{12} = S^{<}(k)&= -\slashed{k} (2\pi) \delta(k^2)\left[-\Theta(-k_0) + n_{F}(|k_0|) \right],\\
\mathbb{S}_{21} = S^{>}(k)&= -\slashed{k} (2\pi) \delta(k^2)\left[-\Theta(+k_0) + n_{F}(|k_0|) \right],
\end{align}
where $n_F(k_0) = (e^{k_0/T} + 1)^{-1}$.

\begin{figure}
	\centering
	\begin{tabular}{  c  c  l  }
	\raisebox{-0.8in}{\includegraphics[height=1.6in]{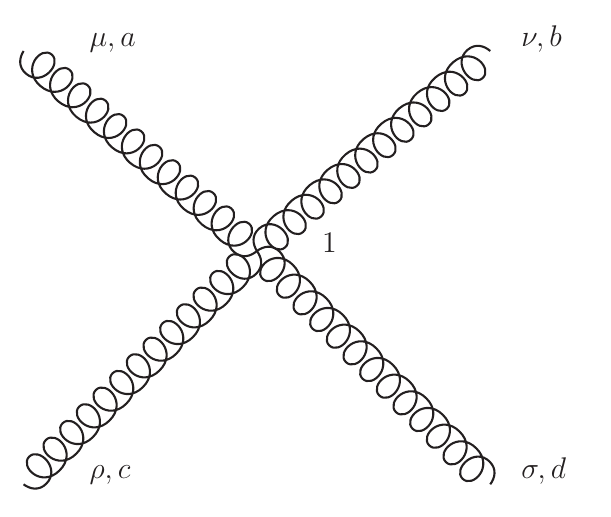}}  & & $ \begin{matrix} = & -ig^2 \big[ f^{abe} f^{cde} (g^{\mu \rho} g^{\nu \sigma} - g^{\mu \sigma} g^{\nu \rho} ) \\ &  + f^{ace} f^{bde} (g^{\mu \nu} g^{\rho \sigma} - g^{\mu \sigma} g^{\nu \rho}) \\ & + f^{ade} f^{bce}(g^{\mu \nu} g^{\rho \sigma} - g^{\mu \rho} g^{\nu \sigma}) \big] \end{matrix} $  \\
	\raisebox{-0.8in}{\includegraphics[height=1.6in]{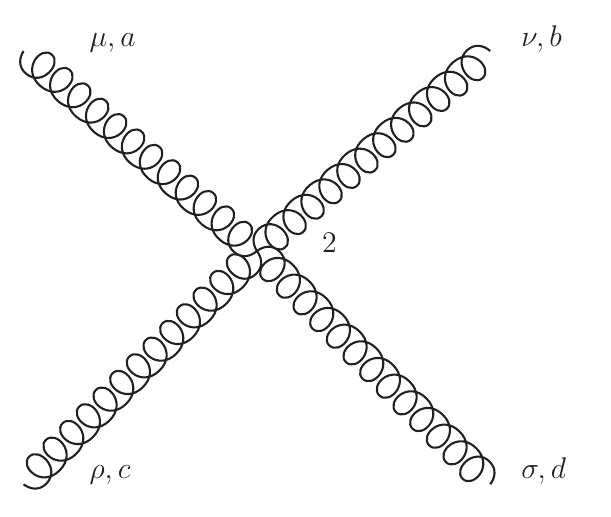}}  & & $ \begin{matrix} = & ig^2 \big[ f^{abe} f^{cde} (g^{\mu \rho} g^{\nu \sigma} - g^{\mu \sigma} g^{\nu \rho} ) \\ &  + f^{ace} f^{bde} (g^{\mu \nu} g^{\rho \sigma} - g^{\mu \sigma} g^{\nu \rho}) \\ & + f^{ade} f^{bce}(g^{\mu \nu} g^{\rho \sigma} - g^{\mu \rho} g^{\nu \sigma}) \big] \end{matrix} $
	\end{tabular}
\caption{Feynman rules associated to the 4-gauge boson vertex, given here for the time-ordered and anti-time ordered branches of the Schwinger-Keldysh contour. The anti-time ordered branch multiplies the vertex factors by $(-1)$, which is naturally included in the notation we adopt in the main text.}
\label{fig:rules-4g-appendix}
\end{figure}

\begin{figure}
	\centering
	\begin{tabular}{  c  c  l  }
	\raisebox{-0.8in}{\includegraphics[height=1.6in]{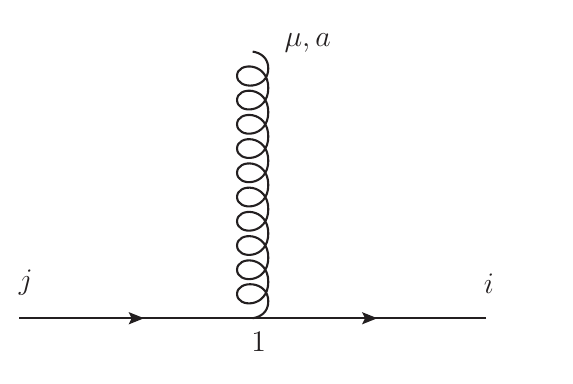}}  &=& $ i g \gamma^\mu \left[T_{\bs{R}}^a\right]_{ij}$  \\
	\raisebox{-0.8in}{\includegraphics[height=1.6in]{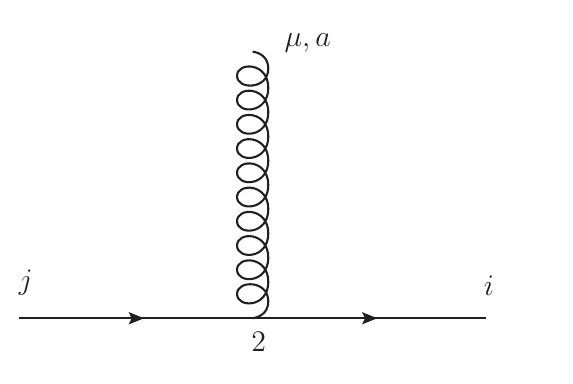}}  &=& $ -i g \gamma^\mu \left[T_{\bs{R}}^a\right]_{ij} $
	\end{tabular}
\caption{Feynman rules associated to the gauge boson-fermion-fermion vertex, given here for the time-ordered and anti-time ordered branches of the Schwinger-Keldysh contour, with the fermions in a representation $\bs{R}$. The anti-time ordered branch multiplies the vertex factors by $(-1)$, which is naturally included in the notation we adopt in the main text.}
\label{fig:rules-g-fermion-appendix}
\end{figure}

\begin{figure}
	\centering
	\begin{tabular}{  c  c  l  }
	\raisebox{-0.8in}{\includegraphics[height=1.6in]{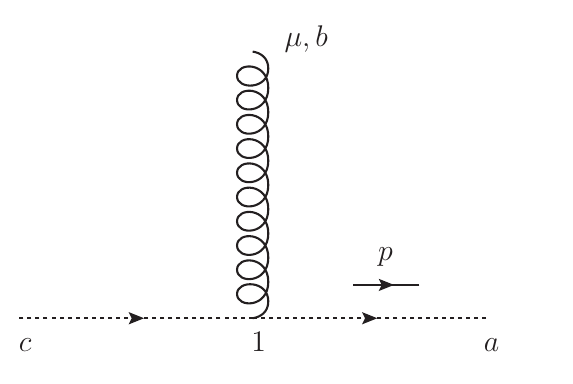}}  &=& $ -g f^{abc} p^\mu $ \\
	\raisebox{-0.8in}{\includegraphics[height=1.6in]{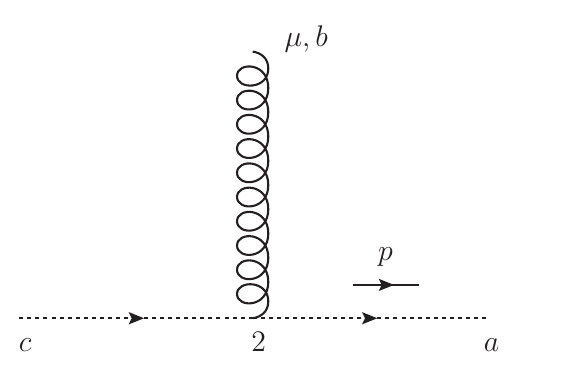}}  &=& $ +g f^{abc} p^\mu $
	\end{tabular}
\caption{Feynman rules associated to the gauge boson-ghost-ghost vertex, given here for the time-ordered and anti-time ordered branches of the Schwinger-Keldysh contour. The anti-time ordered branch multiplies the vertex factors by $(-1)$, which is naturally included in the notation we adopt in the main text.}
\label{fig:rules-g-ghost-appendix}
\end{figure}

For completeness, we also list the anti-commuting scalar field Wightman functions (see e.g., \cite{Hata:1980yr}), 
\begin{align}
C^{<}(k)&= (2\pi) \delta(k^2)\left[\Theta(-k_0) + n_{B}(|k_0|) \right],\\
C^{>}(k)&= (2\pi) \delta(k^2)\left[\Theta(+k_0) + n_{B}(|k_0|) \right].
\end{align}
Finally, we list in Figures~\ref{fig:rules-4g-appendix},~\ref{fig:rules-g-fermion-appendix}, and~\ref{fig:rules-g-ghost-appendix} the remaining Feynman rules that are relevant for the calculations shown in the main text.


\section{Proof of gauge-dependence cancellation at \texorpdfstring{$O(1-\xi)$}{O(1-x)}} \label{app:gauge-invariance}

We start from the diagram $(1)$. Evaluating it, and keeping the linear terms on $(1-\xi)$ only, we can work through the index contractions to find
\begin{align}
&(1)_{(1-\xi)} \nn\\
&= -\frac{i}{2} g^2 N_c \delta^{ad} \D(p)_{I'I} \int_k \D(k)_{J'I'} \D(p-k)_{J'I'} (-1)^{I'+J'} \D(p)_{JJ'} P_{\rho \rho'}(-p) (-i p_0 g_j^{\rho} + i p_j g_{0}^{\rho}) \nonumber \\
& \quad \times (-2)\frac{(1-\xi)}{k^2} \bigg[ (p_0 g_i^{\rho'} - p_i g_0^{\rho'}) (- (p^2)^2 + 2p^2 (p-k)^2 - ((p-k)^2)^2 ) \nonumber \\ & \quad\quad\quad\quad\quad\quad\quad\quad + (p_0 k_i - p_i k_0) \left[  k^{\rho'} ( (p-k)^2  -2 p^2 ) - p^{\rho'} (k^2 - k \cdot p) \right] \bigg].
\end{align}
Because there is a chromoelectric field being contracted through the $\rho,\rho'$ indices, the contribution proportional to $p^{\rho'}$ vanishes, and we can therefore drop it, obtaining
\begin{align}
&(1)_{(1-\xi)} \nn\\
&= -\frac{i}{2} g^2 N_c \delta^{ad} \D(p)_{I'I} \int_k \D(k)_{J'I'} \D(p-k)_{J'I'} (-1)^{I'+J'} \D(p)_{JJ'} P_{\rho \rho'}(-p) (-i p_0 g_j^{\rho} + i p_j g_{0}^{\rho}) \nonumber \\
& \quad \times (-2)\frac{(1-\xi)}{k^2} \bigg[ (p_0 g_i^{\rho'} - p_i g_0^{\rho'}) (- (p^2)^2 + 2p^2 (p-k)^2 - (p-k)^2 (p^2 - 2k \cdot p + k^2) ) \nonumber \\ & \quad\quad\quad\quad\quad\quad\quad\quad + (p_0 k_i - p_i k_0) \left[  k^{\rho'} ( (p-k)^2  -2 p^2 ) \right] \bigg].
\end{align}
Now all terms in the integrand are explicitly proportional to a propagator momentum squared. The ones proportional to $p^2$ will cancel one of the $\D(p)$ propagators and allow this diagram to be put on equal footing as diagrams with only one $\D(p)$ propagator. However, the ones that have no factor of $p^2$ keep their two ``external'' propagators, and therefore any gauge-dependent contribution must be cancelled by a diagram with the same structure. The only other diagram with two $\D(p)$ propagators is (2), the 4-point vertex tadpole.

So we proceed to evaluate the tadpole diagram $(2)$, where if we only keep the $(1-\xi)$-dependent terms, we have
\begin{align}
    (2)_{(1-\xi)} &= (1-\xi)  \D(p)_{I'I}  \D(p)_{JI'} P_{\rho \rho'}(-p) (-i p_0 g_j^{\rho} + i p_j g_{0}^{\rho}) (-ig^2) N_c  \delta^{ad} \nonumber \\
    & \quad \times \int_k (-1)^{I'+1} \D(k)_{I'I'} (-i p_0 g_{i\mu} + ip_i g_{0\mu})\left[ g^{\mu \rho'} - \frac{k^\mu k^{\rho'}}{k^2} \right] \nonumber \\
    &= (1-\xi) g^2 N_c \delta^{ad} \D(p)_{I'I}  \D(p)_{JI'} P_{\rho \rho'}(-p) (-i p_0 g_j^{\rho} + i p_j g_{0}^{\rho}) (-1)^{I'+1} \nonumber \\
    & \quad \times \int_k \D(k)_{I'I'} \left[ (p_0 k_i - p_i k_0) \frac{k^{\rho'}}{k^2} - ( p_0 g_{i}^{\rho'} - p_i g_{0}^{\rho'}) \right] \,. \label{4-pt-gl-bubble}
\end{align}
Comparing this to the contributions from (1) that are not proportional to $p^2$, we find
\begin{align}
(1)_{(1-\xi), \, \not\propto p^2 } &= -\frac{i}{2} g^2 N_c \delta^{ad} \D(p)_{II'} \D(p)_{JJ'} P_{\rho \rho'}(-p) (-i p_0 g_j^{\rho} + i p_j g_{0}^{\rho}) \nonumber \\
& \quad  \int_k \D(k)_{J'I'} \D(p-k)_{J'I'} (-1)^{I'+J'} (-2)\frac{(1-\xi)}{k^2} (p-k)^2  \nonumber \\ & \quad\quad\quad\quad\quad\quad\quad\quad \times  \left[  k^{\rho'} (p_0 k_i - p_i k_0) - (p_0 g_i^{\rho'} - p_i g_0^{\rho'}) (k^2 - 2k\cdot p) \right] \nonumber \\
&= - (1-\xi) g^2 N_c \delta^{ad} \D(p)_{I'I} \D(p)_{JJ'} P_{\rho \rho'}(-p) (-i p_0 g_j^{\rho} + i p_j g_{0}^{\rho}) (-1)^{J'+1}  \nonumber \\
& \quad  \int_k \frac{1}{k^2} \D(k)_{J'I'} \mathbbm{1}_{J'I'}  \left[  k^{\rho'} (p_0 k_i - p_i k_0) - (p_0 g_i^{\rho'} - p_i g_0^{\rho'}) (k^2 - 2k\cdot p) \right] \nonumber \\
&= - (1-\xi) g^2 N_c \delta^{ad} \D(p)_{I'I} \D(p)_{JI'} P_{\rho \rho'}(-p) (-i p_0 g_j^{\rho} + i p_j g_{0}^{\rho}) (-1)^{I'+1}  \nonumber \\
& \quad  \int_k  \D(k)_{I'I'} \left[   (p_0 k_i - p_i k_0) \frac{k^{\rho'}}{k^2} - (p_0 g_i^{\rho'} - p_i g_0^{\rho'}) \right] \,,\label{3-pt-gl-bubble}
\end{align}
where we have dropped the term proportional to $k \cdot p$ in the last equality because it integrates to zero by inversion symmetry $k \to -k$.  The cancellation of \eqref{4-pt-gl-bubble} with \eqref{3-pt-gl-bubble} is obvious at this point: one contribution is just the additive inverse of the other.

Now we have to deal with the remaining pieces, which are all proportional to $p^2$. Concretely, the pieces of interest are given by
\begin{align}
(1)_{(1-\xi), \, \propto p^2} &= i p^2 g^2 N_c \delta^{ad} \D(p)_{I'I} \int_k \D(k)_{J'I'} \D(p-k)_{J'I'}  \D(p)_{JJ'} P_{\rho \rho'}(-p) (-i p_0 g_j^{\rho} + i p_j g_{0}^{\rho}) \nonumber \\
& \quad \times (-1)^{I'+J'} \frac{(1-\xi)}{k^2} \bigg[ (p_0 g_i^{\rho'} - p_i g_0^{\rho'} ) (- p^2 +  (p-k)^2  ) -2 k^{\rho'} (p_0 k_i - p_i k_0) \bigg] \nonumber \\
&= -g^2 N_c \delta^{ad} \int_k \D(k)_{I'I} \D(p-k)_{I'I} (-1)^{I'+1} \D(p)_{JI'} P_{\rho \rho'}(-p) (-i p_0 g_j^{\rho} + i p_j g_{0}^{\rho}) \nonumber \\
& \quad \times \frac{(1-\xi)}{k^2} \bigg[ (p_0 g_i^{\rho'} - p_i g_0^{\rho'} ) (  (p-k)^2 -p^2 ) -2 k^{\rho'} (p_0 k_i - p_i k_0) \bigg] \,.
\end{align}
This now has the same structure as diagrams (5) and (6), so we must evaluate those to compare with this contribution. We start with the $\xi$-dependent part of (5):
\begin{align}
    &(5)_{(1-\xi)} \nn\\
    &= (1-\xi) g^2 N_c \delta^{ad} \D(p)_{JJ'} P_{\rho \rho'}(p) (-i p_0 g_j^{\rho} + i p_j g_{0}^{\rho}) \nonumber \\
    &\quad \times \int_k (-i(p-k)_0 g_{i \mu} + i(p-k)_i g_{0\mu} )\D(p-k)_{I'I} \frac{1}{-ik_0+\epsilon}  \frac{k_0 k_\nu}{k^2} \D(k)_{I'I}  \nonumber \\
    & \quad \times (-1)^{I'+1} \left[ g^{\rho' \mu} (k-2p)^{\nu} + g^{\mu \nu}(p- 2k)^{\rho'} + g^{\nu \rho'} (k+p)^{\mu} \right] \nonumber \\
    &= (1-\xi) g^2 N_c \delta^{ad} \D(p)_{JJ'} P_{\rho \rho'}(p) (-i p_0 g_j^{\rho} + i p_j g_{0}^{\rho}) (-1)^{I'+1} \nonumber \\
    &\quad \times \int_k \D(p-k)_{I'I}\D(k)_{I'I} \frac{1}{k^2} \left[ ((p-k)_0 g_i^{\rho'} - (p-k)_i g_0^{\rho'}) (k^2 - 2 k \cdot p) + (p_0k_i  - p_i k_0)p^{\rho'} \right] \nonumber \\
    &= g^2 N_c \delta^{ad} \int_k \D(p-k)_{I'I}\D(k)_{I'I} (-1)^{I'+1} \D(p)_{JJ'} P_{\rho \rho'}(p) (-i p_0 g_j^{\rho} + i p_j g_{0}^{\rho})  \nonumber \\
    &\quad \times  \frac{(1-\xi)}{k^2}  \left[(p-k)_0 g_i^{\rho'} - (p-k)_i g_0^{\rho'}\right] ( (p-k)^2 - p^2 ) \,,
\end{align}
where in the last line we have dropped the term proportional to $p^{\rho'}$ because it vanishes when contracted with the momentum structure coming from the chromoelectric field.

Similarly for the diagram $(6)$, we find the $\xi$-dependent part is given by
\begin{align}
    (6)_{(1-\xi)} &= - g^2 N_c \delta^{ad} \int_k \D(k)_{I'I} \D(p-k)_{I'I} (-1)^{I'+1}  \D(p)_{JI'} P_{\rho \rho'}(p)  (-i p_0 g_j^{\rho} + i p_j g_{0}^{\rho}) \nonumber \\
    & \quad \times \frac{(1-\xi)}{k^2} \left[ ( k_0 g_i^{\rho'} - k_i g_0^{\rho'}) (2k \cdot p - k^2) + k^{\rho'} (p_0 k_i -k_0 p_i) \right]
\end{align}

These are all the contributions with three propagators, so at the very least we expect a cancellation of the $k^{\rho'}$terms. Adding things up, we have (the prefactor $1/2$ is the symmetry factor).
\begin{align}
    &\frac{1}{2} (1)_{(1-\xi), \, \propto p^2} + \big((5)+(6)\big)_{(1-\xi)} \nn\\
    &= g^2 N_c \delta^{ad}  \D(p)_{JJ'} P_{\rho \rho'}(-p) (-i p_0 g_j^{\rho} + i p_j g_{0}^{\rho})  \nonumber \\
    & \quad \times \int_k \D(p-k)_{I'I}\D(k)_{I'I} (-1)^{I'+1} \nonumber \\
    & \quad \times \frac{(1-\xi)}{k^2} \bigg[ - \frac{1}{2} \left( (p_0 g_i^{\rho'} - p_i g_0^{\rho'} ) (  (p-k)^2 -p^2 ) -2 k^{\rho'} (p_0 k_i - p_i k_0) \right) \nonumber \\
    & \quad \quad \quad \quad \quad \quad + \left[(p-k)_0 g_i^{\rho'} - (p-k)_i g_0^{\rho'}\right] ( (p-k)^2 - p^2 )  \nonumber \\
    & \quad \quad \quad \quad \quad \quad - \left[ ( k_0 g_i^{\rho'} - k_i g_0^{\rho'}) (p^2 - (p-k)^2 ) + k^{\rho'} (p_0 k_i -k_0 p_i) \right] \bigg] \nonumber \\
    &= \frac{1}{2} g^2 N_c \delta^{ad}  \D(p)_{JJ'} P_{\rho \rho'}(-p) (-i p_0 g_j^{\rho} + i p_j g_{0}^{\rho})  \nonumber \\
    & \quad \times \int_k \D(p-k)_{I'I}\D(k)_{I'I} (-1)^{I'+1}  \times \frac{(1-\xi)}{k^2} \left[p_0 g_i^{\rho'} - p_i g_0^{\rho'} \right] \left( (p-k)^2 - p^2 \right) \,.
\end{align}
Due to the last term, the integrand of the above expression has two pieces: one proportional to $p^2$, and another proportional to $(p-k)^2$. The term proportional to $p^2$ has two propagators in the integrand, one with momentum $(p-k)$ and the other with momentum $k$, which thus must be cancelled by the sum of diagrams (3), (4), and (8). The term proportional to $(p-k)^2$ has a loop integral over $k$ that is disconnected from the $p$ momentum flow, so it must be cancelled by contributions from (7) (diagram (9) vanishes).

We focus on the term proportional to $(p-k)^2$ first. A little algebra leads to
\begin{align}
    (7)_{(1-\xi)} = \frac{g^2 N_c}{2} \delta^{ad} \left[p_0^2 g_{ij} + p_i p_j g_{00} \right] \D(p)_{JI} \int_k \frac{(1-\xi)}{k^2} \D(k)_{II} \,.
\end{align}
Now, going back to our result from the three- and four-propagator diagrams, and looking at the piece proportional to $(p-k)^2$, we have
\begin{align}
    &\left(\frac{1}{2} (1)_{(1-\xi), \, \propto p^2} + \big((5)+(6) \big)_{(1-\xi)} \right)_{\propto (p-k)^2} \nn\\
    &=  \frac{1}{2} g^2 N_c \delta^{ad}  \D(p)_{JI'} P_{\rho \rho'}(p) (-i p_0 g_j^{\rho} + i p_j g_{0}^{\rho})  \nonumber \\
    & \quad \times \int_k (p-k)^2\, \D(p-k)_{I'I}\D(k)_{I'I} (-1)^{I'+1}  \frac{(1-\xi)}{k^2} \left[p_0 g_i^{\rho'} - p_i g_0^{\rho'} \right]  \nonumber \\
    &= \frac{g^2 N_c}{2} \delta^{ad} \D(p)_{JI} (- g_{\rho\rho'}) ( p_0 g_j^{\rho} -  p_j g_{0}^{\rho}) \left[p_0 g_i^{\rho'} - p_i g_0^{\rho'} \right] \int_k \frac{(1-\xi)}{k^2} \D(k)_{II} \nonumber \\
    &= -\frac{g^2 N_c}{2} \delta^{ad} \D(p)_{JI} \left( p_0^2 g_{ij} + p_i p_j g_{00} \right) \int_k \frac{(1-\xi)}{k^2} \D(k)_{II} \,,
\end{align}
and thus
\begin{align}
 \left(\frac{1}{2} (1)_{(1-\xi), \, \propto p^2} + \big((5)+(6)\big)_{(1-\xi)} \right)_{\propto (p-k)^2} + (7)_{(1-\xi)} = 0 \,.
\end{align}

Then we consider the remaining contribution from the term proportional to $p^2$
\begin{align}
    &\left(\frac{1}{2} (1)_{(1-\xi), \, \propto p^2} + \big((5)+(6)\big)_{(1-\xi)} \right)_{\propto p^2} \nn\\
    &= \frac{1}{2} g^2 N_c \delta^{ad}  (-p^2) \D(p)_{JI'} P_{\rho \rho'}(p) (-i p_0 g_j^{\rho} + i p_j g_{0}^{\rho})  \nonumber \\
    & \quad \times \int_k \D(p-k)_{I'I}\D(k)_{I'I} (-1)^{I'+1} \frac{(1-\xi)}{k^2} \left[p_0 g_i^{\rho'} - p_i g_0^{\rho'} \right] \nonumber \\
    &= \frac{-ig^2}{2} N_c \delta^{ad} (ip_0 g_{j \rho'} - i p_j g_{0  \rho'}) \left[p_0 g_i^{\rho'} - p_i g_0^{\rho'} \right] \int_k \D(p-k)_{JI}\D(k)_{JI} \frac{(1-\xi)}{k^2} \nonumber \\
    &= \frac{g^2 N_c}{2} \delta^{ad} \left(p_0^2 g_{ij} + p_i p_j g_{00} \right) \int_k \D(p-k)_{JI}\D(k)_{JI} \frac{(1-\xi)}{k^2}\,,
\end{align}
which is expected to be cancelled by diagrams with two propagators. To show this, we need to evaluate diagrams (3), (4), and (8). We start with $(4)$, whose gauge-dependent part is given by
\begin{align}
    (4)_{(1-\xi)} &= - g^2 N_c \delta^{ad} \int_k \big( (p-k)_0^2 g_{ij} +  (p-k)_i (p-k)_j g_{00} \big) \D(p-k)_{JI} \D(k)_{JI} \frac{(1-\xi)}{k^2} \,.
\end{align}
Then we evaluate the linear $(1-\xi)$ gauge-dependent part of the diagram $(3)$:
\begin{align}
(3)_{(1-\xi)} &= g^2 N_c \delta^{ad} \int_k \D(k)_{JI} \D(p-k)_{JI} (1-\xi) \left[ -g_{00} \frac{(p-k)_i (p-k)_j}{(p-k)^2} -g_{ij} \frac{k_0 k_0}{k^2} \right] \nonumber \\
&= -g^2 N_c \delta^{ad} \int_k \D(k)_{JI} \D(p-k)_{JI} \frac{(1-\xi)}{k^2} \left[ k_0^2 g_{ij} + k_i k_j g_{00} \right] \,.
\end{align}
At last, we evaluate the $(1-\xi)$-dependent part of diagram (8), obtaining
\begin{align}
(8)_{(1-\xi)} = -g^2 N_c \delta^{ad} \int_k \D(k)_{JI} \D(p-k)_{JI} \frac{(1-\xi)}{k^2} \left[ (p_0 k_0 - k_0^2) g_{ij} + (p_i k_j - k_i k_j) g_{00} \right] \,.
\end{align}
Finally, adding things up, one finds
\begin{align}
    & \left(\frac{1}{2} (1)_{(1-\xi), \, \propto p^2} + \big((5)+(6)\big)_{(1-\xi)} \right)_{\propto p^2} + \frac{1}{2} \left( (3)_{(1-\xi)} + (4)_{(1-\xi)} \right) + (8)_{(1-\xi)} \nonumber \\
    &= \frac{g^2 N_c}{2} \delta^{ad} \int_k \D(p-k)_{JI} \D(k)_{JI} \frac{(1-\xi)}{k^2} \nonumber \\
    & \quad \quad \quad \bigg[ \left(p_0^2 g_{ij} + p_i p_j g_{00} \right) -  \left[ k_0^2 g_{ij} + k_i k_j g_{00} \right] \nonumber \\
    & \quad \quad \quad \quad - \left( (p-k)_0^2 g_{ij} +  (p-k)_i (p-k)_j g_{00}\right) - 2\left[ (p_0 k_0 - k_0^2) g_{ij} + (p_i k_j - k_i k_j) g_{00} \right] \bigg] \nonumber \\
    &= \frac{g^2 N_c}{2} \delta^{ad} \int_k \D(p-k)_{JI} \D(k)_{JI} \frac{(1-\xi)}{k^2} \nonumber \\
     & \quad \quad \quad \bigg[ \left(p_0^2 - k_0^2 - (p_0^2 - 2p_0 k_0 + k_0^2) - 2p_0 k_0 + 2k_0^2 \right) g_{ij} \nonumber \\
     & \quad \quad \quad \quad + \left( p_i p_j - k_i k_j - (p_i p_j - p_i k_j - p_j k_i + k_i k_j ) - 2(p_i k_j -k_i k_j) \right) g_{00} \bigg] \nonumber \\
     &= 0 \,,
\end{align}
where we have made the symmetry factors of diagrams $(3)$ and $(4)$ explicit, which are $1/2$ for both diagrams.

Therefore, after taking into account all contributions, the result is $R_\xi$ gauge-invariant.
The proof of gauge invariance is now complete.

\section{NLO evaluation} \label{app:collinear-integrals}
Here, we briefly outline two different integration orders that we adopted to check independently our results. This provides a verification that the collinear limit of the integrals was handled correctly in all cases, and also provides two different calculation strategies that could be adopted in the future to carry out this type of integrals.

\subsection{Integration order without regulator}

After some algebra, one can show that diagrams $(1)$, $(f)$, $(g)$, $(3)$, $(4)$, $(5)$, $(5r)$, $(6)$, $(6r)$, $(8)$, $(8r)$ and $(11)$ in their retarded forms, involve the structure:
\begin{align}
\Pi^R(p)\equiv\int_k N(p,k) \left[ D^{\ml{T}}(k) D^{\ml{T}}(p-k) - D^{<}(k) D^{<}(p-k) \right] \,. \label{eq:a1}
\end{align}
From here, we outline a procedure to arrange the propagator structure in a generic form that makes manifest how the different pole structures are handled, such that we can generically use it for the expressions given in Section~\ref{sect:nlo} to obtain the relevant part of the retarded component. 

Using the Feynman rules, we can write the finite temperature part of Eq.~(\ref{eq:a1}), after some algebra and relabelling, as
\begin{align}
&\Pi^R(p)= i \int \frac{\text{d}^3 k}{(2\pi)^3 2  |\mathbf k |} \frac{\text{d}^3 q}{(2\pi)^3 2  |\mathbf q |} n_{B}(|{\bs k}|) \times \nonumber \\
&\bigg\{ \left[ N(p,k) + N(p,p-k) \right] \left[ \frac{ (2\pi)^3 {\delta}^3({\bs p} - {\bs k} - {\bs q})}{p_0 - |{\bs k}| - |{\bs q}| + i 0^+}- \frac{ (2\pi)^3 {\delta}^3({\bs p} - {\bs k} + {\bs q})}{p_0 - |{\bs k}| + |{\bs q}| + i 0^+}  \right] \nonumber \\
&+ \left[ N(p,-k) + N(p,p+k)\right] \left[\frac{(2\pi)^3 {\delta}^3({\bs p} + {\bs k} - {\bs q})}{p_0 + |{\bs k}| - |{\bs q}| + i 0^+} - \frac{(2\pi)^3 {\delta}^3({\bs p} + {\bs k} + {\bs q})}{p_0+|{\bs k}|+|{\bs q}| + i 0^+}\right] \bigg\} \,. \label{eq:trans}
\end{align}
For example, the numerator factor in the case of diagram (1) (gauge boson self-energy, gauge boson loop) is:
\begin{align}
N(p,k)=- \frac{1}{2} \left[ g^{\mu \nu} (5 p^2 - 2 p \cdot k ) + 10 k^{\mu} k^{\nu} \right] \,,
\end{align}
which leads to the retarded object:
\begin{align}
\Pi^{R,(1)}(p)&= i \int \frac{\text{d}^3 k}{(2\pi)^3 2  |\mathbf k |} \frac{\text{d}^3 q}{(2\pi)^3 2  |\mathbf q |} 2 n_{B}(|{\bs k}|) \bigg\{- \frac{1}{2} \left[ g^{\mu \nu} 4 p^2 + 5 k^{\mu} k^{\nu} \right] \bigg\} \times \nonumber \\
&\bigg\{  \left[ \frac{ (2\pi)^3 {\delta}^3({\bs p} - {\bs k} - {\bs q})}{p_0 - |{\bs k}| - |{\bs q}| + i 0^+}- \frac{ (2\pi)^3 {\delta}^3({\bs p} - {\bs k} + {\bs q})}{p_0 - |{\bs k}| + |{\bs q}| + i 0^+}  \right] \nonumber \\
&+ \left[\frac{(2\pi)^3 {\delta}^3({\bs p} + {\bs k} - {\bs q})}{p_0 + |{\bs k}| - |{\bs q}| + i 0^+} - \frac{(2\pi)^3 {\delta}^3({\bs p} + {\bs k} + {\bs q})}{p_0+|{\bs k}|+|{\bs q}| + i 0^+}\right] \bigg\} \,.\label{eq:selfgeneric}
\end{align}
We can write down similar expressions for all the other diagrams. 

\begin{table*}
\begin{center}
\begin{tabular}{c||c|c|c}
\toprule
   $\sigma_1\sigma_2$ & $z_0$ & $z_p$& $z_m$   \\ \midrule
    $++$ & $\Delta E$ & $-|{\bs k}|\tau + \sqrt{{\bs k}^2\tau^2 + \Delta E^2 + 2 \Delta E |{\bs k}|}$ & $-$ \\ \midrule
    $+-$ & $\Delta E$ & $-$ & $-$ \\  \midrule
    $-+$ & $\Delta E$ & $|{\bs k}|\tau + \sqrt{{\bs k}^2\tau^2 + \Delta E^2 - 2 \Delta E |{\bs k}|}$\,, & $|{\bs k}|\tau - \sqrt{{\bs k}^2\tau^2 + \Delta E^2 - 2 \Delta E |{\bs k}|}$ \,, \\
& & if $  \bigg\{ \Delta E/2 \leq |{\bs k}| \leq \Delta E  $ & if $\Delta E/2 \leq |{\bs k}| \leq \Delta E  $\\
& & $ \land \tau \geq \sqrt{\frac{2\Delta E |{\bs k}|-\Delta E^2}{|{\bs k}|^2}} \bigg\} $ & $\land \tau \geq \sqrt{\frac{2\Delta E |{\bs k}|-\Delta E^2}{|{\bs k}|^2}}$\,. \\
& & or $ 0 \leq |{\bs k}| \leq \Delta E/2 \,. $ &  \\ \midrule
$--$ & $\Delta E$ & $|{\bs k}|\tau - \sqrt{{\bs k}^2\tau^2 + \Delta E^2 - 2 \Delta E |{\bs k}|}$ \,, & $|{\bs k}|\tau + \sqrt{{\bs k}^2\tau^2 + \Delta E^2 - 2 \Delta E |{\bs k}|}$ \,, \\  
 &  & if $|{\bs k}|\geq \Delta E \land \tau\geq\sqrt{\frac{2\Delta E |{\bs k}|-\Delta E^2}{|{\bs k}|^2}} $ \,. & if $|{\bs k}|\geq \Delta E \land \tau\geq\sqrt{\frac{2\Delta E |{\bs k}|-\Delta E^2}{|{\bs k}|^2}} $ \,. \\  
 \bottomrule
\end{tabular}
\end{center}
\caption{Summary of the single ($z_{p/m}$) and double ($z_0$) poles, as well as their existence criteria.}
\label{tab:poles}
\end{table*}

For all contributions that can be written in this form, to obtain their contributions to the integrated spectral function $\varrho_E^{++}(p_0=\Delta E)$, we use the three momentum delta function to perform the $\bs q$ integration first. Then, we interchange the order of the loop $\bs k$ integration with the external $\bs p$ integration, and finally we perform the $\bs p$ integration after taking the imaginary part of the retarded correlation function structure shown above, multiplied by $\bs p$-dependent propagators coming from the rest of the diagram. When taking the imaginary part, we apply the residue theorem. These procedures lead to the following schematic arrangement in terms of the contributing poles:
\begin{align}
R^{++}_{0} &\equiv \int_0^{\infty} \text{d} |{\bs k}| \int_{-1}^{1} \text{d} \tau \left[ \text{Res}(G_{0}^{++},z_0) + \text{Res}(G_{0}^{++},z_p) \right], \label{eq:Rpp}\\
R^{+-}_{0} &\equiv  \int_0^{\infty} \text{d} |{\bs k}| \int_{-1}^{1} \text{d} \tau \left[ \text{Res}(G_{0}^{+-},z_0) \right], \label{eq:Rpm} \\
R^{-+}_{0} &\equiv  \int_0^{\Delta E /2} \text{d} |{\bs k}| \int_{-1}^{1} \text{d} \tau \left[ \text{Res}(G_{0}^{-+},z_0) +\text{Res}(G_{0}^{-+},z_p)\right] \label{eq:Rmp}\\
&+ \int_{\Delta E /2}^{\Delta E} \text{d} |{\bs k}|  \bigg\{ \int_{\sqrt{\frac{2\Delta E |{\bs k}|-\Delta E^2}{|{\bs k}|^2}}}^{1} \text{d} \tau  \left[ \text{Res}(G_{0}^{-+},z_0) +\text{Res}(G_{0}^{-+},z_p)-\text{Res}(G_{0}^{-+},z_m)\right]\nonumber  \\
& \phantom{\int_{\Delta E /2}^{\Delta E} \text{d} |{\bs k}| } + \int_{-1}^{\sqrt{\frac{2\Delta E |{\bs k}|-\Delta E^2}{|{\bs k}|^2}}}\text{d} \tau  \left[ \text{Res}(G_{0}^{-+},z_0)\right] \bigg\} \nonumber \\
&+\int_{\Delta E}^{\infty} \text{d} |{\bs k}| \int_{-1}^{1} \text{d} \tau \left[ \text{Res}(G_{0}^{-+},z_0) \right], \nonumber \\
R^{--}_{0}&\equiv  \int_0^{\Delta E} \text{d} |{\bs k}| \int_{-1}^{1} \text{d} \tau \left[ \text{Res}(G_{0}^{--},z_0) \right] \label{eq:Rmm}\\
&+ \int_{\Delta E}^{\infty} \text{d} |{\bs k}|  \bigg\{ \int_{\sqrt{\frac{2\Delta E |{\bs k}|-\Delta E^2}{|{\bs k}|^2}}}^{1} \text{d} \tau  \left[ \text{Res}(G_{0}^{--},z_0) +\text{Res}(G_{0}^{--},z_p)-\text{Res}(G_{0}^{--},z_m)\right]\nonumber\\
& \phantom{\int_{\Delta E /2}^{\Delta E} \text{d} |{\bs k}| } + \int_{-1}^{\sqrt{\frac{2\Delta E |{\bs k}|-\Delta E^2}{|{\bs k}|^2}}}\text{d} \tau  \left[ \text{Res}(G_{0}^{-+},z_0)\right] \bigg\} \,. \nonumber \\
\end{align}
Here we have introduced many new notations that we will explain now. Firstly, we have defined the location of the single ($z_{p/m}$) and double poles ($z_0$) where the integral over ${\bs p}$ gives residue contributions (the double pole comes from having $[D^R(p)]^2$ multiplying the gauge boson self-energy diagrams; in diagrams (5), (5r), (6), (6r) it is only a single pole because there is only a single $D^R(p)$ factor, and there is no corresponding pole in the rest of the diagrams). Their existence criteria, as well as their values as a function of ${\bs k}$ and $\Delta E$ are listed in Table~\ref{tab:poles}. The four $G^{\sigma_1 \sigma_2}_0$ functions originate from the numerator structures that accompany the four single poles in Eq.~(\ref{eq:selfgeneric}), and can be explicitly determined from comparison after performing the integration order as listed above ($\sigma_1$ and $\sigma_2$ denote the different relative sign choices in front of the absolute values $|{\bs q}|$, $|{\bs k}|$ in the four denominators of~\eqref{eq:selfgeneric}). Concretely, for the fermion self-energy these are given by Eq.~(4.15) in Ref.~\cite{Binder:2020efn}. As discussed in the main text, in the case of the gauge boson self-energy the double and single poles are paired, in the sense that they are individually divergent when the momenta become collinear, and only the sum over both residues is collinear-finite. A similar pairing is required for diagrams (5), (5r), (6), (6r), where the sum over poles is also collinear-finite. Because there is no $D^R(p)$ propagator in the rest of the diagrams, their respective spectral functions are safe in the collinear limit (as explained in the main text, the purported collinear divergence appears when two poles become closer, but there is no pair of poles at all in these diagrams, only single poles).

For all terms we then manage to further perform the angular $\tau = \cos\theta$ ($\theta$ denotes the relative angle bewtween ${\bs p}$ and ${\bs k}$) integration analytically. As an illustrative example, for diagram (3) in Figure~\ref{fig:diagrams} we can even perform the last $k$ integral analytically, obtaining a fully analytic expression:
\begin{align}
R_3\left(x=\frac{\Delta E}{T}\right) = - \frac{3 N_c}{2x^2} \big[ & -2 \Re[\text{Li}_2(e^x)]+2 \text{Li}_2[-\cosh(x)+\sinh(x)+1] \\ & + x^2-2 x \ln[e^x-1]+\pi^2 \big] \,.\nonumber
\end{align}
The function is entirely negative, monotonically increasing for increasing $x$, and at large $x$ it satisfies $\lim_{x\rightarrow \infty} R_3(x)=0$.

We note that, for diagrams (5) and (5r) in Figure~\ref{fig:diagrams}, which involve contributions coming from Wilson lines, the remaining $k$ integration needs to be performed with a principal part prescription at the location $k=\Delta E$, in the same way as the result is described in the main text.

\subsection{Integration order with regulator}

Starting from equation~\eqref{eq:integrated-gluon-self-energy-before-integration}, we can work with the full expression (before taking the real part) for the temperature-dependent pieces. We define
\begin{equation}
\int_{\p} L_T = i\int_{\k,\q} \sum_{\sigma_1,\sigma_2}  \frac{\sigma_2 2 n_B(|\k|) }{2|\k|2|\q|} \frac{(-1) N((p_0,\q - \k),(-\sigma_1 |\k|,-\k)) }{( (p_0 + i\epsilon)^2 - (\q - \k)^2 )^2 (p_0 + \sigma_1 |\k| - \sigma_2 |\q| + i0^+)} \,,
\end{equation}
which satisfies
\be
{\rm Re} \{g^2 N_c (N_c^2-1) \int_{\p} L_T\} = \left. \varrho_E^{++}(p_0) \right|_{\rm NLO}^{\rm gauge\ boson+ghost} - \left. \varrho_E^{++}(p_0) \right|_{{\rm NLO}, \, T=0 }^{\rm gauge\ boson+ghost}  \,.\quad\ 
\ee

First, we do the integral over the angle between $\k $ and $\q $, $u \equiv \frac{\k \cdot \q}{kq}$, which only involves the numerator and the retarded propagator with momentum $p$. Explicitly, writing
\begin{equation}
    N((p_0,\q - \k),(-\sigma_1 |\k|,-\k)) = c(p_0,-\sigma_1 k, q, k) + d(p_0,-\sigma_1 k, q, k) u + e(p_0,-\sigma_1 k, q, k) u^2\,,
\end{equation}
we have
\be
\int_{\p} L_T &=& \frac{(-i) (2\pi) (4\pi) }{ (2\pi)^6 } \nn\\
&\times& \int_{0}^\infty \diff k \diff q \sum_{\sigma_1,\sigma_2}  \frac{  \sigma_2 n_B(k)  }{8 k q (p_0 + \sigma_1 k - \sigma_2 q + i0^+)} \int_{-1}^1 \diff u \frac{ c + du + eu^2 }{  ( u - u_0 + i{\rm sgn}(p_0) 0^+)^2  } \,,\quad\quad\quad
\ee
with $u_0 = u_0(p_0,q,k) = (k^2 + q^2 - p_0^2)/(2kq)$. The integral over $u$ can be done straightforwardly. If we introduce a regulator $\delta$ to control the collinear divergence,
\begin{equation}
\begin{split}
    &\int_{-1+\delta}^{1-\delta} \diff u \frac{ c + du + eu^2 }{  ( u - u_0 + i{\rm sgn}(p_0) 0^+)^2  } \\ &=   \bigg[ 2 e - ( e u_0^2 + d u_0 + c  ) \left( \frac{1}{1 - u_0+ i0^+ {\rm sgn}(p_0) } - \frac{1}{-1 - u_0+ i0^+ {\rm sgn}(p_0) } \right) \\
& \quad  + (2e u_0 + d ) \bigg( \ln \left| \frac{1 - \delta - u_0}{-1 + \delta - u_0} \right| - i \pi \Theta(1-\delta -u_0) \Theta(u_0 - (-1 + \delta))   \bigg)  \bigg] \,,
\end{split}
\end{equation}
where we have let $\delta \to 0$ in the non-problematic terms; namely, in every place except the logarithms and the terms needed to compensate for them. For notational simplicity we define
\begin{align}
    {\rm NC} = 2e/(8kq)\,, & & {\rm NL} = (2 e u_0 + d)/(8 k q)\,, & & {\rm ND} = -(e u_0^2 + du_0 + c)/(8p_0) \,.
\end{align}
Then we have
\be
\int_{\p} L_T^{(1)} &=& \frac{(-i) }{ 8\pi^4 } \int_{0}^\infty \!\! \diff k  \diff q \, n_B(k) \sum_{\sigma_1,\sigma_2} \bigg[  \frac{  \sigma_2  {\rm NC}  }{ p_0 + \sigma_1 k - \sigma_2 q + i0^+} \nn\\
&& \quad \quad +  \frac{  \sigma_2 {\rm NL}   }{ p_0 + \sigma_1 k - \sigma_2 q + i0^+} \bigg( \ln \left| \frac{1 - \delta - u_0}{1 - \delta + u_0} \right| - i \pi \Theta(1-\delta -u_0) \Theta(u_0 +1 - \delta)   \bigg) \nn\\
&& \quad \quad +  \sum_{\sigma_1' \sigma_2'} \frac{  \sigma_2 {\rm ND}   }{ p_0 + \sigma_1 k - \sigma_2 q + i0^+} \frac{\sigma_1' \sigma_2'}{p_0 + \sigma_1' k - \sigma_2' q + i0^+}  \bigg] \,.
\ee
Numerically the strategy is to take the real part of this expression and integrate over $q$. The terms proportional to ${\rm NC}$ and ${\rm ND}$ may be evaluated directly using the residue theorem, and then we take $\delta \to 0$ at the end.

For the terms proportional to ${\rm NL}$, there are two cases:
    \begin{enumerate}
        \item $\sigma_1 \sigma_2 = 1$, in which case the logarithmic term can be written as
        \be
        \label{eqn:log}
            && \ln \left| \frac{ (|\k| - |\q|)^2 - p_0^2 + 2\delta |\k| |\q| }{(|\k| + |\q|)^2 - p_0^2 - 2\delta |\k| |\q| } \right|_{|\q| = |\k| + \sigma_2 p_0} \Theta(|\k| + \sigma_2 p_0) \nn\\
            &=& \ln \left| \frac{ 2\delta |\k| |\q| }{(|\k| + |\q|)^2 - p_0^2 - 2\delta |\k| |\q| } \right|_{|\q| = |\k| + \sigma_2 p_0} \Theta(|\k| + \sigma_2 p_0) \,,
        \ee
        where the theta functions come from the fact that the integration variable $|\q|$ is positive. This diverges as $\delta \to 0$, so it must be compensated by the other term with the $\Theta$ functions. For purposes of evaluating the divergent terms, we can add and subtract a term that treats ${\rm NL}$ as a constant, where all $|\q|$ dependence in NL is set to be the value where the singularity occurs $|\q| = \sigma_1 \sigma_2 |\k| + \sigma_2 p_0$. So, we have to evaluate
        \begin{equation}
        \int_0^\infty \diff |\q| \frac{\Theta\big( (|\k| + |\q|)^2 - p_0^2 - 2\delta |\k| |\q| \big) \Theta\big( p_0^2 - (|\k| - |\q|)^2 - 2\delta |\k| |\q| \big) }{ p_0 + \sigma_1|\k| - \sigma_2|\q| } \,.
        \end{equation}
        The Heaviside step functions contain essential information. They bound either momentum in terms of the other as
        \begin{equation}
            \left| (1-\delta) k - \sqrt{p_0^2 - 2\delta k^2 + \delta^2 k^2} \right| < q < (1-\delta) k  + \sqrt{p_0^2 - 2\delta k^2 + \delta^2 k^2} \,,
        \end{equation}
        which means that our integral is actually
        \begin{equation}
        \begin{split}
        \label{eqn:angle}
            \int_{\big| (1-\delta) |\k| - \sqrt{p_0^2 - 2\delta |\k|^2 + \delta^2 |\k|^2} \big|}^{(1-\delta) |\k| + \sqrt{p_0^2 - 2\delta |\k|^2 + \delta^2 |\k|^2}} \frac{\diff |\q|}{p_0 + \sigma_1 |\k| - \sigma_2 |\q|} \\
            = -\sigma_2 \ln \left|\frac{ -\sigma_2 p_0 - \sigma_1 \sigma_2 |\k| +  (1-\delta) |\k| + \sqrt{p_0^2 - 2\delta |\k|^2 + \delta^2 |\k|^2}}{ -\sigma_2 p_0 - \sigma_1 \sigma_2 |\k| + |  (1-\delta) |\k| - \sqrt{p_0^2 - 2\delta |\k|^2 + \delta^2 |\k|^2}| } \right| \\
            = -\sigma_2 \ln \left|\frac{ -\sigma_2 p_0 - \delta |\k| + \sqrt{p_0^2 - 2\delta |\k|^2 + \delta^2 |\k|^2}}{ -\sigma_2 p_0 - |\k| + |(1-\delta) |\k| - \sqrt{p_0^2 - 2\delta |\k|^2 + \delta^2 |\k|^2}| } \right| \,,
        \end{split}
        \end{equation}
        where we have used $\sigma_1 \sigma_2 = 1$. The sign of the integral~(\ref{eqn:angle}) depends on $\sigma_2$ (actually, on $\sigma_2 \,{\rm sgn}(p_0)$, but we take $p_0>0$ throughout), which is an external overall factor. All that remains now is to take the limit. We first consider the case with $\sigma_2 = -1$. It is now clear that Eq.~(\ref{eqn:angle}) is divergent as $\delta \to 0$ when $|\k| > |p_0|$, and finite if the converse inequality is true (we can always choose $\delta$ small enough so that it doesn't affect the inequalities, because $|\k|$ and $|p_0|$ are fixed at this step). Therefore, we can erase the absolute value in the denominator of Eq.~(\ref{eqn:angle}) and obtain the sum of Eqs.~(\ref{eqn:log}) and~(\ref{eqn:angle}) for $|{\bs k}_1|>|p_0|$
        \begin{equation}
        \begin{split}
            \lim_{\delta \to 0} & \ln \left| \frac{ 2\delta  |\k| (|\k| + \sigma_2 p_0) }{( 2|\k| + \sigma_2 p_0)^2 - p_0^2 - 2\delta |\k|(|\k| + \sigma_2 p_0) } \right|  \\ & -  \sigma_2 \ln \left|\frac{ -\sigma_2 p_0 - \delta |\k| + \sqrt{p_0^2 - 2\delta |\k|^2 + \delta^2 |\k|^2}}{ -\sigma_2 p_0 -\delta |\k| - \sqrt{p_0^2 - 2\delta |\k|^2 + \delta^2 |\k|^2}} \right| \,,
        \end{split}
        \end{equation}
        and we can take $\delta \to 0$ in all terms that are not going to 0, obtaining
        \begin{equation}
        \begin{split}
            \lim_{\delta \to 0} & \ln \left| \frac{ 2\delta  |\k| (|\k| - |p_0|) }{ 4 |\k|^2 - 4  |\k| |p_0| } \right|  +  \ln \left|\frac{ 2|p_0| }{ |p_0| -\delta |\k| - \sqrt{p_0^2 - 2\delta |\k|^2 + \delta^2 |\k|^2}} \right| \\
            =\lim_{\delta \to 0} & \ln \left| \frac{ \delta}2 \right| +  \ln \left|\frac{ 2|p_0| }{ |p_0| -\delta |\k| - |p_0| (1 - \delta |\k|^2/p_0^2 )} \right| \\
            = \lim_{\delta \to 0} & \ln \left| \frac{ \delta}2 \right| +  \ln \left|\frac{ 2 }{  \delta (|\k|^2/p_0^2 - |\k|/|p_0|) } \right| = - \ln \left| \frac{|\k|^2}{|p_0|^2} - \frac{|\k|}{|p_0|} \right| \,,
        \end{split}
        \end{equation}
        which is finite. Then we consider $\sigma_2 = 1$. That means that the divergence comes from the numerator, and the limit of the sum is equal to
        \begin{equation}
        \begin{split}
            \lim_{\delta \to 0} & \ln \left| \frac{ 2\delta  |\k| (|\k| + \sigma_2 p_0) }{( 2|\k| + \sigma_2 p_0)^2 - p_0^2 - 2\delta |\k|(|\k| + \sigma_2 p_0) } \right|  \\ &-  \sigma_2  \ln \left|\frac{ -\sigma_2 p_0 - \delta |\k| + \sqrt{p_0^2 - 2\delta |\k|^2 + \delta^2 |\k|^2}}{ -\sigma_2 p_0 - |\k| + | |\k| - |p_0|| } \right| \\
            = \lim_{\delta \to 0} & \ln \left| \frac{ 2\delta  |\k| (|\k| + |p_0|) }{ 4|\k|^2 + 4 |\k| |p_0|  } \right| -  \ln \left|\frac{ -|p_0| - \delta |\k| + \sqrt{p_0^2 - 2\delta |\k|^2 + \delta^2 |\k|^2}}{ -|p_0| - |\k| + | |\k| - |p_0|| } \right| \\
            =\lim_{\delta \to 0} & \ln \left| \frac{ \delta}2 \right| -  \ln \left|\frac{ - \delta |\k| - \delta |\k|^2/|p_0| }{ -|p_0| - |\k| + | |\k| - |p_0|| } \right| = - \ln \left| \frac{ |\k|^2 + |\k| |p_0| }{|p_0| \min(|\k|,|p_0|) } \right|\,,
        \end{split}
        \end{equation}
        which again, is finite. So we see that this should work for all terms.
        \item $\sigma_1 \sigma_2 = -1$ is the remaining case. Now the first term (the term with the logarithm) after integrating over $|\q|$ goes to
        \begin{equation}
        \label{eqn:case2_log}
        \ln \left| \frac{  (|\k| - |\q|)^2 - p_0^2 + 2\delta |\k| |\q| }{(|\k| + |\q|)^2 - p_0^2 - 2\delta |\k| |\q| } \right| \to \ln \left| \frac{ 4 |\k|^2 - 4\sigma_2 p_0 |\k| }{ 2\delta |\k| (|\k| - \sigma_2 p_0) } \right| \Theta(-|\k| + \sigma_2 p_0)  \,.
        \end{equation}
        The term with the $\Theta$-functions becomes after the integral over $|\q|$
        \begin{equation}
        \label{eqn:case2_theta}
            -\sigma_2  \ln \left|\frac{ -\sigma_2 p_0 + |\k| +  (1-\delta) |\k| + \sqrt{p_0^2 - 2\delta |\k|^2 + \delta^2 |\k|^2}}{ -\sigma_2 p_0 + |\k| + |  (1-\delta) |\k| - \sqrt{p_0^2 - 2\delta |\k|^2 + \delta^2 |\k|^2}| } \right| \,,
        \end{equation}
        which only diverges in the denominator as $\delta \to 0$ if $\sigma_2 = 1$ and $|p_0| > |\q|$. This regime is exactly where the Heaviside step function of the first term with the logarithm~(\ref{eqn:case2_log}) is supported. In this regime, the term~(\ref{eqn:case2_theta}) can be simplified as
        \begin{equation}
            - \ln \left|\frac{ 2|\k| }{ \delta |\k| - \delta |\k|^2/|p_0| } \right|\,,
        \end{equation}
        and so the limit of the sum of these two terms is
        \begin{equation}
            \lim_{\delta \to 0} \left( \ln \frac{2}{\delta} - \ln \left| \frac{ 2|\k| }{ \delta |\k| - \delta |\k|^2/|p_0|} \right| \right) = \ln \left| 1 - \frac{|\k|}{|p_0|}\right| \,.
        \end{equation}
        This shows that the collinear divergence cancels out exactly. Therefore, this guarantees that if we evaluate the integrals (numerically or analytically) with $\delta > 0$ and then take $\delta \to 0$ at the end, we will get a well-defined result. 
    \end{enumerate}

An equivalent treatment can be done for diagrams $(5)$, $(5r)$, $(6)$, $(6r)$, by simply modifying the number of propagators $D^R(p)$ that appear. However, we will not apply the methods shown in this section for these diagrams, since we want to evaluate them in dimensional regularization to extract their UV divergence in vacuum, which will be performed in the next section of this Appendix.

\subsection{Contribution from diagrams \texorpdfstring{$(5)$}{(5)}, \texorpdfstring{$(5r)$}{(5r)}, \texorpdfstring{$(6)$}{(6)}, and \texorpdfstring{$(6r)$}{(6r)}} \label{app:3-prop-detail}

Here we give the details of how we evaluate the contribution of diagrams $(5)$, $(5r)$, $(6)$, $(6r)$ to the integrated spectral function
\begin{equation}
\begin{split}
\left. \varrho_E^{++}(p_0) \right|_{\rm NLO}^{5-6} &= i g^2 N_c (N_c^2-1) \tilde{\mu}^\epsilon \int_{\k,\p} \frac{1+2n_B(k)}{2k} \sum_{\sigma_1}   N_{3p}((p_0,\p),(\sigma_1 k, \k)) \\
  & \quad \quad \quad \quad \quad \quad \quad \times {\rm Re}\left\{ \frac{i}{((p_0+i0^+)^2 -\p^2 )((p_0-k_0 + i0^+)^2 - (\p-\k)^2 )} \right\} \\
  & \quad { + g^2 N_c (N_c^2 -1) \pi \int_{\k,\p}   \frac{ \left[ k_0 N^{(5),(6)}(p,k) \right]_{k_0=0} }{\k^2} \ml{P} \left( \frac{1}{p_0^2 - (\p - \k)^2} \right) \delta(p^2) } \,,
\end{split}
\end{equation}
where $N_{3p}(p,k)$ includes the sum of all numerators of diagrams (5), (5r), (6), (6r) and $\tilde{\mu}^2 = \mu^2e^{\gamma_E}/(4\pi)$.  { Here we have factored out of the numerators the factor of $N_c$.} All integrals, $\int_\k$ and $\int_\p$, are in $d = 3-\epsilon$ dimensions.

{ Let us first evaluate the last piece. To that end, note that 
\begin{align}
    \left[ k_0 N^{(5),(6)}(p,k) \right]_{k_0=0} &= p_0 \left[ 2\k^2 - 2 \p \cdot \k - 2(d-1) p_0^2 + 2\p^2 \right] \nonumber \\ 
    &=  p_0 \left[ \k^2 +(\p-\k)^2 - p_0^2 - (2d-3) p_0^2 + \p^2 \right] \, ,
\end{align}
and therefore
\begin{align}
    &\pi \int_{\k,\p}   \frac{ \left[ k_0 N^{(5),(6)}(p,k) \right]_{k_0=0} }{\k^2} \ml{P} \left( \frac{1}{p_0^2 - (\p - \k)^2} \right) \delta(p^2) \nonumber \\
    &= \pi p_0 \int_\p \delta(p^2) \int_\k \left[ \ml{P} \frac{1}{p_0^2 - (\p-\k)^2} - \frac{1}{\k^2} + \ml{P} \frac{\p^2 - (2d-3)p_0^2}{\k^2 (p_0^2 - (\p-\k)^2)} \right] \, .
\end{align}
The first two terms in the integrand vanish in dimensional regularization, and the last one decays as $\k^4$, which means it is convergent in $d=3$ dimensions for the $\k$ integral. Consequently,
\begin{align}
    &\pi \int_{\k,\p}   \frac{ \left[ k_0 N^{(5),(6)}(p,k) \right]_{k_0=0} }{\k^2} \ml{P} \left( \frac{1}{p_0^2 - (\p - \k)^2} \right) \delta(p^2) \nonumber \\
    &= \pi p_0 \frac{(4\pi)(2\pi)}{(2\pi)^6} \int_0^\infty \!\! d|\p| \, \p^2 \delta(p_0^2 - \p^2) \int_0^\infty \!\! d|\k| \, \k^2 \int_{-1}^1 \! du \, \ml{P} \frac{-4p_0^2}{ \k^2 (2 |\p| |\k| u - \k^2 ) } \nonumber \\
    &= - \frac{4p_0^3}{(2\pi)^3} \frac{p_0^2}{2p_0} \int_0^\infty \!\! d|\k| \int_{-1}^1 \! du \frac{1}{2 p_0 |\k|} \ml{P} \frac{1}{u - \k^2/(2 p_0 |\k|)} \nonumber \\
    &= - \frac{p_0^3}{(2\pi)^3} \int_0^\infty \frac{dk}{k} \ln \left| \frac{1 - \frac{k}{2p_0}}{1 + \frac{k}{2p_0}} \right| =  \frac{p_0^3}{(2\pi)^3} \int_0^\infty \frac{dk}{k} \ln \left| \frac{1 + k}{1 - k} \right| = \frac{p_0^3}{(2\pi)^3} \frac{\pi^2}{2} \,.
\end{align}
}

Now, { for the remaining piece,} instead of splitting the $\p$ integral into angular and radial components, we proceed using the standard QFT machinery to evaluate the $\p$ integral. We define $\tilde{p}_0 \equiv p_0 + i0^+$, and $\bar{N}(p,k) = {\rm Im}\{N_{3p}(p,k)\}$. Then we have
\begin{equation}
\begin{split}
&\left. \varrho_E^{++}(p_0) \right|_{\rm NLO}^{5-6} { - \frac{g^2 N_c (N_c^2-1)\pi^2}{2(2\pi)^3}} \\
&= (-1) g^2 N_c (N_c^2-1) \tilde{\mu}^\epsilon \int_{\k} \frac{1+2n_B(k)}{2k} \sum_{\sigma_1} {\rm Re}\left\{ \int_0^1 \diff x \int_\p \frac{i \bar{N}((p_0,\p+x\k),k) }{ ( \p^2 -  (\tilde{p}_0 - x k_0 )^2 ) } \right\} \\
 &= - g^2 N_c (N_c^2-1) \tilde{\mu}^\epsilon \frac{\Omega_{3-\epsilon}}{(4\pi)^{(3-\epsilon)/2}} \Gamma \left( \frac{-1+\epsilon}{2} \right) \int_0^\infty \frac{\diff k k^{1 - \epsilon}}{(2\pi)^{d-1}} \frac{1+2n_B(k)}{2} \\
 & \times \sum_{\sigma_1} {\rm Re}\bigg\{i \int_0^1 \diff x \bigg[ \frac{3-\epsilon}{2} A(p_0,k_0,x) D^{(1-\epsilon)/2} + \frac{-1+\epsilon}{2} B(p_0,k_0,x) D^{(-1-\epsilon)/2} \bigg] \bigg\}\,,
\end{split}
\end{equation}
where $D = - (\tilde{p}_0 - x k_0 )^2$ and 
\be
\bar{N}((p_0,\p+x\k),k) = A(p_0,k_0) \p^2 + B(p_0,k_0,x) + ({\rm terms\ linear \ in \ } \p)  \,.
\ee
Explicit inspection of the coefficients reveals that $A$ does not depend on $x$.

Next we do the integrals over the Feynman parameter $x$. We can define
\begin{align}
    I_1(p_0,k_0) &= p_0^{-3} {\rm Re}\bigg\{ i \int_0^1 \diff x \, D^{(1-\epsilon)/2} \bigg\} \\
    I_2(p_0,k_0) &= p_0^{-3} {\rm Re}\bigg\{ i \int_0^1 \diff x \, D^{(-1-\epsilon)/2} \bigg\} \\
    I_3(p_0,k_0) &= p_0^{-3} {\rm Re}\bigg\{ i \int_0^1 \diff x \, x \, D^{(-1-\epsilon)/2} \bigg\} \\
    I_4(p_0,k_0) &= p_0^{-3} {\rm Re}\bigg\{ i \int_0^1 \diff x \, x^2 \, D^{(-1-\epsilon)/2} \bigg\} \,,
\end{align}
all of which can be done analytically. If we further decompose $B(p_0,k_0,x) = B_2(p_0,k_0) + B_3(p_0,k_0) x + B_4(p_0,k_0) x^2 $, we can write the final result as
\begin{equation}
\begin{split}
&\left. \varrho_E^{++}(p_0) \right|_{\rm NLO}^{5-6} { - \frac{g^2 N_c (N_c^2-1)\pi^2}{2(2\pi)^3}} \\
&= - g^2 p_0^3 \tilde{\mu}^\epsilon \frac{N_c (N_c^2-1)}{(2\pi)^{d-1}} \frac{\Omega_{3-\epsilon}}{(4\pi)^{(3-\epsilon)/2}} \Gamma \left( \frac{-1+\epsilon}{2} \right) \int_0^\infty \diff k k^{1 - \epsilon} \frac{1+2n_B(k)}{2} \\
 & \quad \quad \quad \quad \times \sum_{\sigma_1} \bigg[ \frac{3-\epsilon}{2} A I_1 + \frac{-1+\epsilon}{2} \big( B_2 I_2 + B_3 I_3 + B_4 I_4 \big)  \bigg] \,. \label{eq:D-36}
\end{split}
\end{equation}

Now let us set $T = 0$ and study the integrand as $k \to \infty$. For notational simplicity, let
\begin{equation}
    K(k;\epsilon) =   \frac{1}{2} k^{1 - \epsilon} \sum_{\sigma_1} \bigg[ \frac{3-\epsilon}{2} A I_1 + \frac{-1+\epsilon}{2} \big( B_2 I_2 + B_3 I_3 + B_4 I_4 \big) \bigg] \,. \label{eq:K-appendix}
\end{equation}
One can then show that for $0< \epsilon < 1$
\begin{align}
    &\lim_{k \to \infty} K(k;\epsilon) k^{1+\epsilon} = - \cos \left( \frac{\pi \epsilon}{2} \right) p_0^{-\epsilon} \\
    &\lim_{k \to \infty} \left(K(k;\epsilon) + \cos \left( \frac{\pi \epsilon}{2} \right) p_0^{-\epsilon} k^{-1-\epsilon} \right) k^{1+2\epsilon} = (2 - 2\epsilon + \epsilon^2) \cos \left( \frac{\pi \epsilon}{2} \right) \\
    &\lim_{k \to \infty} \left(K(k;\epsilon) + \cos \left( \frac{\pi \epsilon}{2} \right) p_0^{-\epsilon} k^{-1-\epsilon} - (2 - 2\epsilon + \epsilon^2) \cos \left( \frac{\pi \epsilon}{2} \right) k^{-1-2\epsilon} \right) k^{2} = 0 \,,
\end{align}
effectively demonstrating that we need to extract the possible divergences as $k\to \infty$ at two different rates (but only two). In practice, we add and subtract two explicitly calculable integrals with the same degree of divergence as $K(k;\epsilon)$ so that we can take the limit $\epsilon \to 0$ before performing the integral. Concretely, we calculate
\begin{equation}
\begin{split}
    & \lim_{\epsilon \to 0} \int_0^\infty \diff k\, K(k;\epsilon) \\ &= \lim_{\epsilon \to 0} \int_0^\infty \diff k \bigg[ K(k;\epsilon) + \cos \left( \frac{\pi \epsilon}{2} \right) \frac{k^{1-\epsilon} p_0^{-\epsilon}}{k^2 + p_0^2} - (2 - 2\epsilon + \epsilon^2) \cos \left( \frac{\pi \epsilon}{2} \right) \frac{k^{1-2\epsilon}}{k^2 + p_0^2} \bigg] \\
    & \quad + \lim_{\epsilon \to 0} \int_0^\infty \diff k \bigg[ - \cos \left( \frac{\pi \epsilon}{2} \right) \frac{k^{1-\epsilon} p_0^{-\epsilon}}{k^2 + p_0^2} + (2 - 2\epsilon + \epsilon^2) \cos \left( \frac{\pi \epsilon}{2} \right) \frac{k^{1-2\epsilon}}{k^2 + p_0^2} \bigg] \,,
\end{split}
\end{equation}
so that the first term of the right-hand side is now absolutely convergent for any $0 \leq \epsilon < 1$, and we can simply take $\epsilon \to 0$ there. If there is any divergence whatsoever, it must be in the second term. We evaluate
\begin{equation}
\begin{split}
    \int_0^\infty \diff k \frac{k^{1-\epsilon}}{k^2 + p_0^2} &= \frac{(2\pi)^{2-\epsilon}}{\Omega_{2-\epsilon}} \int \frac{\diff^{2-\epsilon} k_E}{(2\pi)^{2-\epsilon}} \frac{1}{k_E^2 + p_0^2} \\
    &= \frac{(2\pi)^{2-\epsilon}}{\Omega_{2-\epsilon}} \frac{\Gamma(1-(2-\epsilon)/2 )  }{(4\pi)^{1-\epsilon/2}} (p_0^2)^{-(1-(2-\epsilon)/2)} \\
    &= \frac{(2\pi)^{2-\epsilon}}{\Omega_{2-\epsilon}} \frac{\Gamma(\epsilon/2 )  }{(4\pi)^{1-\epsilon/2}} p_0^{-\epsilon}\,,
\end{split}
\end{equation}
which shows that
\begin{equation}
    \begin{split}
        &\int_0^\infty \diff k \bigg[ - \cos \left( \frac{\pi \epsilon}{2} \right) \frac{k^{1-\epsilon} p_0^{-\epsilon}}{k^2 + p_0^2} + (2 - 2\epsilon + \epsilon^2) \cos \left( \frac{\pi \epsilon}{2} \right) \frac{k^{1-2\epsilon}}{k^2 + p_0^2} \bigg] \\
        &= \cos \left( \frac{\pi \epsilon}{2} \right) p_0^{-2\epsilon} \left[ - \frac{(2\pi)^{2-\epsilon}}{\Omega_{2-\epsilon}} \frac{\Gamma(\epsilon/2 )  }{(4\pi)^{1-\epsilon/2}} + (2 - 2\epsilon + \epsilon^2) \frac{(2\pi)^{2-2\epsilon}}{\Omega_{2-2\epsilon}} \frac{\Gamma(\epsilon )  }{(4\pi)^{1-\epsilon}} \right] \\
        &= -p_0^{-2\epsilon} + \mathcal{O}(\epsilon).
    \end{split}
\end{equation}
In a nutshell, all we get from the second term as $d \to 4$ is just $-1$. 

For the first term, since the integral is absolutely convergent, we can take the limit $\epsilon\to0$ in the integrand, and obtain
\begin{equation}
\begin{split}
    & \lim_{\epsilon \to 0} \int_0^\infty \diff k \bigg[ K(k;\epsilon) + \cos \left( \frac{\pi \epsilon}{2} \right) \frac{k^{1-\epsilon} p_0^{-\epsilon}}{k^2 + p_0^2} - (2 - 2\epsilon + \epsilon^2) \cos \left( \frac{\pi \epsilon}{2} \right) \frac{k^{1-2\epsilon}}{k^2 + p_0^2} \bigg] \\
    &= \int_0^\infty \diff k \left[ K(k;0) - \frac{k}{k^2 + p_0^2} \right] \\
    &= \frac{1}{12} \left[ -\pi^2 - 6 \left( -4 + (\ln(2))^2 - 2 {\rm Li}_2(-1/2) + {\rm Li}_2(1/4) \right) \right] { = 2 - \frac{\pi^2}{6}} \,.
\end{split}
\end{equation}

Using
\begin{equation}
    -\frac{\Omega_{3-\epsilon}}{(4\pi)^{(3-\epsilon)/2}} \Gamma \left( \frac{-1+\epsilon}{2} \right) = 1 + \mathcal{O}(\epsilon)\,,
\end{equation}
we find the full contribution at $T=0$ from the 3-propagator diagrams is
\begin{equation}
\begin{split}
    &\left. \varrho_E^{++}(p_0) \right|_{{\rm NLO}, \, T=0 }^{5-6} \\
    &= \frac{ g^2 N_c (N_c^2-1) p_0^3}{  (2\pi)^3} \left[ -1 + { 2 - \frac{\pi^2}{6} + \frac{\pi^2}{2} } \right] + \mathcal{O}(\epsilon)\, \\
    &= \frac{ g^2 N_c (N_c^2-1) p_0^3}{ (2\pi)^3} \left[ { 1 + \frac{\pi^2}{3}} \right] + \mathcal{O}(\epsilon)\,
\end{split}
\end{equation}

Since there are no UV divergences in the terms that come from purely temperature-dependent contributions, the finite $T$ contribution is obtained by taking the $\epsilon \to 0$ limit of $K(k;\epsilon)$, defined in~\eqref{eq:K-appendix}, and then plug it in~\eqref{eq:D-36}. The result is given in the main text.

\section{Time-ordered correlator in vacuum} \label{app:T-ordered-vacuum-5-5r}

{
To explore further aspects of similar-looking correlators and compare the finite vacuum constant piece of our result, which is a Wightman correlation function, with the time-ordered vacuum NLO electric field correlator calculated in Ref.~\cite{Eidemuller:1997bb}, we will show here that by taking the time-ordered version of our correlator on the Schwinger-Keldysh contour and integrating over the momentum ${\bs p}$ (which is equivalent to setting ${\bs y}={\bs x}$), we reproduce the results of Ref.~\cite{Eidemuller:1997bb}. To this end, we calculate
\begin{align}
 \big[g_E^{++}\big]_{I=J=1}(y,x) = \Big\langle \ml{T} & \big[{E}_i(y) \ml{W}_{[( y^0, {\bs y}), (+\infty, {\bs y})]} \big]^a
\big[ \ml{W}_{[(+\infty, {\bs x}),(x^0, {\bs x})]} {E}_i(x) \big]^a \Big\rangle_T \,,  
\end{align}
and compare it with
\begin{align}
\big[g_E^{++}\big]^{\ml{T}}(y,x) \equiv \theta(y^0 - x^0) \big[g_E^{++}\big]^>(y,x) + \theta(x^0 - y^0) \big[g_E^{++}\big]^<(y,x) \,,
\end{align}
which is what we have already determined from the main text.}

{
If there were no Wilson lines, then we would simply have an equality between the two, i.e., $ \big[g_E^{++}\big]_{I=J=1}(y,x) = \big[g_E^{++}\big]^{\ml{T}}(y,x)$, because the time-ordering symbol would amount exactly to reordering the electric field insertions. Therefore, perturbatively, any diagram that does not involve Wilson lines will not generate any difference. So we only need to investigate diagrams that involve gauge boson insertions from the Wilson lines. Furthermore, we will restrict ourselves to looking at the real part of the correlation functions in momentum space to see if a difference appears, because this is the contribution that would appear in the spectral function.

\subsection{2-propagator diagrams}

Following our conventions, the time-ordered correlation function as given by $I=J=1$ for the 2-propagator diagrams. After substituting the corresponding expressions for the vertex factors and propagator structures in Tables~\ref{tab:2prop-structure} and~\ref{tab:2prop-vertex}, we find
\begin{align}
\big[g_E^{++}\big]_{I=J=1}^{\rm 2-propagator}(p_0) &= g^2 N_c (N_c^2-1) \int_{\bs p} \int_k \frac{1}{k_0^2} \big[ \left((\p-\k)^2 - (d-1) p_0^2 \right) D^{\ml T}(p-k) D^{\ml T}(k) \nonumber \\ & \quad \quad \quad \quad \quad \quad \quad \quad \quad \quad \quad + \left( (d-1) p_0^2 - \p^2  \right) D^{\ml T}(p) D^{\ml T}(k)  \big] \,,
\end{align}
which by direct calculation in $d=4-\epsilon$ up to $O(\epsilon)$ terms gives (for $p_0>0$)
\begin{align}
& {\rm Re} \left\{ \big[g_E^{++}\big]_{I=J=1}(p_0) \right\} \\ 
&= \frac{\pi N_c (N_c^2 - 1)}{8} \left(\frac{\Omega_{d-1}}{(2\pi)^{d-1}} \right)^2 p_0^{2d-5} \tilde{\mu}^{4-d} \nonumber \\ & \quad \times \int_0^\infty \diff \tilde{k} \, \tilde{k}^{d-5} \left[ |\tilde{k}-1|^{d-3} \left((1-\tilde{k})^2-(d-1)\right) + |\tilde{k}+1|^{d-3} \left((1+\tilde{k})^2-(d-1)\right) + (d-2) \right] \nonumber \\
&= (N_c^2 -1) \frac{(d-2) \pi \Omega_{d-1}}{2 (4\pi^2) (2\pi)^{d-1}} g^2 p_0^{d-1}  N_c \bigg[ \frac{1}{\epsilon} + \frac{7}{12}  + \frac{1}{2}  \ln \left( \frac{\mu^2}{p_0^2} \right) + \frac{\gamma_E + \psi(3/2)}{2} \bigg] \,,
\end{align}
where $\tilde{k}$ is $k/p_0$. This result is
consistent with the finite pieces we obtained in the calculation of $\big[g_E^{++}\big]^>(y,x)$.

\subsection{3-propagator diagrams}

Here the focus will be on diagrams $(5)$ and $(5r)$, since these are the only remaining diagrams that involve Wilson lines. However, and again for convenience, we will include diagrams $(6)$ and $(6r)$ in the relevant numerator. Starting from the propagator structures and vertex factors in Tables~\ref{tab:3prop-structure} and~\ref{tab:3prop-vertex}, we obtain that in vacuum for $p_0>0$, 
\begin{align}
\big[g_E^{++}\big]_{I=J=1}^{\rm 3-propagator}(p_0) &= \frac{g^2 N_c (N_c^2-1)}{2} \nn \\ & \quad \times \int_{\bs p} \int_k \left[ \frac{1}{k_0+i0^+} + \frac{1}{k_0-i0^+} \right] \frac{\bar{N}(p,k)}{(k^2 + i0^+) ((p-k)^2 + i0^+) (p^2 + i0^+) } \,,
\end{align}
where $\bar{N} = 2p_0 \left[ (d-1)k_0 p_0 + 2\k^2 - 2\p \cdot \k + 2\p^2 - 2(d-1)p_0^2 \right]$. It is convenient for our purposes to rearrange $\bar{N}$ as follows:
\begin{align}
    \bar{N}(p,k) &= \tilde{N}(p,k) - 4 p_0 (p_0^2 - \p^2) \\
    \tilde{N}(p,k) &=  2p_0 \left[ (d-1)k_0 p_0 + 2\k^2 - 2\p \cdot \k - 2(d-2)p_0^2 \right] \,.
\end{align}
Then, the contribution from the piece of the numerator that is proportional to $p^2 = p_0^2 - \p^2$ is simply given by 
\begin{align}
    & \frac12 \int_{\bs p} \int_k \left[ \frac{1}{k_0+i0^+} + \frac{1}{k_0-i0^+} \right] \frac{-4 p_0 p^2}{(k^2 + i0^+) ((p-k)^2 + i0^+) (p^2 + i0^+) } \nn \\
    &= \int_{\bs p} \int_k \mathcal{P} \frac{1}{k_0} \frac{-4 p_0}{(k^2 + i0^+) ((p-k)^2 + i0^+)} \nn \\ &= -4p_0 \frac{\Gamma\left(\frac{3-d}{2} \right)^2}{(4\pi)^{d-1}} \int_{-\infty}^{\infty} \frac{\diff k_0}{2\pi} \frac{(-k_0^2-i0^+)^{\frac{d-3}{2}} (-(p_0-k_0)^2 - i0^+)^{\frac{d-3}{2}} }{k_0} \nn \\
    &= -4 p_0 (-i)^{2d-6} \frac{\Gamma\left(\frac{3-d}{2} \right)^2}{2\pi (4\pi)^{d-1}} \int_{0}^{\infty} \diff k_0 \, k_0^{d-4} \left[ |p_0-k_0|^{d-3} - |p_0+k_0|^{d-3} \right] \nn \\
    &{}_{{\rm DR}, \, d \to 4} = \frac{p_0^3}{(2\pi)^3} \,,
\end{align}
where in the last line we have taken the result of the previous line in dimensional regularization, and then set $d=4$.

The other contribution requires a more complicated treatment. Using Georgi and Feynman parameters, we can write
\begin{align}
   &\frac12 \int_{\bs p} \int_k \left[ \frac{1}{k_0+i0^+} + \frac{1}{k_0-i0^+} \right] \frac{\tilde{N}(p,k)}{(k^2 + i0^+) ((p-k)^2 + i0^+) (p^2 + i0^+) } \nn \\
   &= \int_{\p} \frac{1}{p^2 + i0^+}  \int_0^1 \diff x \int_0^\infty \!\! \diff \lambda \, \int_k \Bigg( \frac{\tilde{N}(p,k)}{(\lambda k_0 + (1-x)k^2 + x(p-k)^2 + i0^+)^3} \nn \\ & \quad \quad \quad \quad \quad \quad \quad \quad \quad \quad \quad \quad \quad \quad \quad \quad - \frac{\tilde{N}(p,k)}{(-\lambda k_0 + (1-x)k^2 + x(p-k)^2 + i0^+)^3} \Bigg) \,.
\end{align}
The integral over $k$ is now a textbook loop integral, and the resulting integral over $\lambda$ can be done explicitly in terms of polynomial and Hypergeometric functions.
One then obtains
\begin{align}
   &\frac12 \int_{\bs p} \int_k \left[ \frac{1}{k_0+i0^+} + \frac{1}{k_0-i0^+} \right] \frac{\tilde{N}(p,k)}{(k^2 + i0^+) ((p-k)^2 + i0^+) (p^2 + i0^+) } \nn \\
   &= \int_{\p} \frac{1}{p^2 + i0^+}  \int_0^1 \diff x \Bigg[ \frac{i p_0 (1-d)}{(4\pi)^{d/2}} \Gamma \! \left( \frac{4-d}{2} \right) 4 x p_0 \frac{{}_2F_1 \! \left( \frac12, \frac{4-d}2,\frac32, \frac{-x^2 p_0^2}{x(1-x) \p^2-xp_0^2-i0^+} \right) }{(x(1-x)\p^2 - xp_0^2 - i0^+)^{2-d/2}} \nn \\
   & \quad \quad \quad \quad \quad \quad \quad \quad + \frac{i p_0^2 (d-1) }{2 (4\pi)^{d/2}} \Gamma \! \left( \frac{4-d}{2} \right) \frac{4}{(x(1-x)\p^2 - xp_0^2 - i0^+)^{2-d/2} } \nn \\
   & \quad \quad \quad \quad \quad \quad \quad \quad + \frac{i p_0^2 (d-1) }{2 (4\pi)^{d/2}} \Gamma \! \left( \frac{6-d}{2} \right) 8 x^2 p_0^2 \frac{{}_2F_1 \! \left( \frac12, \frac{6-d}2,\frac32, \frac{-x^2 p_0^2}{x(1-x) \p^2-xp_0^2-i0^+} \right) }{(x(1-x)\p^2 - xp_0^2 - i0^+)^{3-d/2}} \nn \\ 
   & \quad \quad \quad \quad \quad \quad \quad \quad + \frac{(-i)}{2(4\pi)^{d/2}} \Gamma \! \left( \frac{6-d}{2} \right) 4 x p_0 \left[ \left( 2(d-1)(x-2)+4 \right) p_0^3 - 4x(1-x) p_0 \p^2 \right] \nn \\ &  \quad \quad \quad \quad \quad \quad \quad \quad  \quad \quad \quad \quad \quad \quad \quad \quad  \quad \quad \quad \quad \quad \quad \times \frac{{}_2F_1 \! \left( \frac12, \frac{6-d}2,\frac32, \frac{-x^2 p_0^2}{x(1-x) \p^2-xp_0^2-i0^+} \right) }{(x(1-x)\p^2 - xp_0^2 - i0^+)^{3-d/2}} \Bigg] \,.
\end{align}
Using the Pfaff transformations of the Hypergeometric functions, and taking $d \to 4$, we arrive at
\begin{align}
   &\frac12 \int_{\bs p} \int_k \left[ \frac{1}{k_0+i0^+} + \frac{1}{k_0-i0^+} \right] \frac{\tilde{N}(p,k)}{(k^2 + i0^+) ((p-k)^2 + i0^+) (p^2 + i0^+) } \nn \\
   &= \int_{\p} \frac{i}{p^2 + i0^+}  \int_0^1 \diff x \frac{1}{(4\pi)^2} \bigg\{ -12 x p_0^2 \frac{\partial}{\partial a} \left[ {}_2F_1 \! \left( a,1,\frac32,\frac{-x^2 p_0^2}{x(1-x)(p^2+i0^+)} \right)  \right]_{a=0} \nn \\
   & \quad \quad \quad \quad \quad \quad \quad \quad  \quad \quad - \frac{16 x p_0^4 + 8 p_0^2 \p^2 x^2 (1-x)}{x(1-x) (p^2 + i0^+) } {}_2F_1 \! \left( 1,1,\frac32,\frac{-x^2 p_0^2}{x(1-x)(p^2+i0^+)} \right) \bigg\} \nn \\
   &= \frac{1}{16 \pi^4} \int_0^\infty \!\! \diff |\p| \frac{i |\p|^2 p_0^2 }{p_0^2 - |\p|^2 + i0^+} \nn \\
   & \quad \quad \quad \quad \quad \quad   \times \Bigg[ 3 \int_0^1 \!\! \diff x \int_0^1 \!\! \diff y \frac{x}{\sqrt{1-y}} \ln \left( \frac{x(1-x) (p_0^2 - |\p|^2 + i0^+) + y x^2 p_0^2 }{x(1-x)(p_0^2 - |\p|^2 + i0^+)} \right) \nn \\
   & \quad \quad \quad \quad \quad \quad \quad \quad \quad  - 2  \int_0^1 \!\! \diff x \int_0^1 \!\! \diff y \frac{x}{\sqrt{1-y}} \frac{2 p_0^2 + x(1-x) |\p|^2}{x(1-x) (p_0^2 - |\p|^2 + i0^+) + y x^2 p_0^2 } \Bigg] \,,
\end{align}
where in the last line we have used an integral representation of the hypergeometric function:
\begin{align}
    {}_2F_1 \! \left(a,1,\frac32,D \right) = \frac12 \int_0^1 \frac{\diff y}{\sqrt{1-y}} (1 - Dy)^{-a} \,.
\end{align}

Now we take the real part of this expression, as it is all we need to compare with our results from the Wightman correlations. The remaining integrals in $|{\bs p}|$ can then be carried out by complex contour integration, yielding
\begin{align}
    &{\rm Re} \left\{ \int_0^\infty  \!\! \diff |\p| \frac{i |\p|^2}{p_0^2-|\p|^2+i0^+} \frac{2p_0^2 + x(1-x) |\p|^2 }{x(1-x)(p_0^2 - |\p|^2 + i0^+) + y x^2 p_0^2} \right\} \nn \\
    &\quad \quad = \frac{\pi p_0}{2} \left( \frac{2+x(1-x)}{yx^2} - \sqrt{\frac{1-x+yx}{1-x}} \frac{2+x(1-x)+yx^2}{yx^2} \right) \\
    &{\rm Re} \left\{ \int_0^\infty  \!\! \diff |\p| \frac{i |\p|^2}{p_0^2-|\p|^2+i0^+} \ln \left( - \frac{x(1-x) (p_0^2 - |\p|^2 + i0^+) + y x^2 p_0^2 }{p_0^2} \right) \right\} \nn \\
    &\quad \quad= \frac{\pi p_0}{2} \left( \ln \left| y x^2 \frac{\sqrt{\frac{1-x+yx}{1-x}}+1}{\sqrt{\frac{1-x+yx}{1-x}}-1} \right| - 2 \sqrt{\frac{1-x+yx}{1-x}} \right) \\
    &{\rm Re} \left\{ \int_0^\infty  \!\! \diff |\p| \frac{i |\p|^2}{p_0^2-|\p|^2+i0^+} \ln \left( - \frac{x(1-x) (p_0^2 - |\p|^2 + i0^+) }{p_0^2} \right) \right\} \nn \\
    &\quad \quad= \frac{\pi p_0}{2} \left( \ln \left|4 x (1-x) \right| - 2 \right) \,,
\end{align}
where we have split the logarithmic term into two pieces and integrate them separately.
Putting everything together, we find that
\begin{align}
    & {\rm Re} \left\{ \big[g_E^{++}\big]_{I=J=1}^{\rm 3-propagator}(p_0) \right\} \nn \\ &= g^2 N_c (N_c^2 - 1) \frac{p_0^3}{(2\pi)^3} \Bigg[ 1 + \frac{1}{2} \int_0^1 \!\!\! \diff x \int_0^1 \!\!\! \diff y \frac{x}{\sqrt{1-y}} \nn \\
    & \quad \quad \quad \quad \quad \quad \quad \quad   \quad \quad \quad \quad  \times \bigg( 3 - 3 \sqrt{\frac{1-x+yx}{1-x}}  + \frac32 \ln \left| \frac{y x}{1-x} \frac{\sqrt{\frac{1-x+yx}{1-x}}+1}{\sqrt{\frac{1-x+yx}{1-x}}-1} \right| \nn \\ 
    & \quad \quad \quad \quad \quad \quad \quad \quad  \quad \quad \quad \quad \quad   + \sqrt{\frac{1-x+yx}{1-x}} \frac{2+x(1-x)+yx^2}{yx^2} - \frac{2+x(1-x)}{yx^2} \bigg) \Bigg] \nn \\
    &= g^2 N_c (N_c^2 - 1) \frac{p_0^3}{(2\pi)^3} \Bigg[ 1 + \frac{\pi^2}{3} \Bigg] \,.
\end{align}
This reproduces the result of Ref.~\cite{Eidemuller:1997bb}, and that of our main text.

\subsection{Conversion of previous results in position space to momentum space}

To reassure the reader that the result of the previous section matches that of Ref.~\cite{Eidemuller:1997bb}, we convert their results into momentum space. In their work, they calculated
\begin{align}
\mathcal{D}_{\mu\nu\rho\sigma}(z) &= \left\langle 0 \left| \ml{T} \left( F_{\mu \nu}^a(z)  \ml{W}_{[z,0]}^{ab} F_{\rho \sigma}^b(0) \right)  \right| 0 \right\rangle \nn \\
& = \left[ g_{\mu \rho} g_{\nu \sigma} - g_{\mu \sigma} g_{\nu \rho} \right] \left( D(z^2) + D_1(z^2) \right) \nn \\ & \quad + \left[ g_{\mu \rho} z_{\nu} z_{\sigma} - g_{\mu \sigma} z_{\nu} z_{\rho} - g_{\nu \rho} z_{\mu} z_{\sigma} + g_{\nu \sigma} z_{\mu} z_{\rho} \right] \frac{\partial D_1}{\partial z^2} \,,
\end{align}
for an arbitrary spacetime separation $z^{\mu}$. Up to NLO, Ref.~\cite{Eidemuller:1997bb} found
\begin{align}
    D(z^2) &= \frac{N_c^2-1}{\pi^2 z^4} \left[ \frac{\alpha N_c}{\pi} \left(-\frac{1}{4} L + \frac38 \right) \right] \\
    D_1(z^2) &= \frac{N_c^2-1}{\pi^2 z^4} \left[ 1 + \frac{\alpha N_c}{\pi} \left( \left( \frac{\beta_1}{2 N_c} - \frac14 \right) L + \frac{\beta_1}{3N_c} + \frac{29}{24} + \frac{\pi^2}{3} \right) \right] \,,
\end{align}
where $L = \ln (e^{2\gamma_E} \mu^2 z^2/4 )$, $\beta_1 = (11 N_c - 4 N_f)/6$, and $\mu$ is the $\overline{\rm MS}$ renormalization scale.

Our correlator of interest is $\mathcal{D}_{0i0i}(z_0,{\bs z}=0)$. After some algebra, one arrives at
\begin{align}
   \mathcal{D}_{0i0i}(z_0) = 3\frac{N_c^2-1}{\pi^2 z_0^4} \left[ 1 + \frac{g^2}{4\pi^2}\left( \left(  \frac{11}{12} N_c - \frac{1}{3} N_f \right) L + N_c \left( \frac79 + \frac{\pi^2}{3} \right) + \frac19 N_f \right) \right] \,,
\end{align}
where now $L = \ln (e^{2\gamma_E} \mu^2 z_0^2/4 )$. One can then use the Fourier transforms
\begin{align}
    \int_{-\infty}^\infty \diff t\, e^{i p_0 t} \frac{1}{t^4} &= \frac{\pi}{6} p_0^3 \,{\rm sgn}(p_0) \\
    \int_{-\infty}^\infty \diff t\, e^{i p_0 t} \frac{\ln(t^2)}{t^4} &= -\frac{\pi}{18} p_0^3 \,{\rm sgn}(p_0) \left( -11 + 6 \gamma_E + 6 \ln |p_0| \right) \,,
\end{align}
which, upon substitution into our correlator of interest, yield (for $p_0>0$):
\begin{align}
   \mathcal{D}_{0i0i}(p_0) = \frac{ N_c^2-1}{(2\pi)^3} p_0^3 \left[ 4\pi^2 + g^2 \! \left( \! \left( \frac{11}{12} N_c - \frac{1}{3} N_f \right) \! \ln \! \left( \frac{\mu^2}{4 p_0^2}\right) + \left( \frac{149}{36} + \frac{\pi^2}{3} \right) \! N_c  - \frac{10}{9} N_f \right)  \right] \,,
\end{align}   
which exactly matches our vacuum result in~\eqref{eq:spectral-full-result}.
}

\bibliographystyle{jhep}
\bibliography{main.bib}
\end{document}